\documentclass[]{aa}  

\usepackage{adjustbox}
\usepackage{threeparttable}
\usepackage{graphicx,url,twoopt,natbib}
\usepackage[varg]{txfonts}           
\usepackage{hyperref}                

\hypersetup{
  colorlinks=true,   
  urlcolor=blue,     
  linkcolor=blue     
}

\bibpunct{(}{)}{;}{a}{}{,}    
\makeatletter
\newcommand{\bibnote}[2]{\global\@namedef{#1note}{#2}}
\newcommand{\biblink}[2]{\global\@namedef{#1link}{#2}}
\makeatother

\usepackage{xcolor}
\usepackage{graphicx}
\usepackage{xspace}
\usepackage{cancel}
\usepackage{dblfloatfix}

\newcommand{\xmm}{\textsl{XMM-Newton}\xspace}

\newcommand{\nustar}{\textsl{NuSTAR}\xspace}

\usepackage{txfonts}
\begin{document}

   \title{Common patterns in pulse profiles of High Mass X-ray Binaries}
   \titlerunning{Common patterns in pulse profiles of HMXBs}  

   \author{J.~Alonso-Hernández
          \inst{1}
          \and
          F.~F\"urst\inst{2}
          \and
          P.~Kretschmar\inst{3}
          \and 
          Isabel Caballero \inst{4}
          \and
          A. M. Joyce \inst{5}
          }

   \institute{Departamento de Física y Astronomía, Universidad Complutense de Madrid, Avenida de Séneca 2, Ciudad Universitaria, 28040 Madrid, Spain
         \and
             Quasar Science Resources S.L for European Space Agency (ESA), European Space Astronomy Centre (ESAC), Camino Bajo del Castillo s/n, 28692 Villanueva de la Cañada, Madrid, Spain
         \and 
            European Space Agency (ESA), European Space Astronomy Centre (ESAC), Camino Bajo del Castillo s/n, 28692 Villanueva de la Cañada, Madrid, Spain   
        \and
            Aurora Technology B.V. for European Space Agency, European Space Astronomy Centre (ESA/ESAC),  Madrid, Spain
        \and
            Dr. Karl Remeis-Observatory \& Erlangen Centre for Astroparticle Physics, Sternwartstr. 7, 96049 Bamberg, Germany
             }

   \date{Received XXX; accepted XXX}

 
  \abstract
   {X-ray pulsars are binary systems which consist of a neutron star in orbit with a mass donor (companion). In these systems the neutron star accretes matter from the companion star, which creates accretion columns or hot spots on the neutron star surface and gives rise to pulsations in the X-ray light-curve. The pulse profiles carry information about the accretion and magnetic field geometry. Here we present a study and classification of energy resolved pulse profiles of a sample of X-ray pulsars, focusing on high-mass X-ray binaries.}
   {Our goal is to perform a classification of X-ray pulsars based on their observed pulse profiles and look for correlations between this classification and their principle physical observables. The analysis pipeline is available online. }
   {We analysed the pulse profiles of a sample of X-ray pulsars using data obtained with the X-ray Multi-Mirror Mission (\xmm) and the Nuclear Spectroscopic Telescope Array (\nustar). 
   We fit the energy resolved pulse profiles with a Fourier series of up to five harmonics. We then use the energy evolution of the different Fourier components to classify the pulse profiles into groups. We investigate relationships between the pulse profile properties and other observables of the systems (e.g., orbital period, magnetic field strength, and luminosity) to study the extreme physics of these systems.}
   {The sources were divided into three groups by a classification based on the shape, the dominance of the fitted Fourier harmonics and their respective evolution with energy. We do not find a conclusive correlation between the pulse profile shapes or groups and other parameters of the systems. 
However, a weak trend was found when comparing our classification to the sources' location in the spin period-orbital period diagram. Further studies are required to confirm this trend.
   }
   {Despite the large variety of pulse profiles of the X-ray pulsars, we found that with our approach clear categories emerge which we use to classify their behavior as function of energy. As we do not find a clear relationship between our classification scheme and other parameters, like the luminosity, the magnetic field strength, or the orbital and spin periods, we conclude that X-ray pulse profiles are influenced by other hidden variables.}

   
   \maketitle
%

\section{Introduction}\label{sec:intro}

Accreting X-ray pulsars are X-ray binary systems in which a highly magnetized neutron star, with a magnetic field strength typically between $10^{11-13}$\,G, accretes matter from its companion star. The mass donors are mainly stars with masses $>10 M_\sun$ \citep{liu_2006}, making the systems High Mass X-ray Binaries (HMXBs), but some also exist in low- or intermediate-mass X-ray binaries (LMXB, IMXB) with companion star of at a few solar masses or less \citep{liu_2007}. 

The mass transfer can take place in different ways, depending on the nature of the mass donor and the size of the binary orbits. For lower mass systems and some HMXBs with neutron stars in very tight orbits, mass transfer will take place via Roche-lobe overflow \citep{campana+disalvo_2018,brumback2020}. In a large group of HMXBs the mass donor will be an early type star losing copious amounts of matter through line-driven winds. Neutron stars in close orbits around their companions can accrete efficiently from this wind and become bright X-ray sources. For a recent review of these systems see \citet{martinez-nunez2017a}. Some systems may mix Roche-lobe overflow and wind accretion \citep{el_mellah_2019}. Finally, there is the sub-class of ``Be X-ray Binaries'' (BeXRBs) in which the companion is a Be or Oe type star, i.e., an early-type star with a large circumstellar disk \citep{nixon+pringle_2020}. The neutron star in these systems can be on a wide, eccentric orbit. When the neutron star is close enough to the mass donor and other conditions are met, X-ray outbursts can occur \citep{negueruela+okazaki_2001,okazaki+negueruela_2001}. In many systems, both regular weaker outbursts and irregular giant outbursts have been observed. For an overview of BeXRBs and their behaviour see \citet{reig2011}.

Regardless of the mode of accretion, at a certain distance from the neutron star -- typically two to three orders of magnitude larger than the neutron star radius \citep[see][for a recent overview of interactions at the magnetosphere]{kretschmar_2019b} -- the magnetic field will begin to dominate the movement of the accreted plasma, which will then follow the field lines towards the magnetic poles. Assuming a dipolar magnetic field, two accretion columns at the magnetic poles of the neutron star will be produced. If there is a misalignment between the magnetic and the rotation axis the neutron star, we will observe the system as an X-ray pulsar, as the X-ray emitting region rotates in and out of our line of sight \citep{pringle+rees_1972}. 

In the simplest picture of a dipolar magnetic field with axial symmetric emission pattern, we would expect pulse profiles with almost sinusoidal shapes, and such profiles can be found in accreting millisecond pulsars \citep{poutanen_2009}, which have lower magnetic fields. But for the majority of accreting X-ray pulsars the profiles are more complex,  frequently showing multi-peaked patterns, especially at low energies and often exhibiting a strong energy dependence \citep[e.g.][for early examples]{kanno_1980}.
Typically, each source shows its own unique behavior, making the pulse profiles ``fingerprints'' of their sources, stable over long periods \citep[see, e.g., Fig. 9 in][]{kretschmar_2021X}. But there are also cases of significant variations in observed profiles of transient sources during an outburst, where the pulse profiles can dramatically change their looks and behavior \citep[see, e.g.,][]{brumback2021,camero-arranz_2007,cappallo_2019}.

The changes in profile shapes with energy indicate corresponding changes in the emission geometry of the accretion columns. Studying this energy dependence contains information about the physical processes inside the accretion column, like Comptonization, Bremsstrahlung and cyclotron radiation \citep{becker2005a,becker2007, wolff2016a}. For disk-fed pulsars there will also be reprocessing of soft X-rays in the large accretion disk \citep{hickox2004, hickox_2005, brumback2020, brumback2021}. In addition, emission lines of heavy elements such as iron, produced by fluorescence stimulated in the accretion disk or in the circumstellar material, can produce significant effects at the corresponding line energies in the spectra and pulse profiles.

Attempts to relate theoretically predicted beam patterns to pulse profiles of accreting X-ray pulsars have been undertaken since the early days of X-ray astronomy \citep[e.g.][]{basko+sunyaev_1975,kanno_1980,Wang+Welter_1981}. Over time, the critical importance of including gravitational light bending effects \citep{riffert+meszaros_1988,riffert_1993} and allowing for asymmetrical emission \citep{leahy_1991,kraus_1995} in such model efforts has been demonstrated. Such approaches have been used by various authors to try and model the emission geometry for specific sources, \citep[e.g.,][]{bulik_1995,leahy_2004,caballero_2011,sasaki_2012,iwakiri2019a}, frequently limited to specific observations or energy ranges. 

As described in detail in \citet{falkner2018a}, a full description of the profiles based on the physical properties of the neutron star and the conditions within the accretion column is extremely complex. Multiple non-linear effects have to be taken into account: plasma density and temperature, energy- and magnetic-field dependent cross-sections in radiation transport, relativistic beaming, and gravitational light bending.
Since, despite all efforts, the a priori constraints on these effects are very limited, large degeneracies between different parameters influencing the profiles can appear, leading to multiple, geometrically different solutions \citep[e.g.,][]{caballero_2011}.
The picture is somewhat different for sources where the emission can be described by surface hotspots without significant columns, for example in accreting millisecond pulsars. In this case the accurate analytical approximation to the Schwarzschild metric at $R\ge 2r_\mathrm{g}$ derived by \citet{beloborodov_2002} can be used \citep{poutanen+beloborodov_2006}, as demonstrated by the recent Neutron star Interior Composition Explorer (NICER) results \citep{riley2019}.

A different and complementary approach to the characterisation of pulse profiles in accreting X-ray pulsars is a decomposition of the overall profile into simple, possibly energy-dependent components without any a priori connection to the complex emission and radiation transport physics. While such an approach can not yield direct insights into the underlying physics, it avoids any entanglement in possibly degenerate parameter dependencies and is conceptually not that different from the widespread approach to describe broadband spectra of X-ray sources with simple phenomenological models. Such an approach can then also be used in an easy and automatic manner across multiple sources, focusing on broad similarities and differences between their pulse profiles.

As has been demonstrated before, e.g., by \citet{finger1999} or \citet{camero-arranz_2007}, describing pulse profiles based on their sinusoids or Fourier frequencies provides a fairly precise fit of the pulse profile and its precision mainly depends on the number of harmonics of the fit.
In this paper, we describe a framework to study the pulse profiles of different X-ray pulsars in a range of energies with the smallest energy steps possible, via a simple combination of sinusoids. Based on the properties of the components we classify the sources into different groups and investigate any dependence of the pulse profile characteristics with other properties of the system, like orbital period, luminosity, or the magnetic field of the neutron star. 

We apply the framework to a small initial catalog of a few sources. As the analysis software is public, this might be used as a basis for future observations and research related to this topic. We hope that the procedure and broad characterisation provided in this way may serve to reduce the parameter space for any physical modelling effort, by highlighting common trends and how they may be related to other physical properties.

The remainder of this paper is structured as follows: in Sect.~\ref{sec:analysis} we explain the selection of the observations, the data analysis for timing corrections, the pulse period search, how the observed pulse profiles are obtained, and the fitting process. In addition we give an overview of the pipeline framework and explain the pipeline outputs. In Sect.~\ref{sec:results} we present the results give possible explanations for the emerging grouping of our source sample. Finally, in Sect.~\ref{sec:conclusions} we conclude the paper and provide an outlook for future work.

\begin{table*}

\renewcommand{\arraystretch}{1.2}
\centering

\caption{Observations of X-ray pulsars used in this study. The columns shows the name of the observed source, the identification number of the observation and the date of the observation in Modified Julian Date and in the time reference system of each telescope. The sources are ordered by their spin periods ($P_\text{spin}$), which are shown in Table~\ref{tab:parameters}.}
\label{tab:observations}


\begin{tabular}{l c c c c c c c c}
\hline\hline 
Source & \multicolumn{2}{c}{  OBS ID} &  \multicolumn{2}{c}{Date of observation (MJD)} 	   \\  
 & \xmm & \nustar & \xmm & \nustar  \\
\hline
SMC~X-1 & 0784570201 & 30202004002 &  57639.930 & 57639.899  \\
4U~1901+03 & & 90501324002 & & 58615.752   \\
           & & 90502307002 & & 58549.308  \\
           & & 90502307004 & & 58584.946\\
           & & 90501305001 & & 58531.121   \\
V~0332+53   & & 80102002002 & & 57223.415 \\
           & & 80102002004 & & 57275.946\\
           & & 80102002010 & & 57299.990\\
Cen~X-3 & 0400550201 & 30101055002 & 53898.938	& 57356.758   \\
LMC~X-4 	& 0771180101 & 30102041002 & 57325.124 & 57325.048  \\
GRO J2058$+$42	& & 90501313004   & &  58584.008    \\
Vela~X-1 &0841890201 & 30501003002 &  58606.925	& 58606.868   \\
1E1145.1$-$6141 & 0841000101	&30501002002 & 58687.528 & 58687.471  \\
IGR~J17252$-$3616 & 0405640901 & &  54006.609  \\
GX~301-2 & 0555200401 & 30101042002 & 55024.104 & 57299.333	  \\
IGR~J16393$-$4643 & & 30001008002 & & 56834.113	  \\
\hline
\end{tabular} 
\renewcommand{\arraystretch}{1.0}
\end{table*}

\section{Data analysis and scripts}\label{sec:analysis}

To simply and automatize the comparison between the pulse profile of a sample of sources, a pipeline was designed to obtain the normalized pulse profile resolved at different energy ranges\footnote{The pipeline is publicly available in a Github repository at \url{https://github.com/piexpiex/X-ray-pulsars-Pulse-Profile}.}. The pipeline uses the Astropy \citep{astropy_2013} and Stingray \citep{huppenkothen_2019} Python packages.

\subsection{Observations}

Our analysis is based on data taken with \xmm and \nustar, two of the most sensitive telescopes in their respective energy bands.

\xmm \citep[X-ray Multi-Mirror Mission,][]{jansen2001} is an X-ray telescope of the European Space Agency (ESA). \xmm data used in this project was obtained from the EPIC-pn camera \citep{pnref}, which provides coverage in the energy band from 0.15\,keV to 12\,keV, with a spectral resolution of $\sim$ 80\,eV. Most data used here was taken in Timing or Burst mode, which provide a  timing resolution of 0.03\,ms and 0.007\,ms, respectively. The only exceptions are observations of LMC~X-4 (ObsID 0771180101) and IGR\,J17252$-$3616 (ObsID 0405640901), which were taken in Small Window mode.
\xmm data were reduced with the standard software SAS v18.0 and screened for background flaring. Given the brightness of the sources, we did not find the need to filter out any flaring periods. We carefully checked for pile-up and found that this also was not an issue, in particular because we are only concerned with wide energy bands. For the timing mode data, we extracted the source light curve from an 8\,column wide region centered on the brightest column. No background light curves were extracted, as the source photon dominate the whole active CCD.  For the imaging mode data we used a circular region with a radius of $30''$ centered on the brightest pixel to extract the source data.

\nustar \citep[Nuclear Spectroscopic Telescope Array,][]{harrison2013} is the first focusing high energy X-ray telescope. \nustar consists of two co-aligned telescopes with a 3--79\,keV energy band bandbass. The spectral resolution (FWHM) varies from 400\,eV at 10\,keV to 900\,eV at 68\,keV with a timing resolution of 2\,$\mu$s. The \nustar data were reduced following the standard procedure using \texttt{nustardas v1.8.0} with CALDB~v20201217. Source photons were extracted from circular regions centered on the brightest pixel with a typical radius between 90$''$--120$''$, depending on the source brightness. We did not extract background photons, as the sources are all orders of magnitudes brighter than the background.

All observations used are listed in Table~\ref{tab:observations}. We have mainly selected observations of targets observed simultaneously between \xmm and \nustar to allow us to study the pulse profile evolution over a wider energy range. Additionally we concentrated on targets with a high signal-to-noise ratio to obtain the best possible results. The sample is far from complete and can be easily extended with the pipeline presented here. For this paper we concentrate on the small sample as a proof of concept.

\subsection{Obtaining and fitting pulse profiles}
   
Pulse profiles are usually presented in units of count-rate, i.e., counts per second. However, because we want to compare different sources with largely different count-rates and are mainly interested in the changing shape as function of energy, we proceeded to perform a normalization of the pulse profile (P($\Phi$)) by subtracting its mean value ($\left< P \right>$) and dividing by the measured standard deviation ($\sigma$), obtaining the following equation for the normalized pulse profile ($P^{*}$):
   
\begin{equation} \label{eq:1}
    P^{*}(\Phi) = \dfrac{ P(\Phi) - \left<P\right> }{\sigma}
\end{equation}
   
Due to the phase periodicity of the pulse profiles, we described each profile by its Fourier series, in this case, the Fourier terms are expressed as harmonics and the fit is resolved by the Levenberg-Marquardt algorithm.
   
\begin{equation} \label{eq:2}
    P'(\Phi) = \sum^{n}_{j} A_{j} \cdot\sin( j\cdot\Phi + F_{j} )
\end{equation}
   
Where $P'(\Phi)$ is the fitted pulse profile, $n$ is the number of terms in the Fourier series, $A_{j}$ is the $j$th harmonic amplitude and $ F_{j}$ is the initial phase gap for the respective harmonic. In our case, five harmonics were used for the fits owing to the fact that allowed us to fit correctly the pulse profiles without a high number of harmonics. However, the harmonics with larger amplitudes, and therefore the most important for the fit, were the first three harmonics.

\begin{figure*}[b!]%
    \centering
    \includegraphics[width=0.47\linewidth]{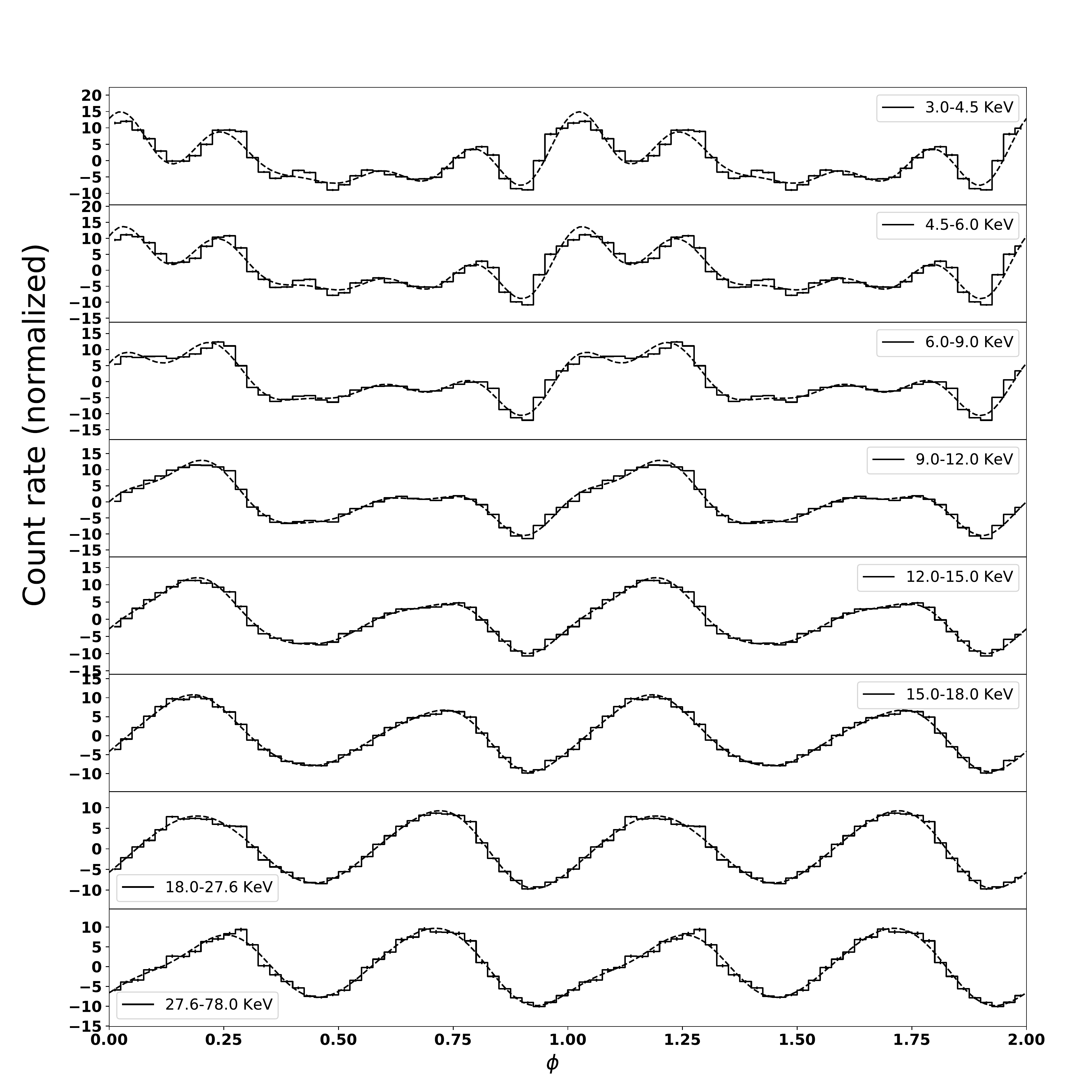} %
    \qquad
    \includegraphics[width=0.47\linewidth]{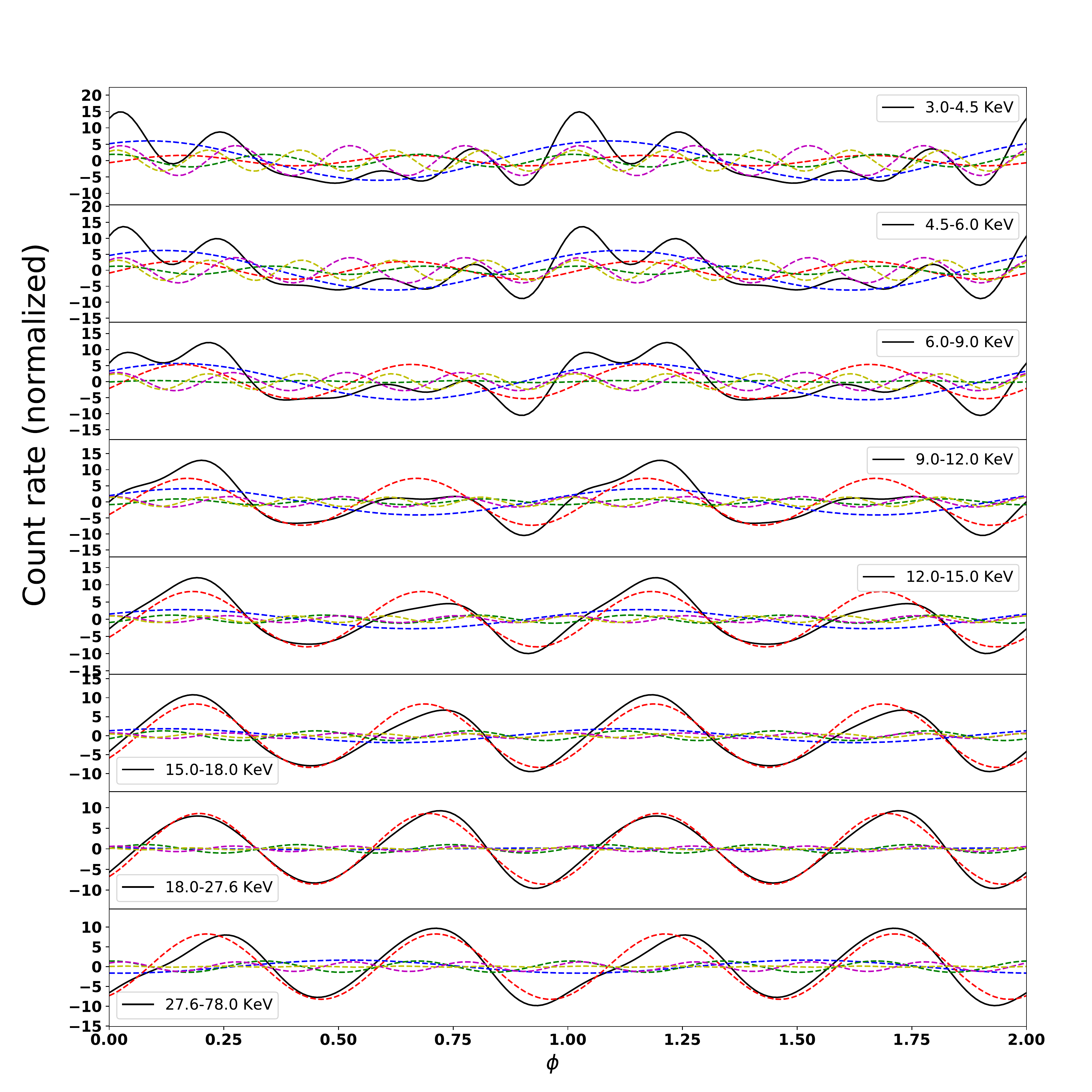} 
    \caption{Left: Normalized pulse profiles of Vela~X-1 at different energy ranges based on \nustar data, the measured profiles are represented by solid lines whereas the Fourier fits are represented by dashed lines, the boundaries of the energy ranges are [3, 4.5, 6, 9, 12, 15, 18, 27.6 and 78] keV. Right: The black solid line represents the overall approximate pulse profile and the colour dashed lines represent the different Fourier components. Blue: 1st, red: 2nd, green: 3rd, purple: 4th and yellow: 5th.}%
    \label{fig:pulse_profiles}%
\end{figure*}

\begin{figure*}[b!]
	\centering
	\includegraphics[width=\linewidth]{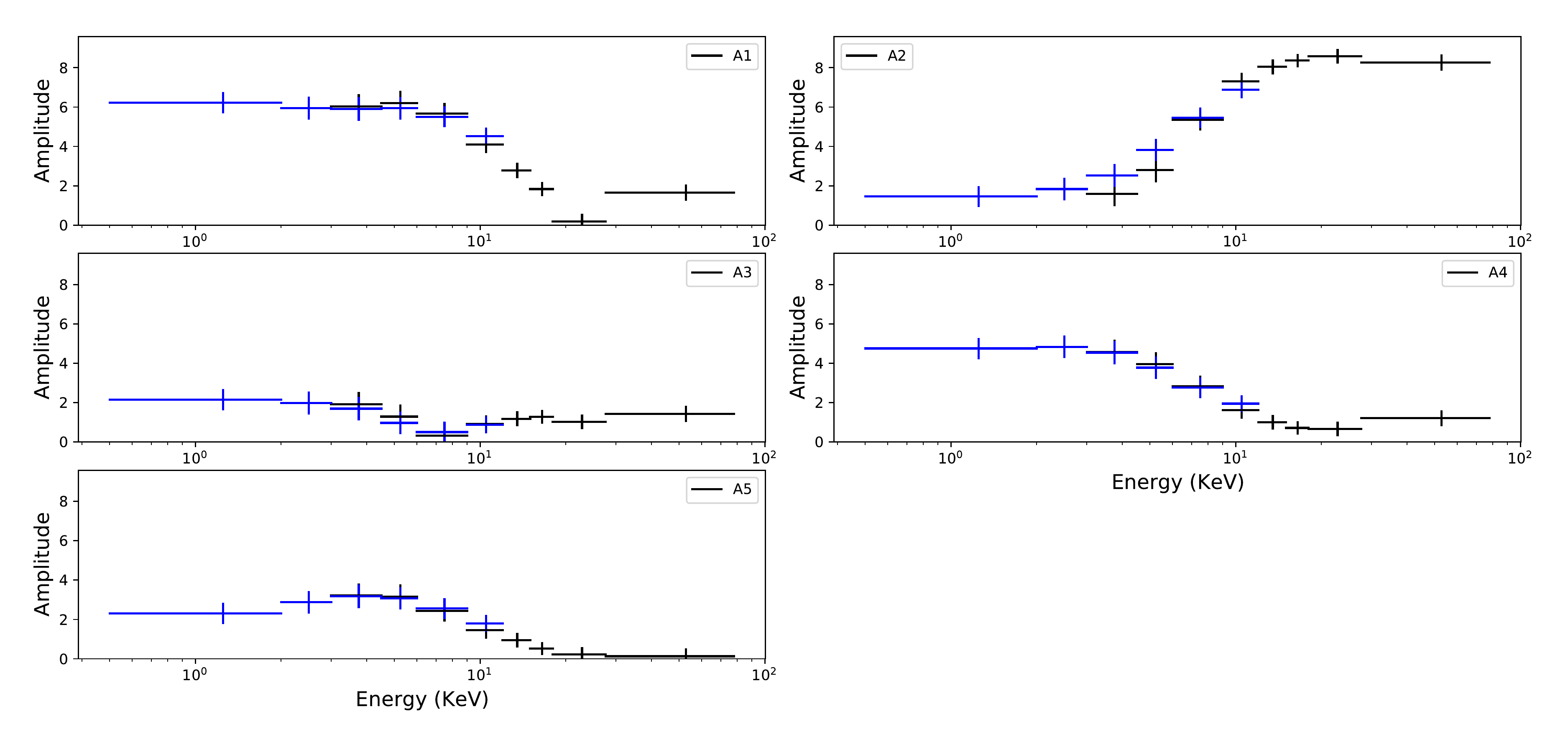}
	\caption{Variation of the harmonic amplitudes for Vela~X-1, the boundaries of the energy ranges are [0.5, 2, 3, 4.5, 6, 9, 12, 15, 18, 27.6 and 78] keV. \xmm data is represented by blue markers and \nustar data is represented by black markers.}
	\label{fig:fourier_variation}
\end{figure*}

Our pipeline performs the following steps for data reduction and extraction:

\begin{enumerate}
    \item \textbf{Timing corrections:} selection of the Good Time Intervals (GTI), definition of a common start time $T_0$ for each dataset, and the correction of the arrival times for each photon to the Binary System barycenter using the correspondening ephemeris.
    \item \textbf{Search of the pulse period} we use the epoch folding statistic around the expected pulse period \citep{leahy_1983}, we selected each pulse period as the maximum value of the epoch folding distribution, its uncertainty is estimated as the half of the Full Width at Half Maximum ($\mathrm{FWHM/2}$) of a Lorentzian curve fitted to the distribution. In addition, the $Z^{2}$ statistic is used to check the obtained value. We did not consider pulse period derivates due to the short periods of the observations and the aims of this study.
    \item \textbf{Energy resolved pulse profiles} we bin the photons according to their energy into $N$ energy bands and calculated a pulse profile for each energy band, based on the pulse period obtained in the previous step. 
    \item The pulse profiles are \textbf{normalized} as in Eq. \ref{eq:1} and \textbf{fitted by Fourier series} as in Eq. \ref{eq:2} (Fig.~\ref{fig:pulse_profiles}).
    \item \textbf{Extract parameter}: we extract the amplitude for each harmonic of the Fourier series as function of energy, to study the energy dependence and compare sources with each other (Fig.~\ref{fig:fourier_variation}).
\end{enumerate}

Fig. \ref{fig:pulse_profiles} shows an example of the energy resolved pulse profiles, in this case the pulse profiles correspond to the \nustar observation of Vela X-1. There is a change in the shape of the pulse profile with the energy and that produces a drastic change in the amplitudes of the harmonics.

\begin{figure*}[b!]
	\centering
	\includegraphics[width=\linewidth]{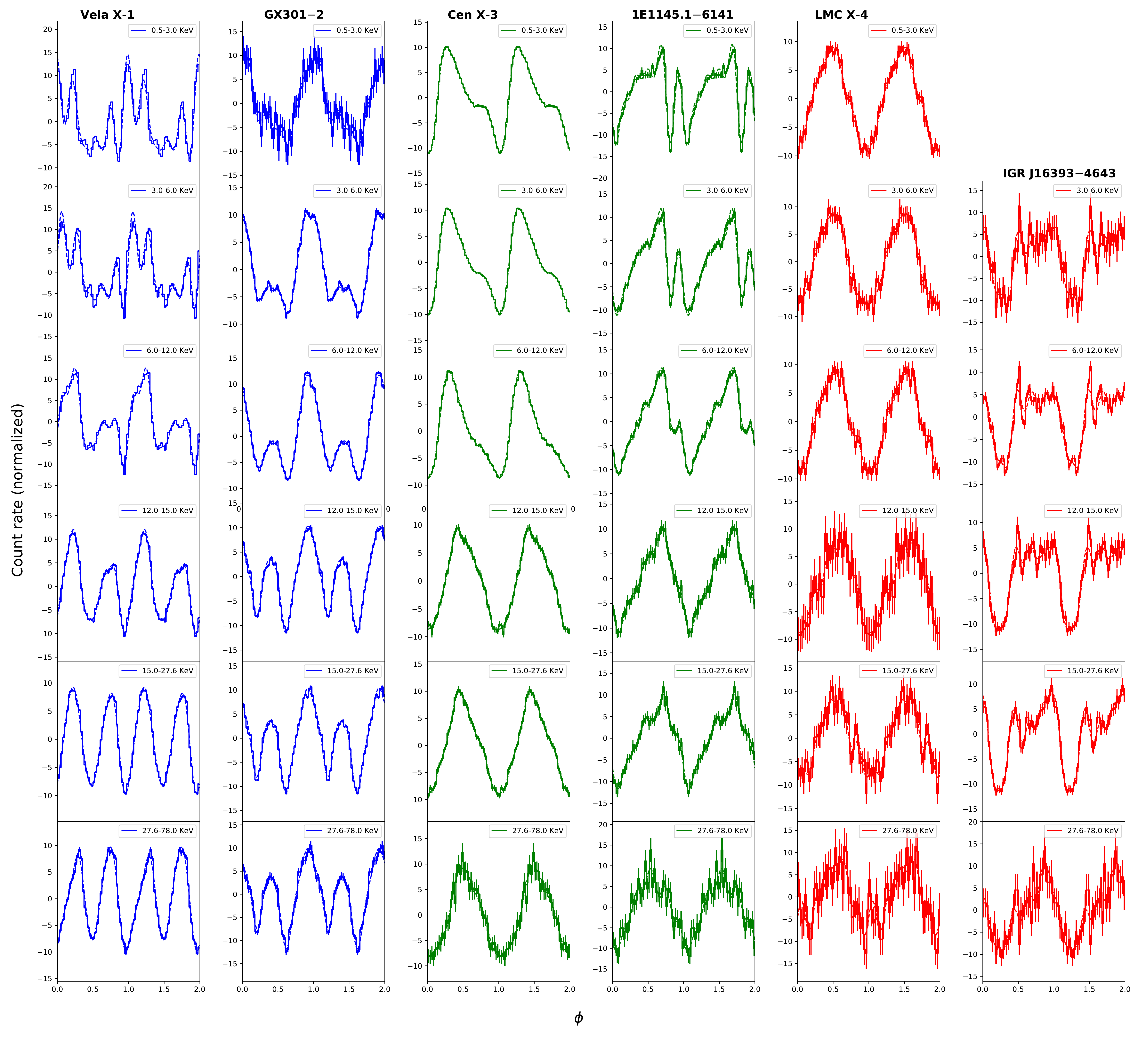}
	\caption{Comparison of the normalized pulse profiles of different sources of the sample studied at different energy ranges. The sources are in order from left to right: Vela X$-1$ (type 2), GX $301-2$ (type 2), Cen X$-3$ (type 1), 1E1145.1$-$6141 (type 1), LMC X$-4$ (type 0) and IGR J16393$-$4643 (type 0). The measured profiles are represented by solid lines whereas the Fourier fits are represented by dashed lines, the boundaries of the energy ranges are [0.5, 3, 6, 12, 15, 27.6 and 78] keV. The pulse profiles have been phase shifted by hand to align the maximums of the pulse profiles of both observations (in cases with observations of both telescopes) and coloured according to their respective pulse profile types in order to better show a comparison between them.}
	\label{fig:pulses_comparation}
\end{figure*}

\begin{figure*}[b!]
	\centering
	\includegraphics[width=\linewidth]{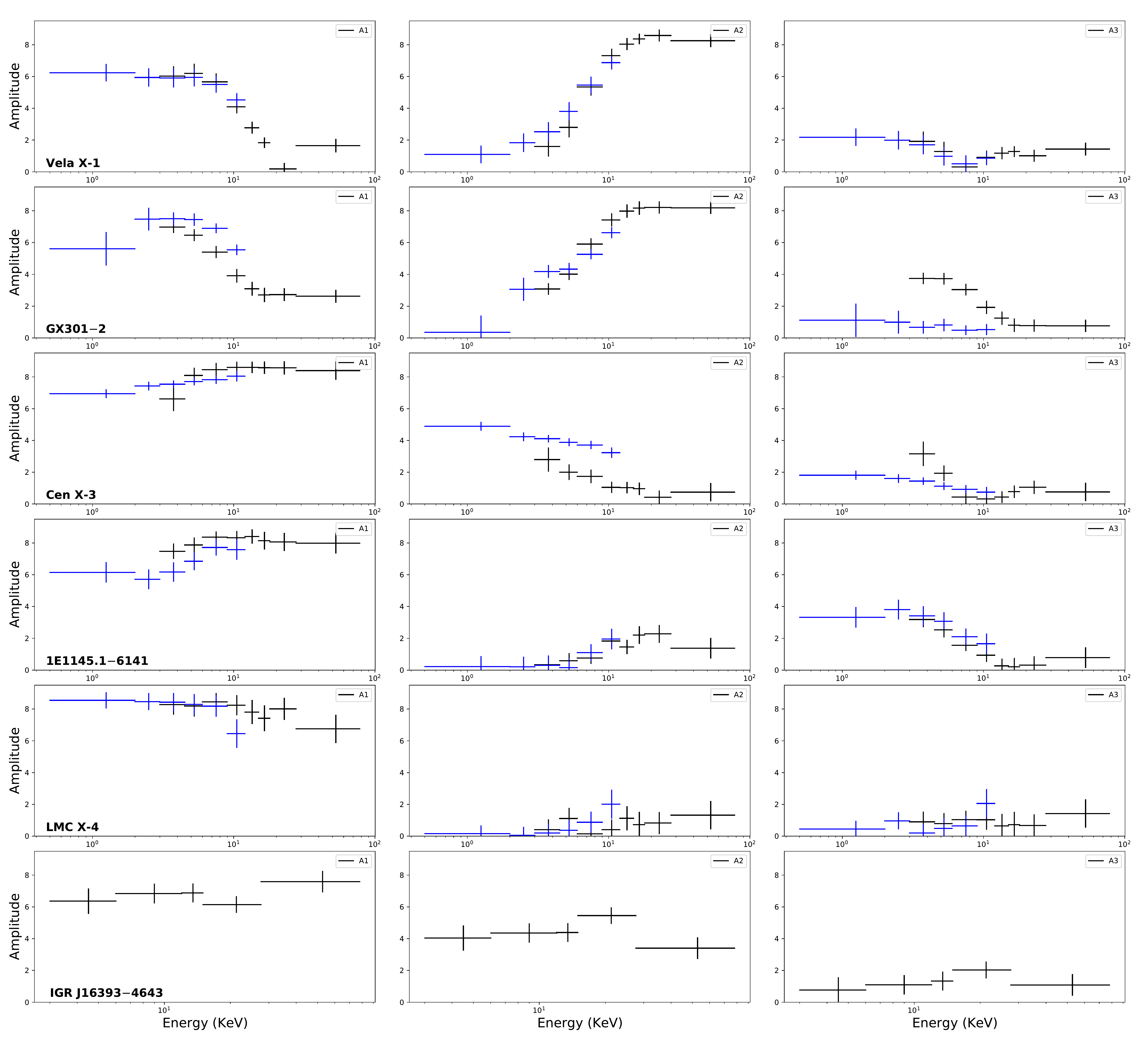}
	\caption{Comparison of the variation of the first three harmonics (typically those with the largest amplitude) of the Fourier series with as function of energy, in order from top to bottom: Vela X-1 (type 2), GX $301-2$ (type 2), Cen X-3 (type 1), 1E1145.1$-$6141 (type 1), LMC X-4 (type 0) and IGR J16393$-$4643 (type 0). the boundaries of the energy ranges are [0.5, 2, 3, 4.5, 6, 9, 12, 15, 18, 27.6 and 78] keV with the exception of IGR J16393$-$4643 whose energy ranges are [3, 6, 12, 15, 27.6 and 78] keV. Blue symbols represent  \xmm data and black symbols \nustar data.}
	\label{fig:amplitudes_comparation}
\end{figure*}

These changes are also shown in Fig.~\ref{fig:fourier_variation}, which shows the variation of the amplitudes of the Fourier harmonics, the first two harmonics are dominant (like for most of the sources). Moreover, they show a clear change as function of energy, with the amplitude of the first harmonic decreasing, while the amplitude of the second harmonic is increasing. Higher order harmonics do not usually show a clear trend as function of energy and might have larger uncertainties, due to the fact that they fit small effects of the pulse profiles. 

The shape of the peaks is given by the difference between the phases and the intensities of the harmonics. In some cases where the dominant harmonics are in phase ($\Delta  \Phi_{0} \simeq 0$) and the two first harmonics have a large amplitude, two intense and thin peaks appear, contrary to the cases where the peaks are not in phase ($\Delta \Phi \,  \bcancel{\simeq} \,  0$), only one wide or truncated peak or some irregular peaks are visisble.

The results presented in this work have been obtained with a number of 40 bins in the pulse profiles phases. However, a change in this parameter does not lead to a significant change neither in the pulse profile shapes nor harmonic amplitudes, even in the cases with complex pulse profiles shapes or low signal-to-noise ratio.

\section{Results}\label{sec:results}

We present our results of the classification of X-ray pulsars in pulse profile categories and the possible relationships between their pulse profiles and important physical observables.

\subsection{Pulse profile classification}

The sources were classified according to their pulse profile shape, the amplitude and phase of their Fourier components and their evolution with energy.
In some cases the shape of the pulse vary drastically between lower energies and higher energies. Even though, in a few cases the shape only shows small variations in the different energy ranges. Therefore, we performed the classification manually by observing common trends in the different sources.

The most decisive aspect of our classification is the variation of the amplitudes of the Fourier harmonics with the energy, there are sources with similar trends and pulse profiles, whereas other sources present some of these characteristics rather different from those common in their respective groups. The groups were named according to the harmonic that presents the largest increase in amplitude with energy and dominates at higher energies. This naming scheme is completely phenomenlogical and might not reflect physical similiarties.

\begin{table*}[t]

\renewcommand{\arraystretch}{1.3}
\centering

\caption{Parameters of the observables of X-ray pulsars sample used in this study, the sources are ordered by their spin periods ($P_\text{spin}$).} 
\label{tab:parameters}

\begin{adjustbox}{max width=\textwidth}
\begin{threeparttable}[b]

\begin{tabular}{l r c c c c c c}
\hline\hline 
Source & \multicolumn{1}{c}{$L_{X}$} & $B$ & $P_{orb} $ & \multicolumn{2}{c}{$P_\text{spin} $ \tnote{(*)} }  & Accretion type & Pulse profile type  \tnote{(*)} \\  
 & \multicolumn{1}{c}{$[\mathrm{erg\cdot s^{-1}}]$} & $[10^{12} \mathrm{G}]$ & $[\mathrm{d}]$ & $ (\xmm)[\mathrm{s}]$ & $ (\nustar)[\mathrm{s}]$ &  &   \\  
\hline 
 SMC X-1 \tnote{(a)} & $(1-50)\cdot 10^{37}$ & $4.56\pm 0.20$ & 3.89  & $0.7002 \pm 0.0007$  & $0.70025 \pm 0.00013$ & Accretion disk & 2  \\
 4U~1901+03 \tnote{(b)} & $4.84^{+0.03}_{-0.02} \cdot 10^{37}$ & $0.4\pm 0.1$ &  $22.5$ & & $2.7608 \pm 0.0008$ & BeXRB & 2  \\
 & $1.483^{+0.002}_{-0.002} \cdot 10^{38}$ & \textbf{---} & \textbf{---}  &  & $2.7616 \pm 0.0008$ & \textbf{---} & 1 \\
 &  $7.87^{+0.03}_{-0.03} \cdot 10^{37}$ & \textbf{---} &  \textbf{---} &    & $2.7622 \pm 0.0008$ & \textbf{---} & 2 \\
 &  $1.242^{+0.010}_{-0.005} \cdot 10^{38}$ & \textbf{---} &  \textbf{---} &   & $2.7629 \pm 0.0008$ & \textbf{---} & 1 \\
 V~0332+53 \tnote{(c)} & $(1.24-1.74)\cdot 10^{38}$ & $2.44\pm 0.05$ & 33.9  &  & $4.3759 \pm 0.0015$ & BeXRB & 2 \\
 &  $(5.7-9.5)\cdot 10^{37}$ & \textbf{---} &  \textbf{---} &    & $4.3759 \pm 0.0017$ & \textbf{---} & 0 \\
 & $(5.3-14)\cdot 10^{36}$ & \textbf{---} &  \textbf{---} &   & $4.3767 \pm 0.0016$ & \textbf{---} & 2 \\
 Cen X-3 \tnote{(d)} & $(1-5) \cdot 10^{37} $ &$ 2.9-3.2$ &  2.09 & $4.802 \pm 0.005$ & $4.797 \pm 0.003$ & Accretion disk & 1  \\
 LMC X-4 \tnote{(e)} & $(2-5)\cdot 10^{38}$ & $10-30$ & 1.41  & $13.48 \pm 0.05$ & $13.49 \pm 0.03$ & Accretion disk & 0   \\
 GRO J2058$+$42	\tnote{(f)}& $(4.0-4.4)\cdot 10^{37}$ & $1.1-1.6$& $53.9 \pm 1.2\, / 110\pm 3$  &   & $194 \pm 3$ & BeXRB & 2 \\
 Vela X-1 \tnote{(g)} & $ (1-10)\cdot 10^{36}$ & $2.60 \pm 0.01$ &  $ 8.96$ &  $283 \pm 4$ & $283 \pm 6$ & Wind-fed & 2 \\
 1E1145.1-6141 \tnote{(h)}&  $ (1-10)\cdot 10^{36}$ &  & $14.4$  &  $297 \pm 6$ & $297 \pm 8$ & Wind-fed & 1 \\
 IGR~J17252$-$3616 \tnote{(i)} & $(6-16)\cdot 10^{35}$ & $0.71\pm 0.03$ &  $9.74$ &  $412 \pm 15$ & & Wind-fed & 0  \\
 GX301-2 \tnote{(j)} & $(8-20) \cdot 10^{36}$ & $3.89 \pm 0.43$  & 41.5  & $680 \pm 30$ & $690 \pm 20$ & Wind-fed & 2 \\
 IGR~J16393$-$4643 \tnote{(k)} & $(5-20) \cdot 10^{36}$ & $2.5 \pm0.1$ &  $4.24$ &   & $900 \pm 20$ & Wind-fed & 0\\
\hline
\end{tabular} 
\begin{tablenotes}
\item[(*)] The $P_\text{spin}$ were obtained with the epoch folding statistic and their uncertainties as a FWHM/2 of the fitted Lorentzian distribution for each observation. The pulse profile types were obtained by qualitative classification in the groups described in this section.
\item[(a)] Ephemeris from \cite{falanga_2015,brumback2020}, other values from \cite{pike_2019A,pradhan_2020}.
\item[(b)] Ephemeris from \cite{sootome_2011}, luminosities from Coley et al. priv comm., magnetic field from \cite{reig_2016}.
\item[(c)] Ephemeris from \cite{Doroshenko_2016}, other values from \cite{Vybornov_2018}.
\item[(d)] Ephemeris from M.H.Finger (HEAD 2010 poster), other values from \cite{Shtykovsky_2017}.
\item[(e)] Ephemeris from \cite{levine_2000}, other values from \cite{Shtykovsky_2017}.
\item[(f)] Ephemeris not determined, other values from \cite{Mukerjee_2020,Corbet_1997,Wilson_1998}.
\item[(g)] Ephemeris from \cite{kreykenbohm_2008}, other values from \cite{sidoli_2015,kretschmar_2019}.
\item[(h)] Ephemeris from \cite{ray_2002}, other values from \cite{Ferrigno_2007}.
\item[(i)] Ephemeris from \cite{manousakis_2011} other values from \cite{manousakis_2012,zurita_2006}.
\item[(j)] Ephemeris from \cite{Doroshenko_2010}, other values from \cite{nishimura_2014}.
\item[(k)] Ephemeris not determined, other values from \cite{bodaghee_2016,islam_2015}.

\end{tablenotes}

\end{threeparttable}
\end{adjustbox}

\renewcommand{\arraystretch}{1.0}

\end{table*}

  \textbf{Type 0:} The most important characteristic of this type is that it only shows little variations or some noise in certain energy ranges. In this case, the harmonics do not exhibit clear variations with the energy and the relationship between the amplitudes remains almost constant in all the energy ranges that we have studied.
  
  The three sources classified as type 0 present these characteristics and pulse profile shapes with some similarity (truncated peak for IGR~J16393$-$4643 and V~0332+53 (OBS ID=80102002004) or a thick peak for LMC~X-4 and IGR~J17252$-$3616).
  
  \textbf{Type 1:} this group presents a dominance of the 1st harmonic, especially at higher energies where the profile shows a single peak. However, at lower energies these sources present a great variety of shapes. 

  In this case the first harmonic increases its amplitude with energy  while the second and third harmonic decrease their amplitude. 4U~1901+03 (OBS ID=90502307002 and OBS ID=90501305001) and Cen X-3 show these characteristics. However, 1E1145.1$-$6141 has also an increase in the amplitude of the 2nd harmonic, although the 1st harmonic is dominant. 
  
  Moreover, for this group, most of the sources have an important change in the shape of their pulse profile in the range $\sim 3$--$15$\,keV. 

  \textbf{Type 2:} This group of sources is described by the dominance of the first two harmonics. In addition, they present an increase in the 2nd harmonic amplitude and in most of the sources a decrease in the 1st harmonic amplitude with  energy, which causes a two-peaked structure in most of the sources at higher energies, with the exception of GRO~J2058$+$42.
  
  At low energies they show complex structure with two or more peaks like SMC X-1, Vela X-1, V~0332+53 (OBS ID=80102002002 and OBS ID=80102002010) and GX301-2, even though  4U~1901+03 (OBS ID=90501324002 and OBS ID=90502307004) only shows one peak in each pulse profile. On the other hand, the case of GRO~J2058$+$42 is more complicated, it shows an increase in both first and second harmonics and shows only one peak at higher energies, although it shows complex shapes at lower energies as well. 
  
  Furthermore, most of the sources in this group have an important change in the shape of their pulse profile in the range $\sim 3$--$30$\,keV.


Fig.~\ref{fig:pulses_comparation} shows a sample of energy resolved pulse profiles, two per each type, is remarkable that type 0 sources show little variations in all the energy range, whereas the type 1 and 2 have notorious changes in the different energy ranges.

On the other hand, the variations in the amplitudes of their respective first three harmonics are reflected in the Fig.~\ref{fig:amplitudes_comparation}, the variations of the sources of type 1 and 2 have identifying tendencies, whereas the sources of type 0 have the common characteristic of not presenting large variations in the different energy steps.

\subsection{Luminosity and magnetic field}

We studied possible correlations between the pulse profile category, the luminosity, and the magnetic field strength of each source, using the data shown in Tab.~\ref{tab:parameters}.

\begin{figure}[h!]
	\centering
	\includegraphics[width=\linewidth]{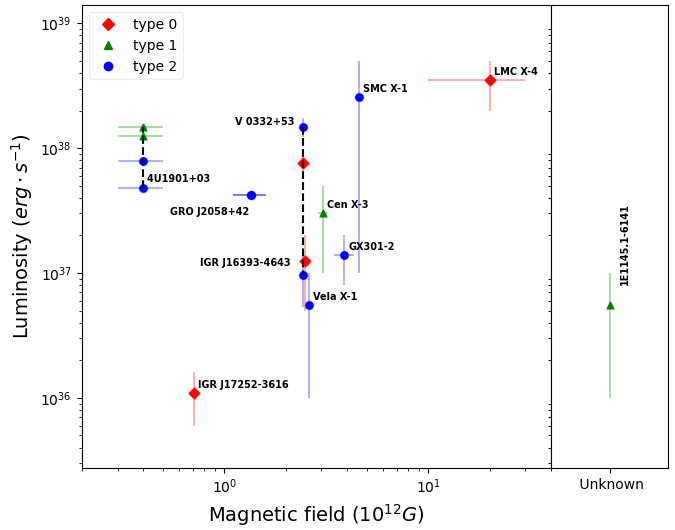}
	\caption{Luminosity ($L_{X}$) versus magnetic field strength ($B$) diagram. Shape and colour of marker indicates the pulse profile group.}
	\label{fig:L_B}
\end{figure}

\begin{figure*}[b]
	\centering
	\includegraphics[width=\linewidth]{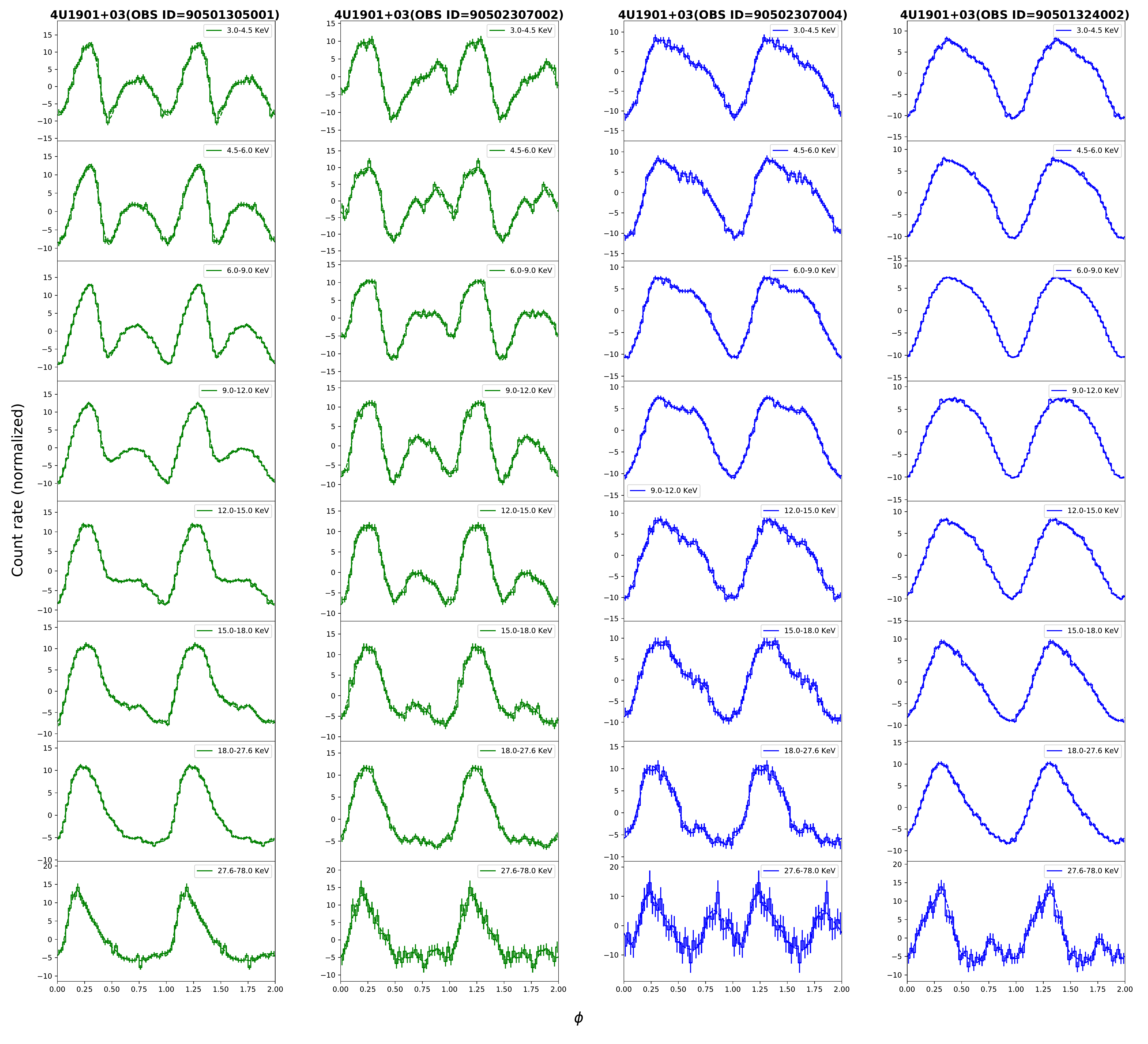}
	\caption{Comparison of the normalized pulse profiles of the source 4U 1901$+$03 studied at different energy ranges in different stages during a flare period. The observations are in temporal order from left to right (1st: The observation with OBS ID=90501305001 and a luminosity value of $1.242^{+0.010}_{-0.005} \cdot 10^{38} \mathrm{erg\cdot s^{-1}}$ with a pulse profile type 1, 2nd: The observation with OBS ID=90502307002 and a luminosity value of $1.483^{+0.002}_{-0.002} \cdot 10^{38} \mathrm{erg\cdot s^{-1}}$ with a pulse profile type 1, 3rd: The observation with OBS ID=90502307004 and a luminosity value of $7.87^{+0.03}_{-0.03} \cdot 10^{37} \mathrm{erg\cdot s^{-1}}$ with a pulse profile type 2 and 4th: The observation with OBS ID=90501324002 and a luminosity value of $4.84^{+0.03}_{-0.02} \cdot 10^{37} \mathrm{erg\cdot s^{-1}}$ with a pulse profile type 2). The measured profiles are represented by solid lines whereas the Fourier fits are represented by dashed lines. The boundaries of the energy ranges are [0.5, 3, 6, 12, 15, 27.6 and 78] keV. The pulse profiles have been shifted by hand to align the minima of the pulse profiles and coloured according to their respective pulse profile types in order to better show a comparison between them.}
	\label{fig:pulses_comparation_4U}
\end{figure*}

We plot our results in Fig. \ref{fig:L_B}, which shows the Luminosity-magnetic field strength diagram, where it is not possible to identify any group of sources, the sources in each pulse profile category are very dispersed and no clear groups emerge. Therefore, it does not seem to be a relationship between these parameters and our classification. Nevertheless, this result is interesting in itself due to the fact that it would indicate an independence between the pulse profile and also the accretion column geometry and these two important parameters.

The fact that sources with different pulse profiles (for example IGR J16393$-$4643, Cen X-3 and GX301$-$2) have similar values of luminosity and magnetic field strength and with little uncertainties is particularly interesting. In addition, sources with similar profiles (for example Vela X-1 and GX301$-$2 at energies between $12-78$ KeV) have different values of luminosity and magnetic field strength.

However, the case of the source 4U~1901$+$03 is especially interesting; it was observed in February 2019 during a flare with a variation in its luminosity. This change was accompanied by a change in its pulse profile, going from type 1 during the luminosity peak of the flare (the observation with OBS ID=90502307002 and a luminosity value of $1.483^{+0.002}_{-0.002} \cdot 10^{38} \mathrm{erg\cdot s^{-1}}$ and the observation with OBS ID=90501305001 and a luminosity value of $1.242^{+0.010}_{-0.005} \cdot 10^{38} \mathrm{erg\cdot s^{-1}}$) to type 2 after the luminosity decrease (the observation with OBS ID=90502307004 and a luminosity value of $7.87^{+0.03}_{-0.03} \cdot 10^{37} \mathrm{erg\cdot s^{-1}}$ and the observation with OBS ID= 90501324002 and a luminosity value of $4.84^{+0.03}_{-0.02} \cdot 10^{37} \mathrm{erg\cdot s^{-1}}$).

The pulse profiles are shown in Fig. \ref{fig:pulses_comparation_4U}, it is important that the changes in the pulse profiles occurred in all energy ranges, most notably in the lower energy steps. Furthermore, it is especially interesting that the trend of the pulse profile with energy is also modified. Despite showing a change in the pulse profile shape in the whole energy range during the flare, it does not implicate a correlation between the change in the pulse profile and the change in the luminosity, this change could be more related to a modification in the accretion column geometry besides a change in disposition of hot material that is around the neutron star during the flare.

The source V~0332+53 also showed a change its pulse profile during a flare in 2015. However, in this case some observations have a low signal-to-noise ratio and their pulse profiles are strongly influenced by noise, which means that their types are not well-defined. Therefore, the relationship between the pulse profile type and the luminosity is less clear.

\begin{figure*}[t]
	\centering
	\includegraphics[width=\linewidth]{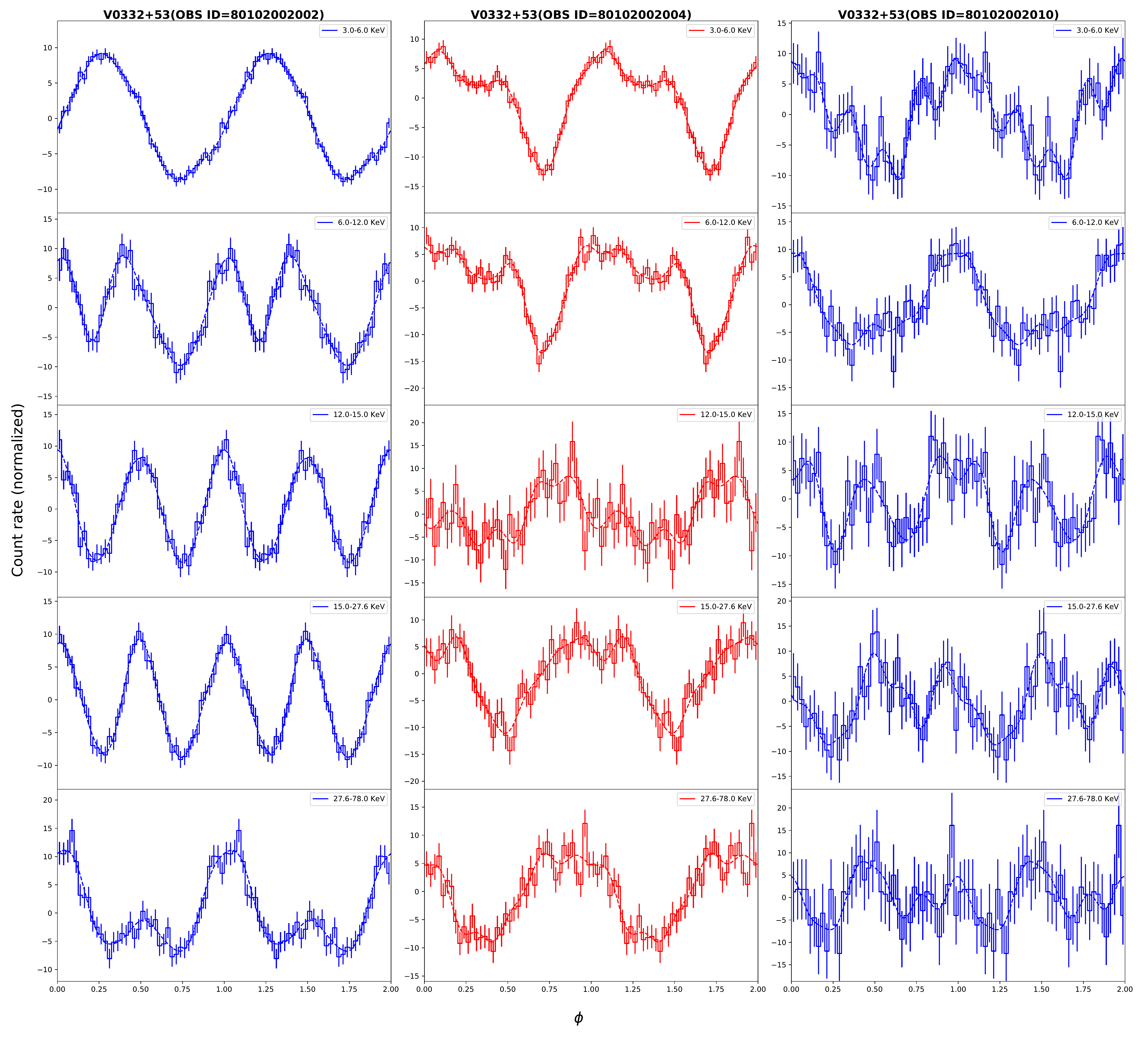}
	\caption{Comparison of the normalized pulse profiles of the source V~0332+53 studied at different energy ranges in different stages during a flare period, the observations are in temporal order from left to right (1st: The observation with OBS ID=80102002002 and a luminosity value of $(1.24-1.74) \cdot 10^{38} \mathrm{erg\cdot s^{-1}}$ with a pulse profile type 2, 2nd: The observation with OBS ID=80102002004 and a luminosity value of $(5.7-9.5) \cdot 10^{37} \mathrm{erg\cdot s^{-1}}$ with a pulse profile type 0 and 3rd: The observation with OBS ID=80102002010 and a luminosity value of $(5.3-14) \cdot 10^{36} \mathrm{erg\cdot s^{-1}}$ with a pulse profile type 2). The measured profiles are represented by solid lines whereas the Fourier fits are represented by dashed lines. The boundaries of the energy ranges are [0.5, 3, 6, 12, 15, 27.6 and 78] keV. The pulse profiles have been shifted by hand to align the minima of the pulse profiles and coloured according to their respective pulse profile types in order to better show a comparison between them.}
	\label{fig:pulses_comparation_V0}
\end{figure*}

Nevertheless, we have presented three of those observations in this paper. The most interesting feature of this case is that the first and the third observation (OBS ID=80102002002 and OBS ID=80102002010) show two peaks or complex profiles at higher energies, which we have classified as pulse profile type 2 according to the parameters previously explained despite their different luminosity values. They have the same type even though they show clear differences in shape and in the sizes of the peaks. Whereas the other observation (OBS ID=80102002004) with a luminosity value between the other two observations shows only one peak in each energy range, which corresponds to a pulse profile type 0.
An overview of the V~0332+53 energy resolved pulse profiles is shown in \ref{fig:pulses_comparation_V0}.

It is important to discuss two results from our analysis of V~0332+53. Firstly, our classification system is not perfect, as these two pulse profiles classified as type 2 (OBS ID=80102002002 and OBS ID=80102002010) have different pulse profile shapes in some energy ranges, although they have been classified as the same type. However, the goal of the system is to perform systematic classifications based on general parameters and their changes throughout the entire energy range. On the other hand, luminosity is not a determining factor in the characteristics of the pulse profile even in the same source (with the parameters that this involves).

It is likely that other non-observable parameters associated with the accretion rate and the accretion column geometry changed during these flares producing the respective variations in the luminosities and the pulse profiles. This could be due to a common accretion rate factor with energy dependence between the different types, e.g., a change in the accretion flow that could produce a change in the luminosity and that modifies the incidence of material on the surface of the neutron star.

\subsection{Spin period and orbital parameters}

We also studied possible correlations between the pulse profile category and the spin and orbital periods, using data shown in Tab.~\ref{tab:parameters}.

\begin{figure}[h!]
	\centering
	\includegraphics[width=\linewidth]{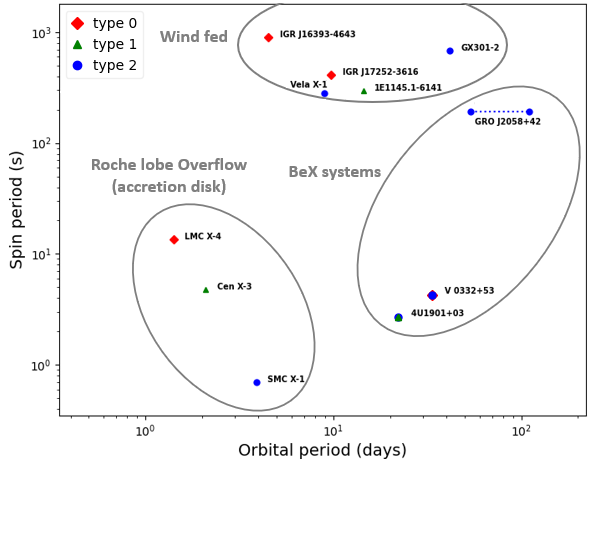}
	\caption{Spin period versus orbital period for the source sample. Shape and colour of marker indicates the pulse profile group and the gray lines indicate the regions of the different types of HMXBs by their accretion method.}
	\label{fig:Porb_S}
\end{figure}

We plot our results in Fig.~\ref{fig:Porb_S}, which shows the spin period-orbital period, also known as a ``Corbet diagram'' \citep{corbet_1986}. In this diagram the different classes of HMXBs are separated into three distinct groups according to their accretion mechanism: stellar wind fed, Roche lobe overflow, and BeXRBs.

Although the sample of sources is small, the pulse profile types are distributed over different accretion mechanisms and no clear relation to the rotation and spin periods is obvious. However, this indicates that the accretion mechanism does not have a decisive role in the shape of the accretion column and emission pattern.

However, there is an indication that type 0 sources cluster in the upper left part of the diagram (except for V~0332+53 (OBS ID=80102002004. Even though, this source presents a low SNR and its type is not well-defined), with relatively fast orbital periods and relatively slow spin periods. This distribution could indicate that these sources are somewhat older and have a simpler magnetic field geometry leading to a more symmetric pulse profile. However, with only three sources any such interpretation is currently speculation and would need to be supported by further data from more sources.

The pulse profiles will also depend on the inclination angle of the neutron star spin axis to our line-of-sight. However, there is no known method to directly measure this angle, or even the inclination of the orbital plane of the binary system \cite{Tomsick_2010}. We therefore cannot investigate this dependence currently.

\section{Conclusions}\label{sec:conclusions}

In this paper we have presented a framework to study and classify energy resolved pulse profiles based on a decomposition into their Fourier components. We have applied this framework to a small sample of X-ray pulsars, for which a wide energy coverage between \xmm and \nustar is available. The framework is publicly available and can be used to create energy resolved and decomposed pulse profiles for any  event data.

In our sample, we find a range of diverse shapes and energy variation of the pulse profiles. Based on the results of the amplitudes and phases of the Fourier components we identified three distinct groups. We investigated correlations between these groups and other physical parameters of the sources, like luminosity, magnetic field strength and spin and orbital period, but could not identify any clear connections.

We only find an interesting, but not yet statistically significant connection, for profile type 0, which seems to be prevalent for sources with a combination of relatively short orbital periods and slow spin periods (Fig.~\ref{fig:Porb_S}). Type 0 sources are characterised by little to no energy dependence of their pulse profiles, indicating that there is no energy dependence in the emission region or geometry. How this could be related to spin or orbital periods is, however,  beyond the scope of this paper.

In 4U\,1901+03 we find that the source changes from profile type 1 to 2 with decreasing luminosity during the outburst. This clearly indicates a significant change in the accretion column and emission geometry as function of accretion rate and luminosity. \citet{becker2012} propose that such a change could be related to the accretion column switching from a Coulomb dominated deceleration regime to a radiation dominated region, i.e., the source becoming super-critical at high luminosities. With the formation of a radiation dominated shock, we expect a significant change in the emission geometry, as observed here.

However, a more quantitative comparison between expected pulse profiles as function of accretion regime and the data is very complex. This complexity is mainly due to the many different physical effects involved, on the micro- well as macro-scale. Small changes in emission height or in-falling velocity can have large effects on the observed pulse profile due to the strong light-bending around the neutron star \citep[\textit{cf.}][]{falkner2018a}.

Our framework provides a simple and straightforward way to decompose pulse profiles into their fundamental components, which will help further theoretical efforts to model the profiles based on first principle physics. By identifying the dominant harmonics and contributions in the pulse profiles, we provide information for the models where most of the energy dependent change occur.

By increasing the sample size with new and archival observations of accretion pulsars in the future, we can populate Fig.~\ref{fig:L_B} and \ref{fig:Porb_S} further and investigate in more detail correlation with physical parameters. With more data it will also be possible to refine the definition of the pulse profile types and identify a more general classification scheme. We believe that our proposed classification can be a useful tool in understanding and sorting X-ray pulsars and in looking for common physical properties within the accretion column.

\begin{acknowledgements}
This research has made use of data and software provided by the High Energy Astrophysics Science Archive Research Center (HEASARC), which is a service of the Astrophysics Science Division at NASA/GSFC as well as NASA’sAstrophysics Data System Bibliographic Service (ADS).
This work made use of data from the NuSTAR mission, a
project led by the California Institute of Technology, managed by the Jet Propulsion Laboratory, and funded by the
National Aeronautics and Space Administration and used the NuSTAR Data Analysis Software (NuSTARDAS) jointly developed by the ASI Science Data Center (ASDC, Italy) and the California Institute of Technology (USA).
This research has also made use of data obtained with XMM-Newton, an ESA science mission with instruments and contributions directly funded by ESA Member States.
AMJ acknowledge partial funding from the European Space Agency (ESA) under partnership agreement 4000133194/20/NL/MH/hm
between ESA and FAU Erlangen-N\"urnberg.
\end{acknowledgements}


\appendix
\section{Pulse profiles and harmonic amplitudes}


\begin{figure*}%
    \centering
    \includegraphics[width=0.47\linewidth]{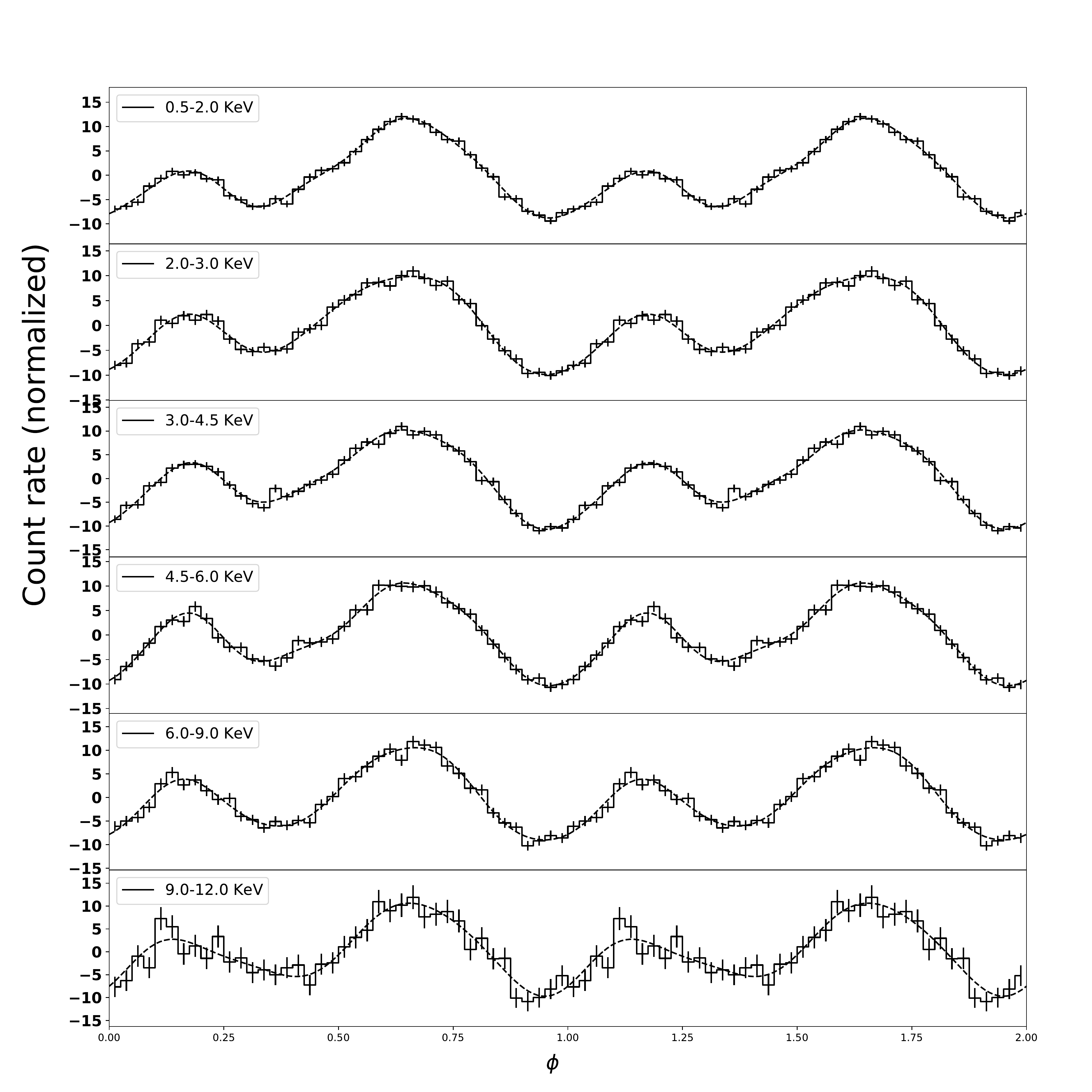} %
    \qquad
    \includegraphics[width=0.47\linewidth]{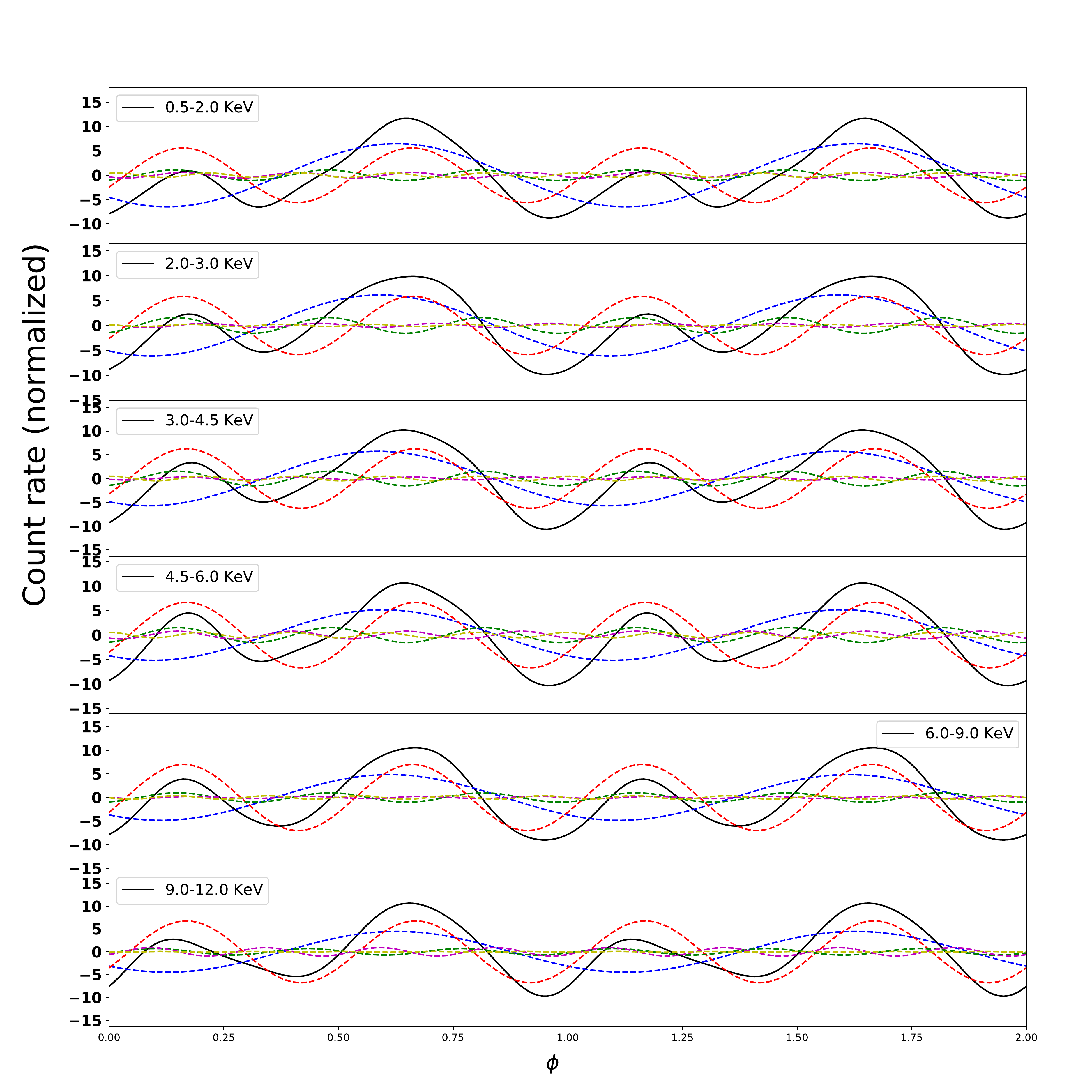} 
    \caption{Left: Normalized pulse profiles of SMC~X-1 at different energy ranges based on \xmm data, the measured profiles are represented by solid lines whereas the Fourier fits are represented by dashed lines, the boundaries of the energy ranges are [0.5, 2, 3, 4.5, 6, 9, 12] keV. Right: The black solid line represents the overall approximate pulse profile and the colour dashed lines represent the different Fourier components. Blue: 1st, red: 2nd, green: 3rd, purple: 4th and yellow: 5th.}%
    \label{fig:SMC_X-1_XMM_pulse_profiles}%
\end{figure*}

\begin{figure*}%
    \centering
    \includegraphics[width=0.47\linewidth]{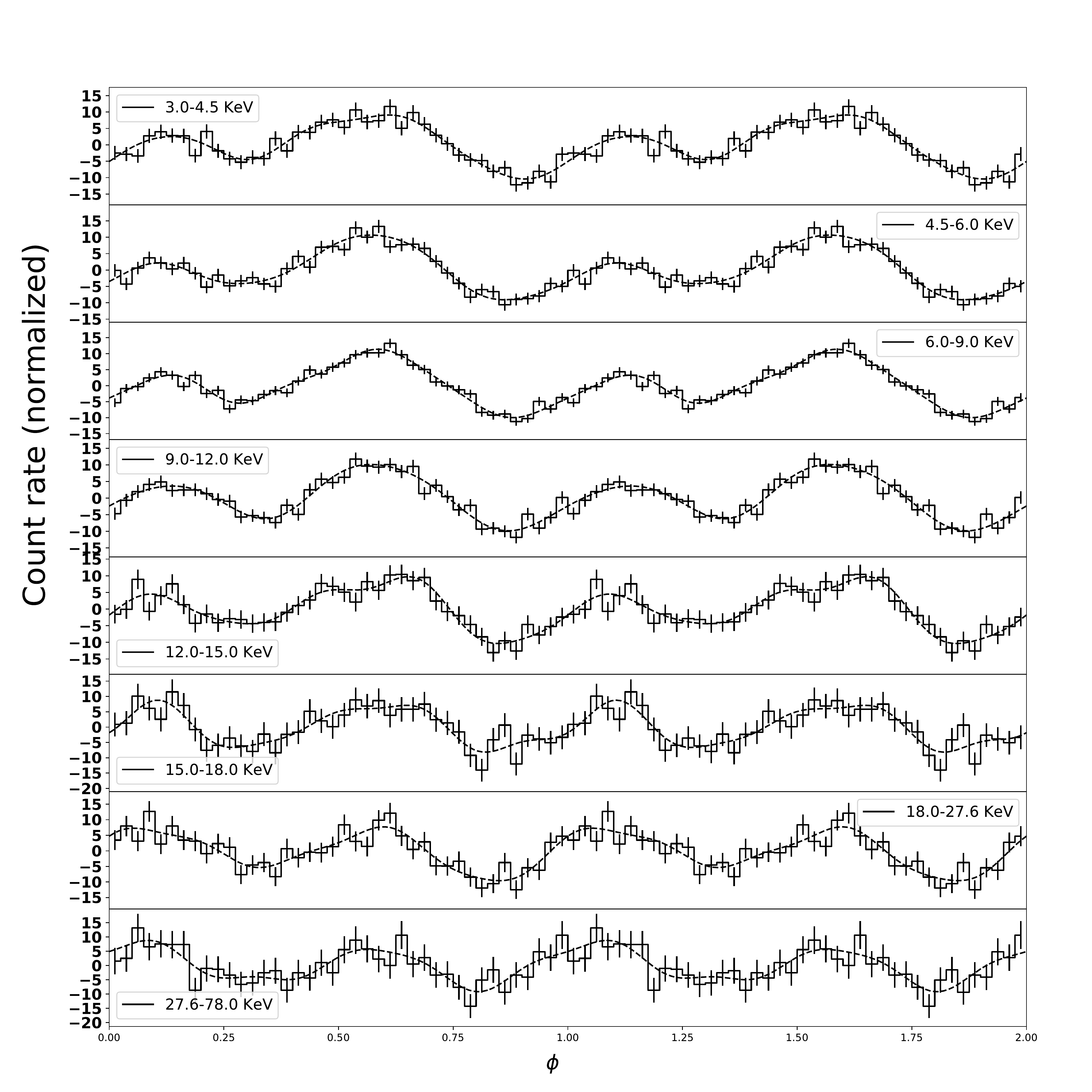} %
    \qquad
    \includegraphics[width=0.47\linewidth]{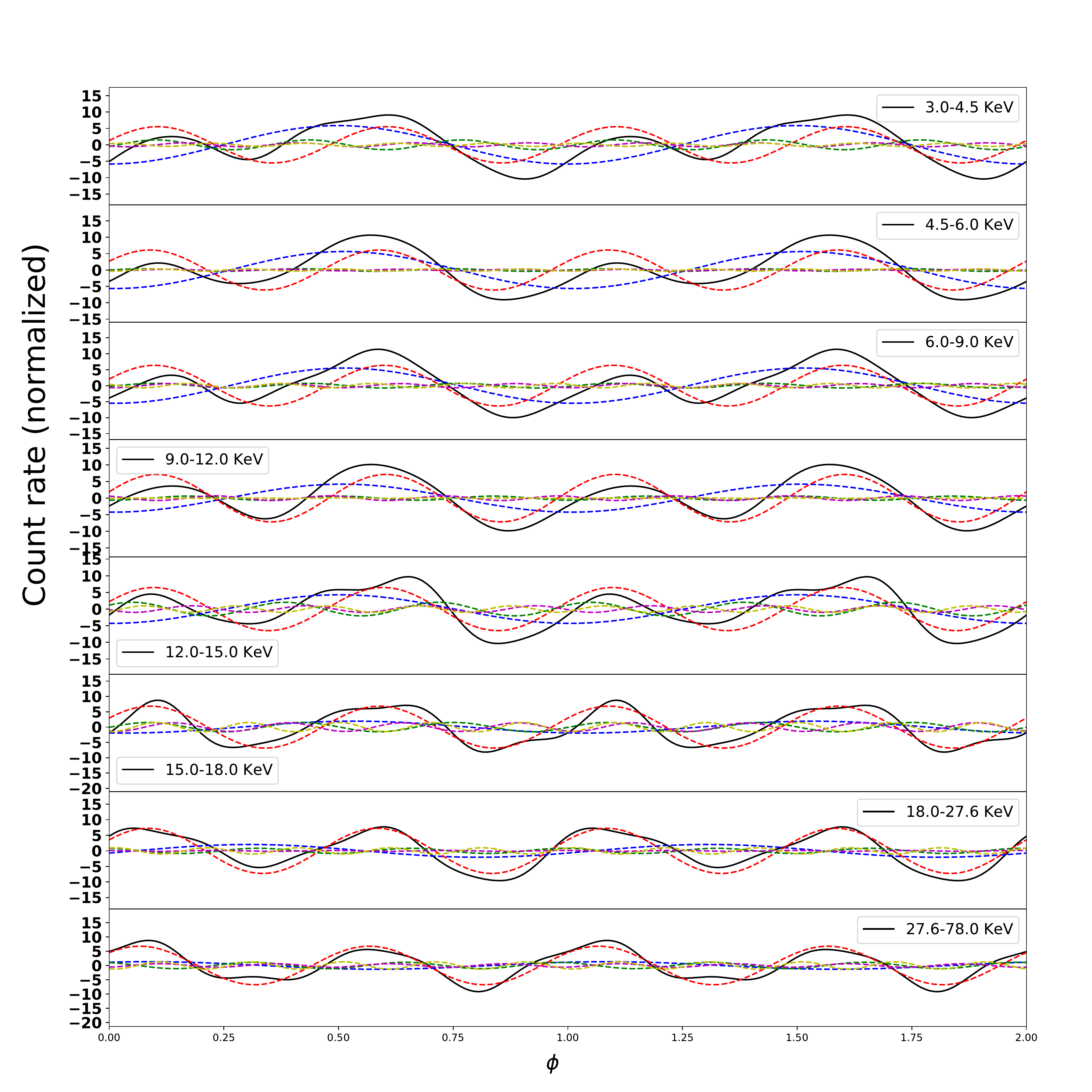} 
    \caption{Left: Normalized pulse profiles of SMC~X-1 at different energy ranges based on \nustar data, the measured profiles are represented by solid lines whereas the Fourier fits are represented by dashed lines, the boundaries of the energy ranges are [3, 4.5, 6, 9, 12, 15, 18, 27.6 and 78] keV. Right: The black solid line represents the overall approximate pulse profile and the colour dashed lines represent the different Fourier components. Blue: 1st, red: 2nd, green: 3rd, purple: 4th and yellow: 5th.}%
    \label{fig:SMC_X-1_NUSTAR_pulse_profiles}%
\end{figure*}

\begin{figure*}
	\centering
	\includegraphics[width=\linewidth]{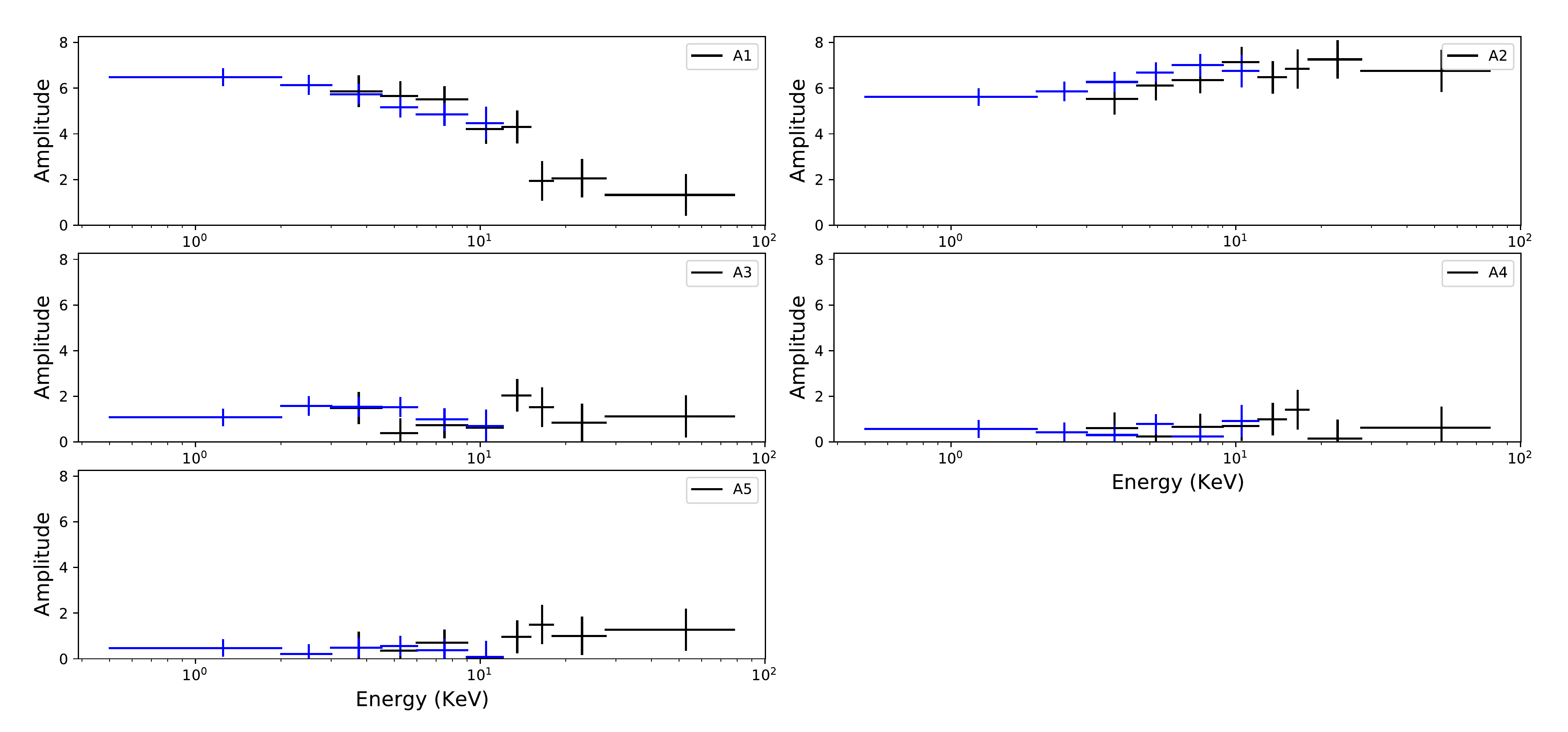}
	\caption{Variation of the harmonic amplitudes for SMC~X-1, the boundaries of the energy ranges are [0.5, 2, 3, 4.5, 6, 9, 12, 15, 18, 27.6 and 78] keV. \xmm data is represented by blue markers and \nustar data is represented by black markers.}
	\label{fig:fig:SMC_X-1_fourier_variation}
\end{figure*}


\begin{figure*}%
    \centering
    \includegraphics[width=0.47\linewidth]{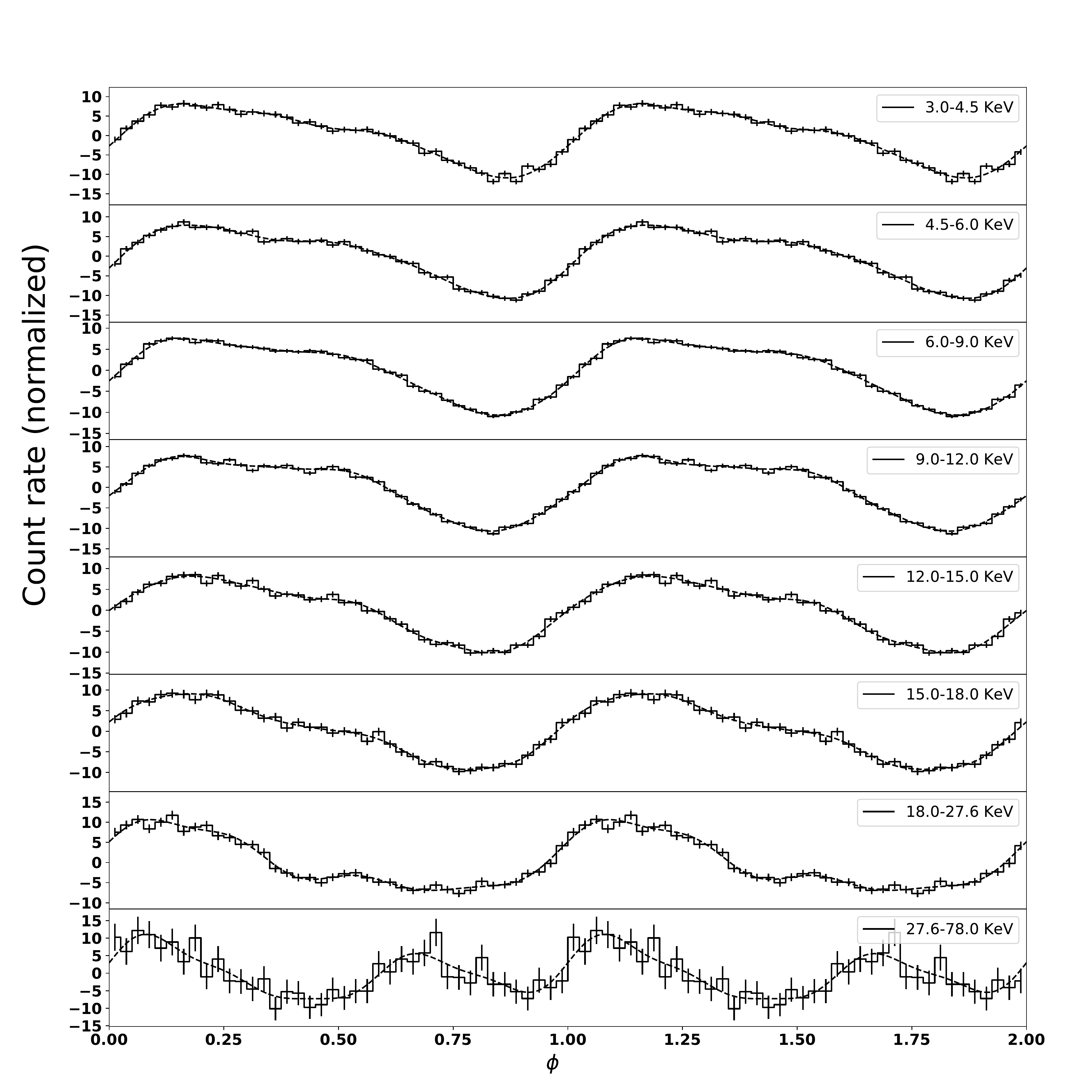} %
    \qquad
    \includegraphics[width=0.47\linewidth]{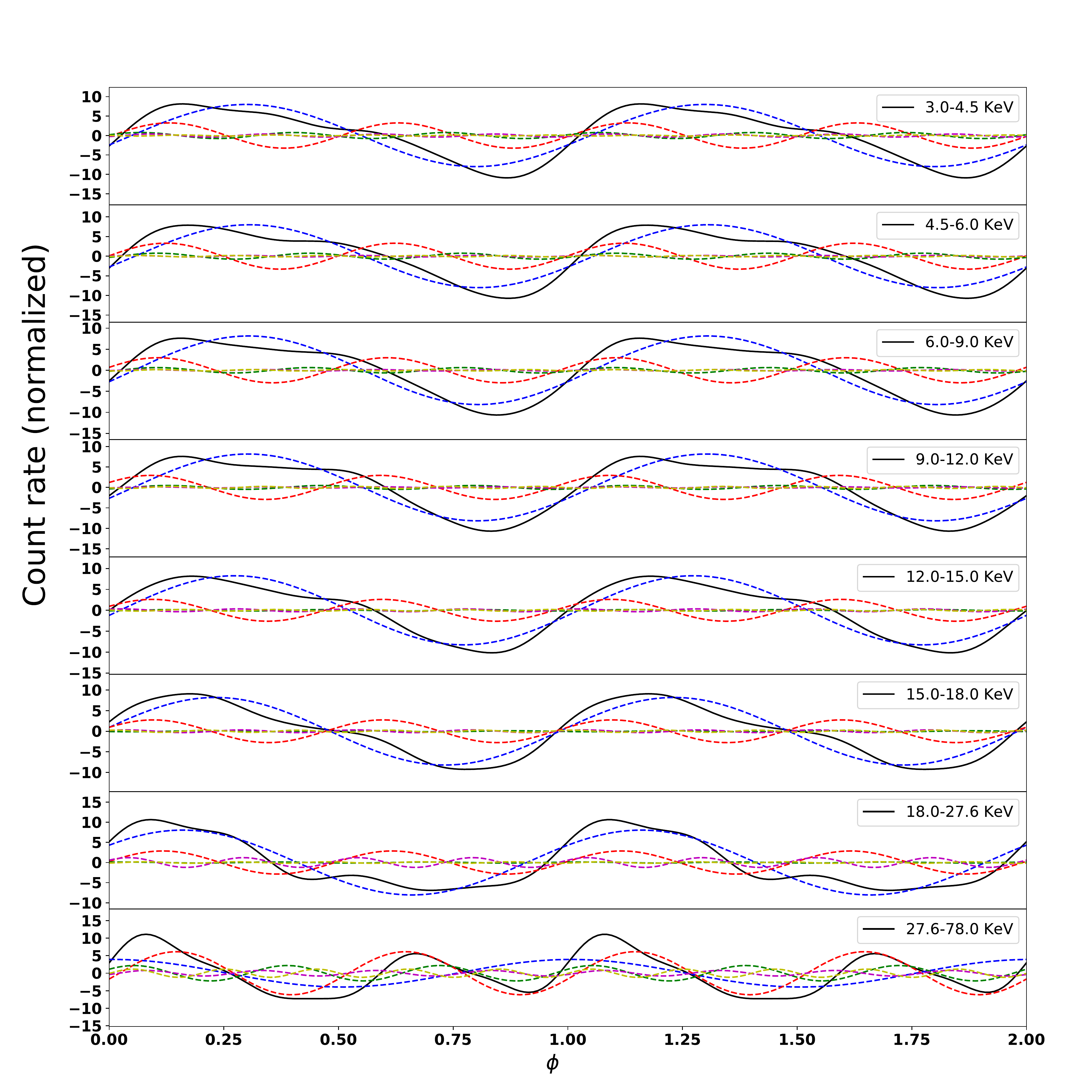} 
    \caption{Left: Normalized pulse profiles of 4U~1901+03 (the observation with OBS ID=90502307004 and a luminosity value of $7.87^{+0.03}_{-0.03} \cdot 10^{37} \mathrm{erg\cdot s^{-1}}$) at different energy ranges based on \nustar data, the measured profiles are represented by solid lines whereas the Fourier fits are represented by dashed lines, the boundaries of the energy ranges are [3, 4.5, 6, 9, 12, 15, 18, 27.6 and 78] keV. Right: The black solid line represents the overall approximate pulse profile and the colour dashed lines represent the different Fourier components. Blue: 1st, red: 2nd, green: 3rd, purple: 4th and yellow: 5th.}%
    \label{fig:4U1901+03_90502307004_NUSTAR_pulse_profiles}%
\end{figure*}

\begin{figure*}
	\centering
	\includegraphics[width=\linewidth]{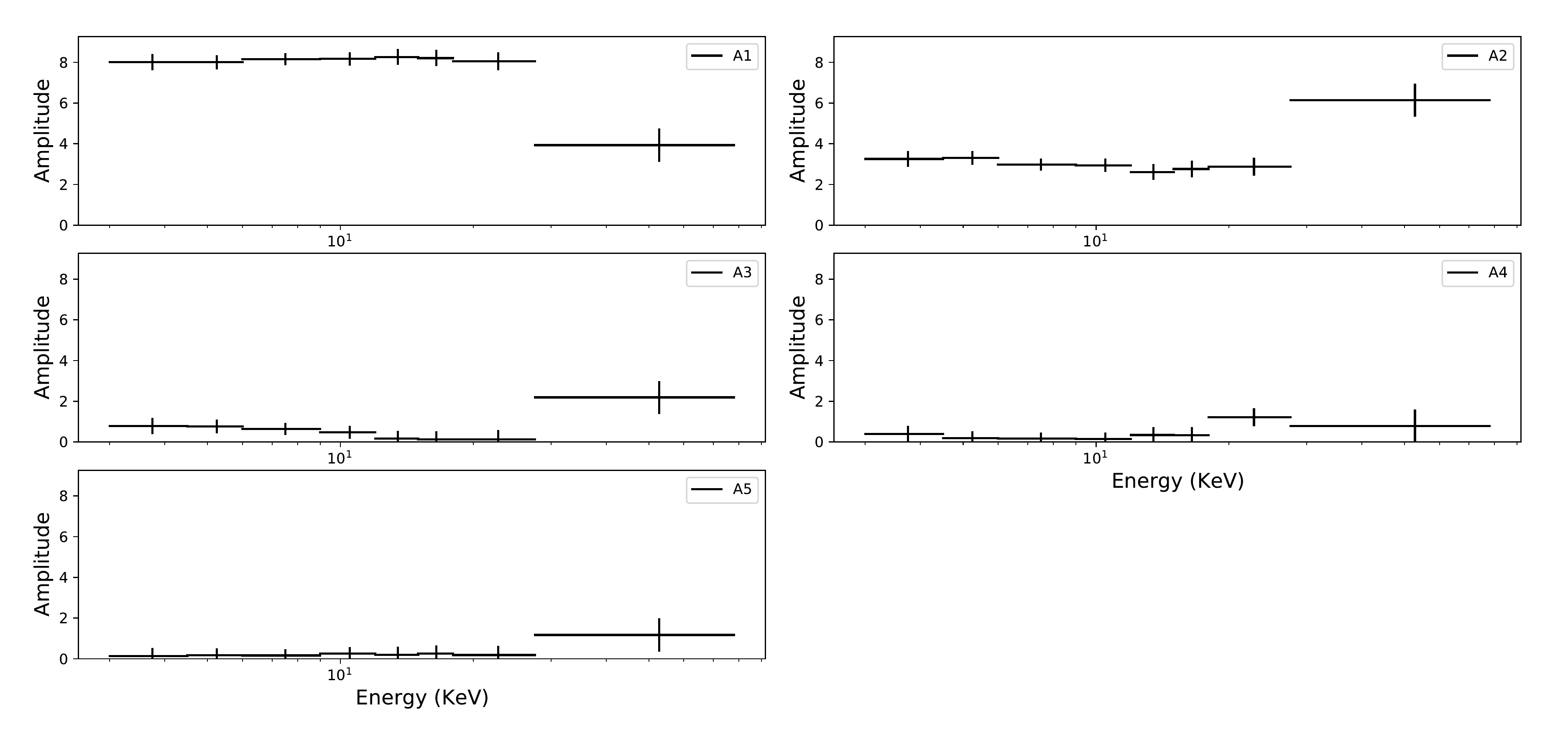}
	\caption{Variation of the harmonic amplitudes for 4U~1901+03 (the observation with OBS ID=90502307004 and a luminosity value of $7.87^{+0.03}_{-0.03} \cdot 10^{37} \mathrm{erg\cdot s^{-1}}$), the boundaries of the energy ranges are [3, 4.5, 6, 9, 12, 15, 18, 27.6 and 78] keV. Only \nustar data is represented by black markers.}
	\label{fig:4U1901+03_90502307004_fourier_variation}
\end{figure*}

\begin{figure*}%
    \centering
    \includegraphics[width=0.47\linewidth]{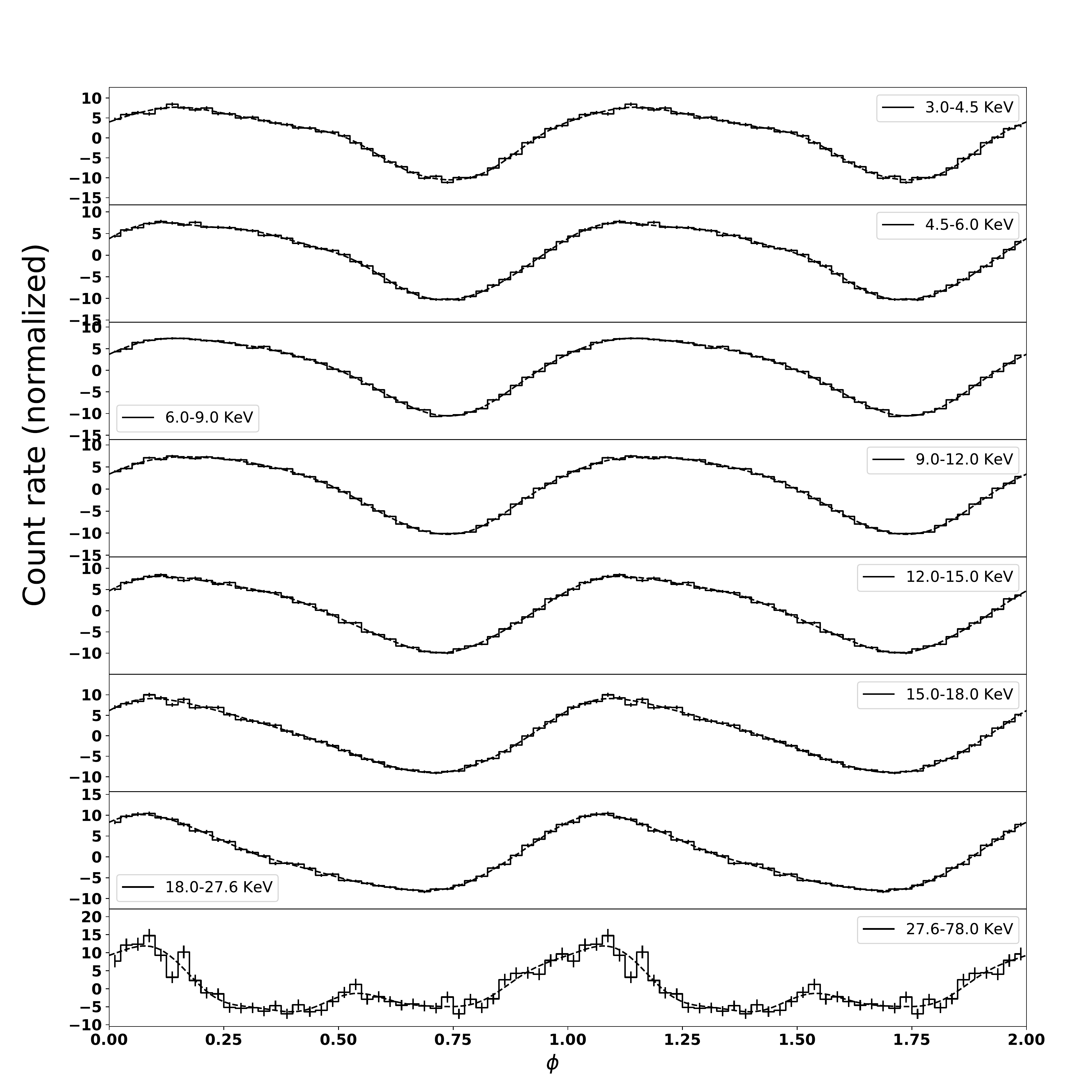} %
    \qquad
    \includegraphics[width=0.47\linewidth]{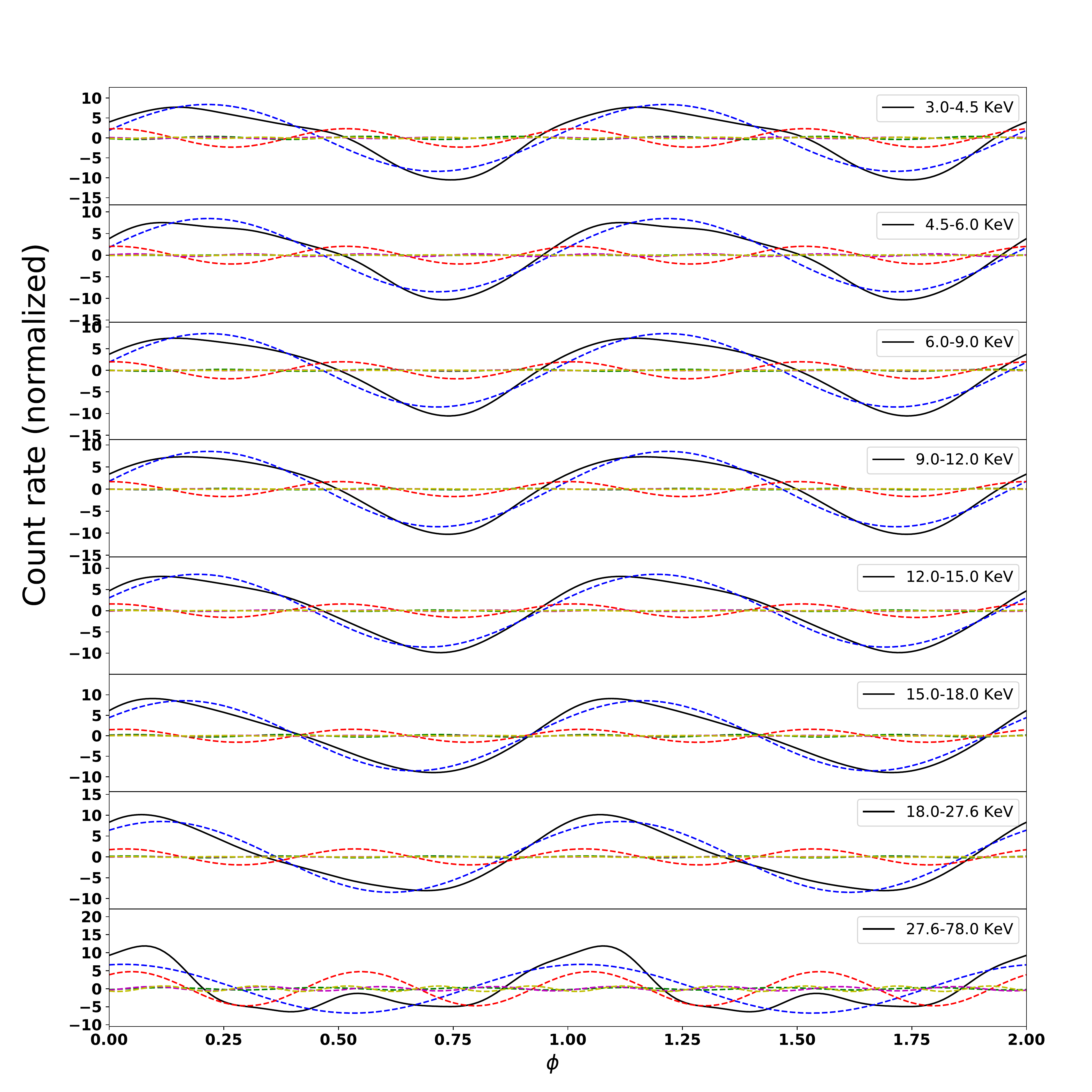} 
    \caption{Left: Normalized pulse profiles of 4U~1901+03 (the observation with OBS ID= 90501324002 and a luminosity value of $4.84^{+0.03}_{-0.02} \cdot 10^{37} \mathrm{erg\cdot s^{-1}}$) at different energy ranges based on \nustar data, the measured profiles are represented by solid lines whereas the Fourier fits are represented by dashed lines, the boundaries of the energy ranges are [3, 4.5, 6, 9, 12, 15, 18, 27.6 and 78] keV. Right: The black solid line represents the overall approximate pulse profile and the colour dashed lines represent the different Fourier components. Blue: 1st, red: 2nd, green: 3rd, purple: 4th and yellow: 5th.}%
    \label{fig:4U1901+03_90501324002_NUSTAR_pulse_profiles}%
\end{figure*}

\begin{figure*}
	\centering
	\includegraphics[width=\linewidth]{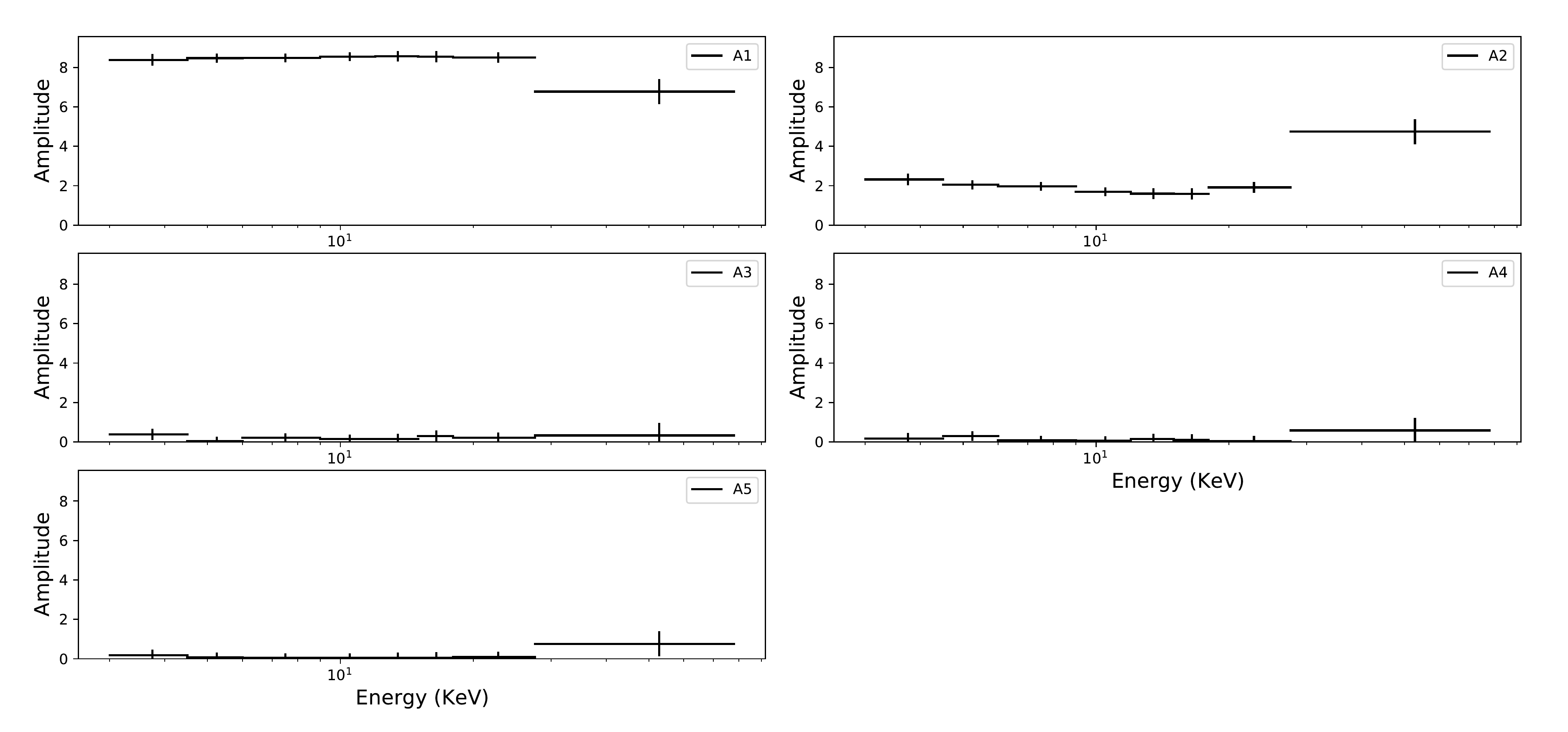}
	\caption{Variation of the harmonic amplitudes for 4U~1901+03 (the observation with OBS ID= 90501324002 and a luminosity value of $4.84^{+0.03}_{-0.02} \cdot 10^{37} \mathrm{erg\cdot s^{-1}}$), the boundaries of the energy ranges are [3, 4.5, 6, 9, 12, 15, 18, 27.6 and 78] keV. Only \nustar data is represented by black markers.}
	\label{fig:4U1901+03_90501324002_fourier_variation}
\end{figure*}

\begin{figure*}%
    \centering
    \includegraphics[width=0.47\linewidth]{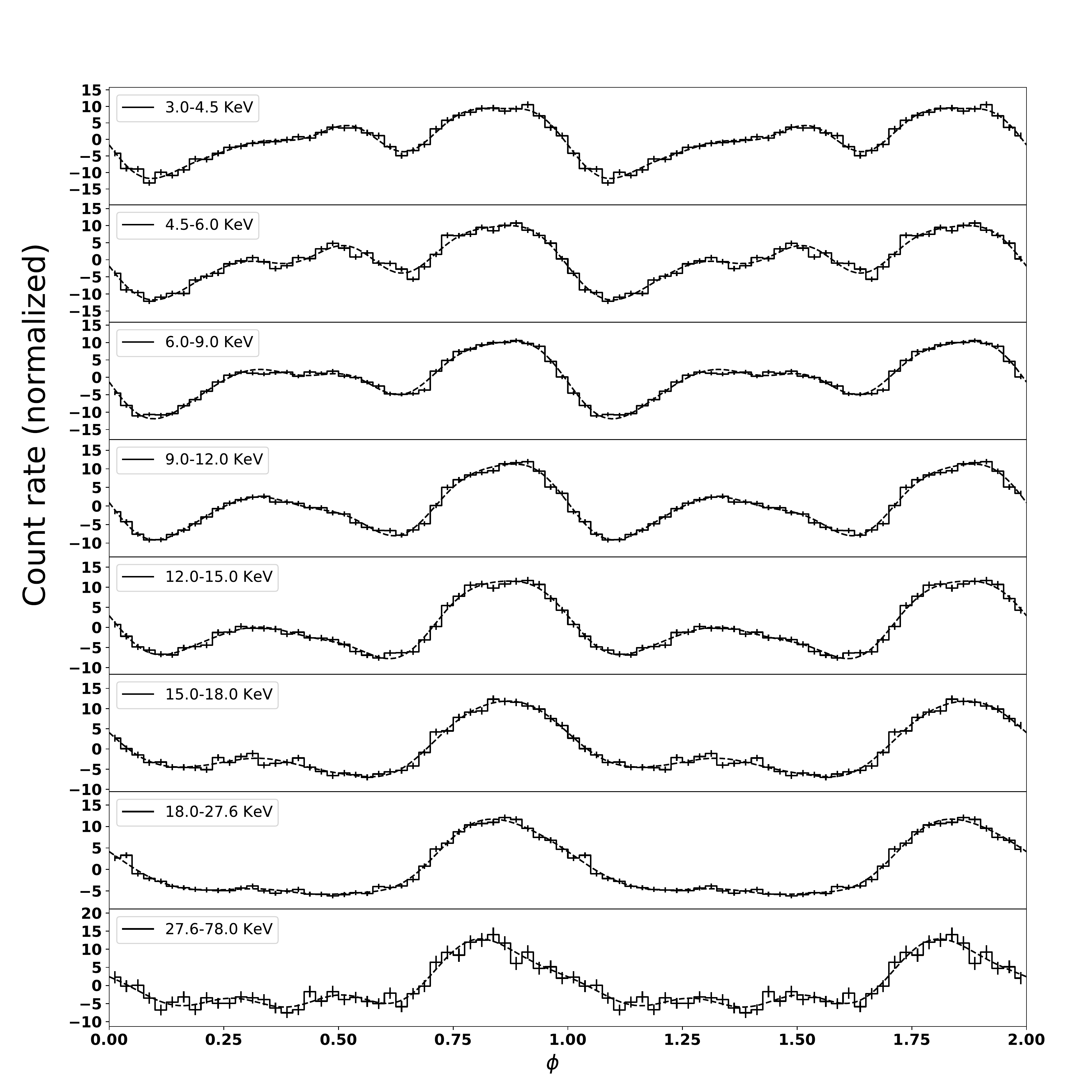} %
    \qquad
    \includegraphics[width=0.47\linewidth]{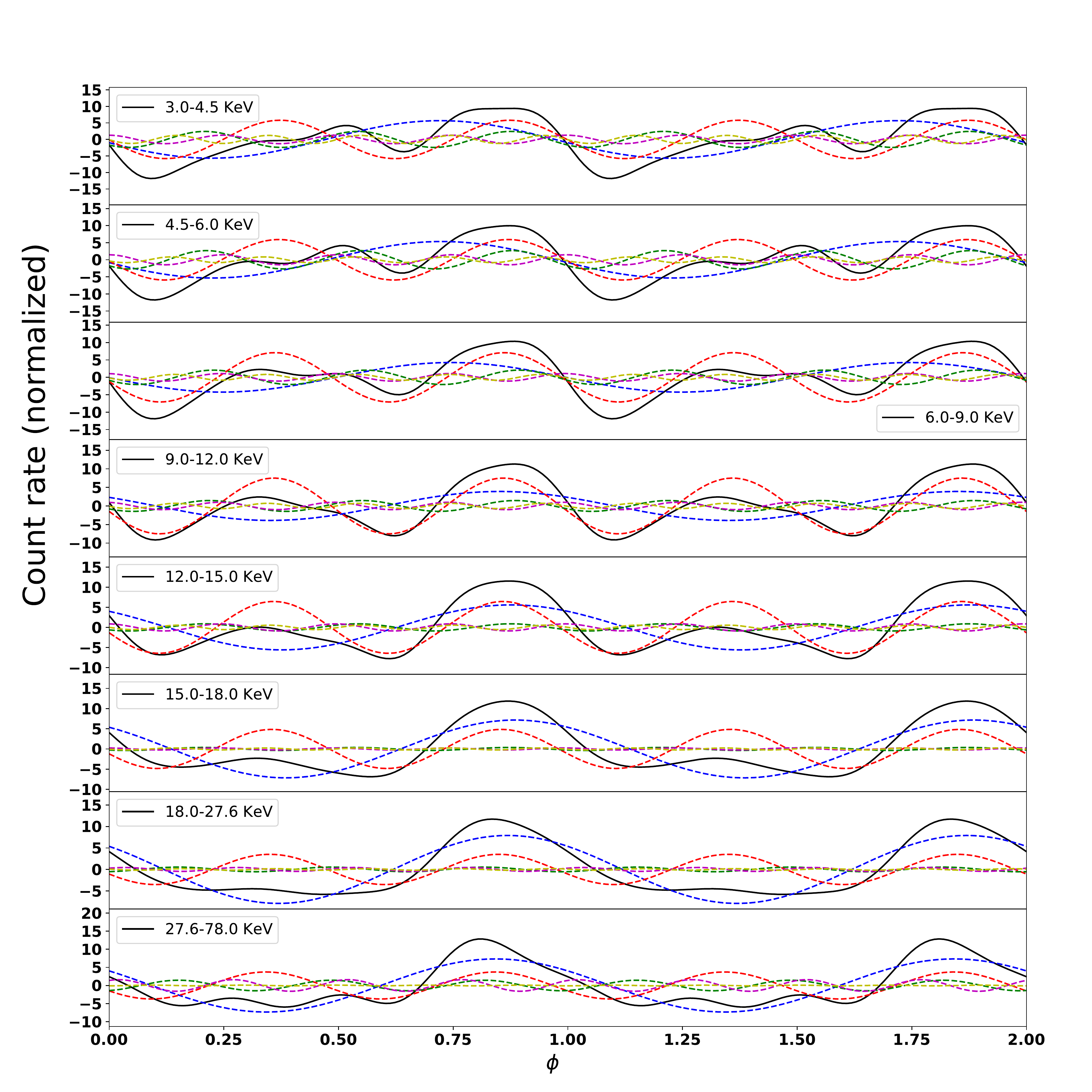} 
    \caption{Left: Normalized pulse profiles of 4U~1901+03 (the observation with OBS ID=90502307002 and a luminosity value of $1.483^{+0.002}_{-0.002} \cdot 10^{38} \mathrm{erg\cdot s^{-1}}$) at different energy ranges based on \nustar data, the measured profiles are represented by solid lines whereas the Fourier fits are represented by dashed lines, the boundaries of the energy ranges are [3, 4.5, 6, 9, 12, 15, 18, 27.6 and 78] keV. Right: The black solid line represents the overall approximate pulse profile and the colour dashed lines represent the different Fourier components. Blue: 1st, red: 2nd, green: 3rd, purple: 4th and yellow: 5th.}%
    \label{fig:4U1901+03_90502307002_NUSTAR_pulse_profiles}%
\end{figure*}

\begin{figure*}
	\centering
	\includegraphics[width=\linewidth]{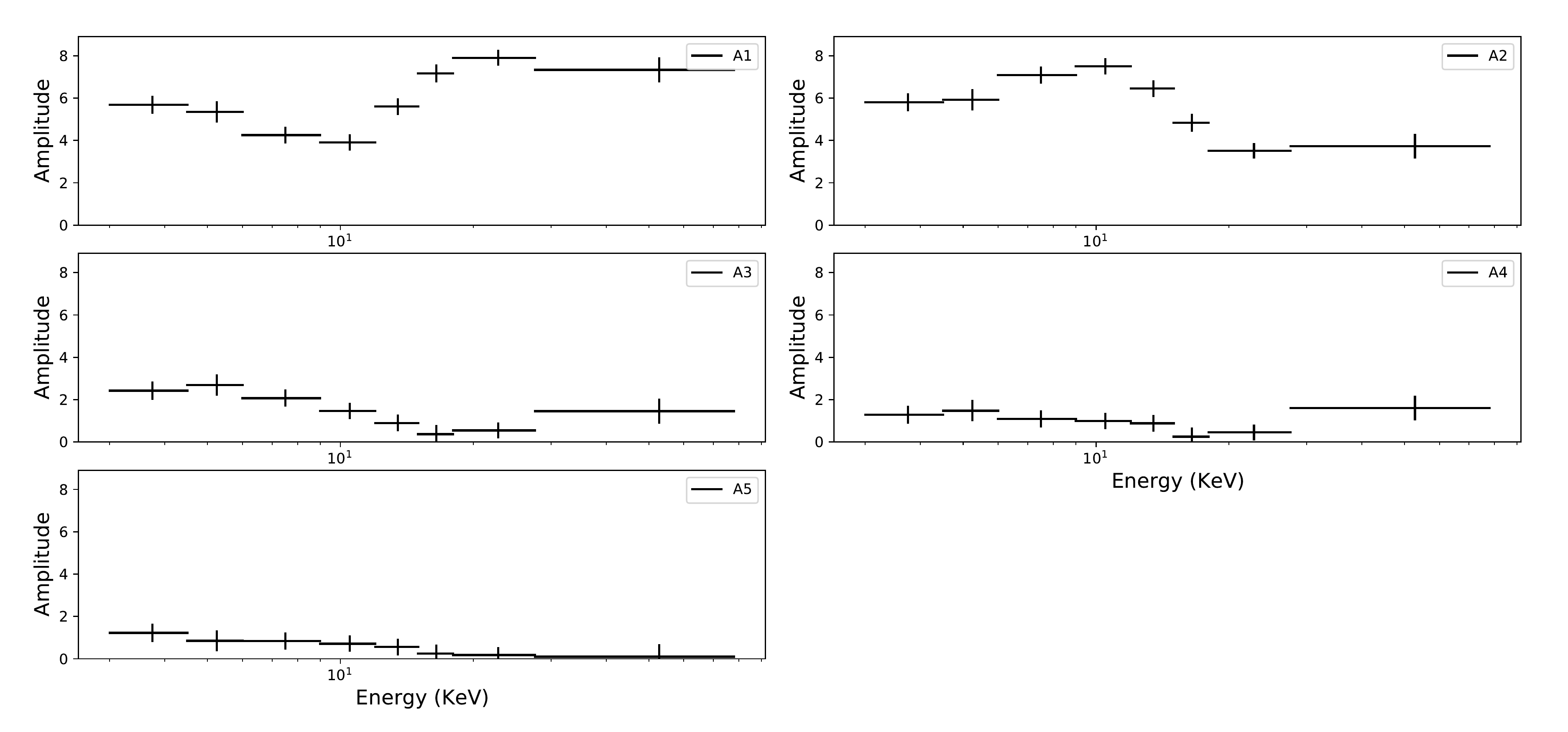}
	\caption{Variation of the harmonic amplitudes for 4U~1901+03 (the observation with OBS ID=90502307002 and a luminosity value of $1.483^{+0.002}_{-0.002} \cdot 10^{38} \mathrm{erg\cdot s^{-1}}$), the boundaries of the energy ranges are [3, 4.5, 6, 9, 12, 15, 18, 27.6 and 78] keV. Only \nustar data is represented by black markers.}
	\label{fig:4U1901+03_90502307002_fourier_variation}
\end{figure*}

\begin{figure*}%
    \centering
    \includegraphics[width=0.47\linewidth]{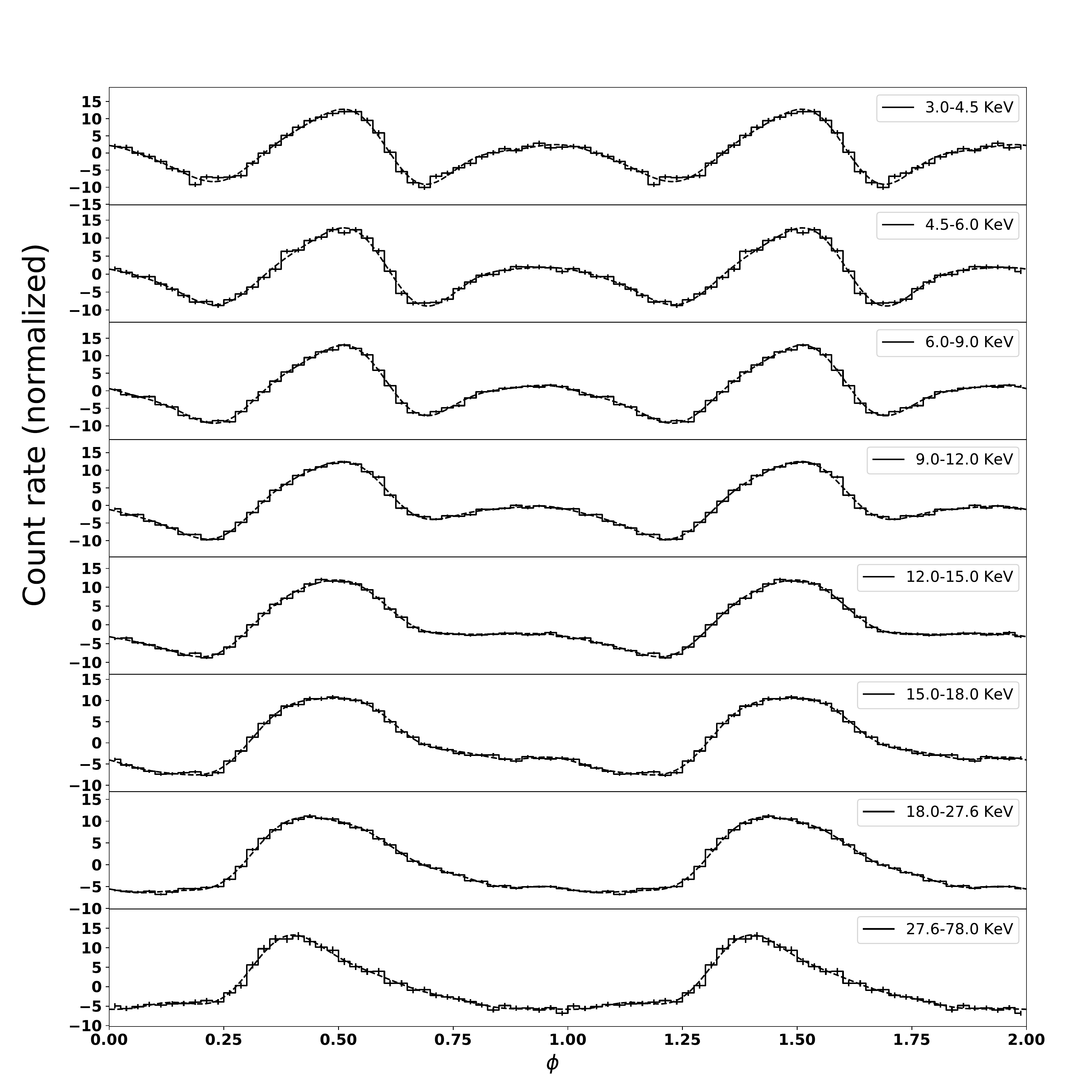} %
    \qquad
    \includegraphics[width=0.47\linewidth]{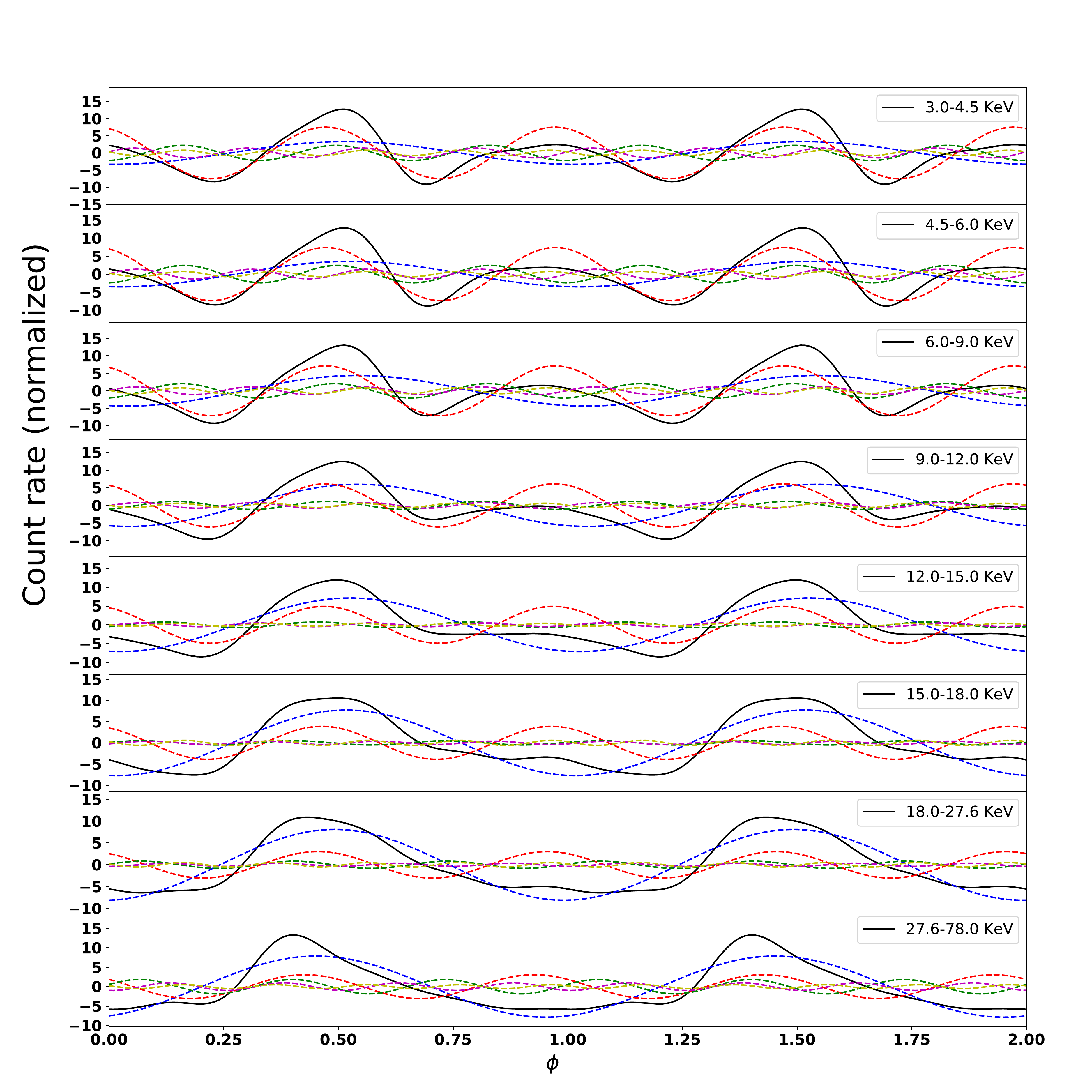} 
    \caption{Left: Normalized pulse profiles of 4U~1901+03 (the observation with OBS ID=90501305001 and a luminosity value of $1.242^{+0.010}_{-0.005} \cdot 10^{38} \mathrm{erg\cdot s^{-1}}$) at different energy ranges based on \nustar data, the measured profiles are represented by solid lines whereas the Fourier fits are represented by dashed lines, the boundaries of the energy ranges are [3, 4.5, 6, 9, 12, 15, 18, 27.6 and 78] keV. Right: The black solid line represents the overall approximate pulse profile and the colour dashed lines represent the different Fourier components. Blue: 1st, red: 2nd, green: 3rd, purple: 4th and yellow: 5th.}%
    \label{fig:4U1901+03_90502307002_NUSTAR_pulse_profiles}%
\end{figure*}

\begin{figure*}
	\centering
	\includegraphics[width=\linewidth]{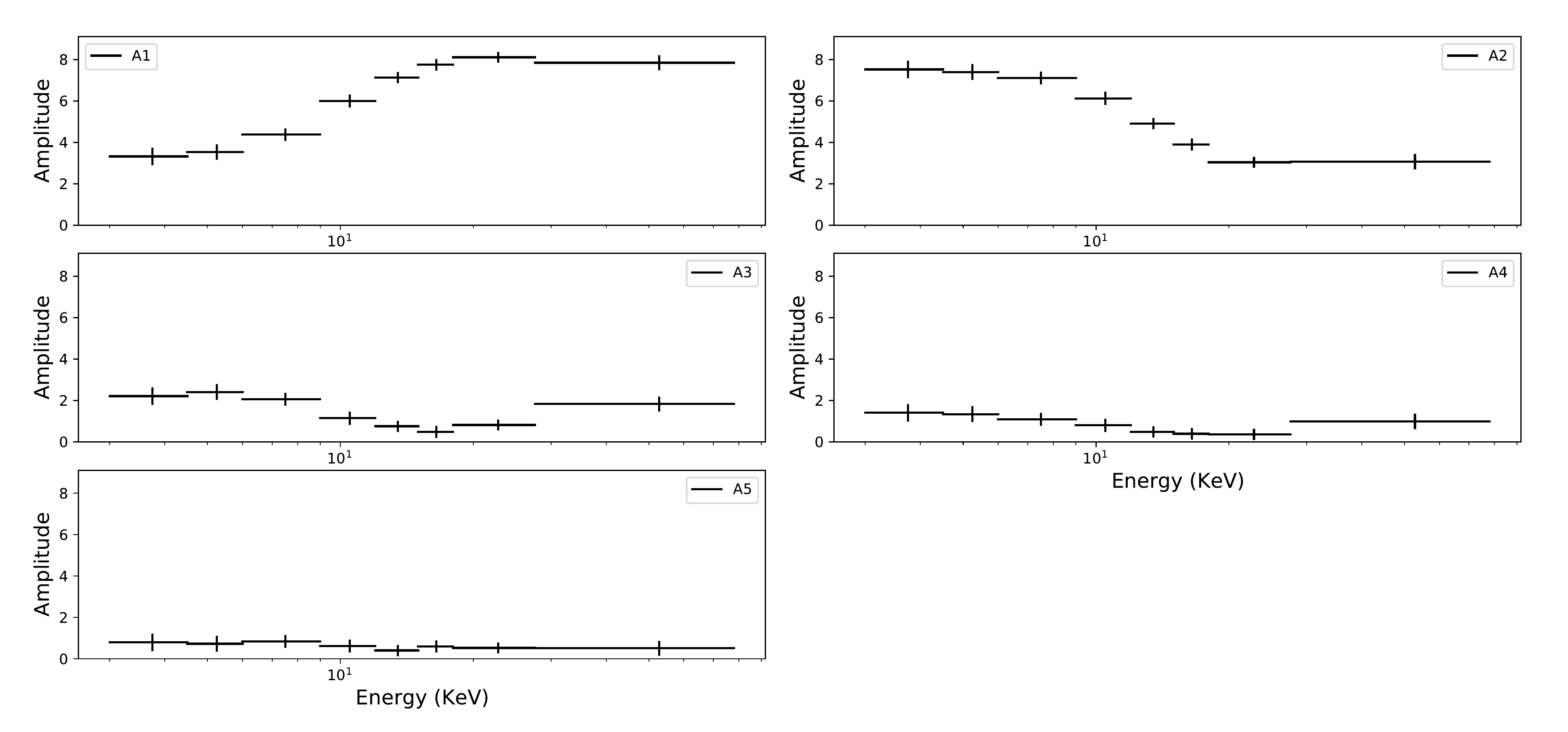}
	\caption{Variation of the harmonic amplitudes for 4U~1901+03 (the observation with OBS ID=90501305001 and a luminosity value of $1.242^{+0.010}_{-0.005} \cdot 10^{38} \mathrm{erg\cdot s^{-1}}$), the boundaries of the energy ranges are [3, 4.5, 6, 9, 12, 15, 18, 27.6 and 78] keV. Only \nustar data is represented by black markers.}
	\label{fig:4U1901+03_90502307002fourier_variation}
\end{figure*}


\begin{figure*}%
    \centering
    \includegraphics[width=0.47\linewidth]{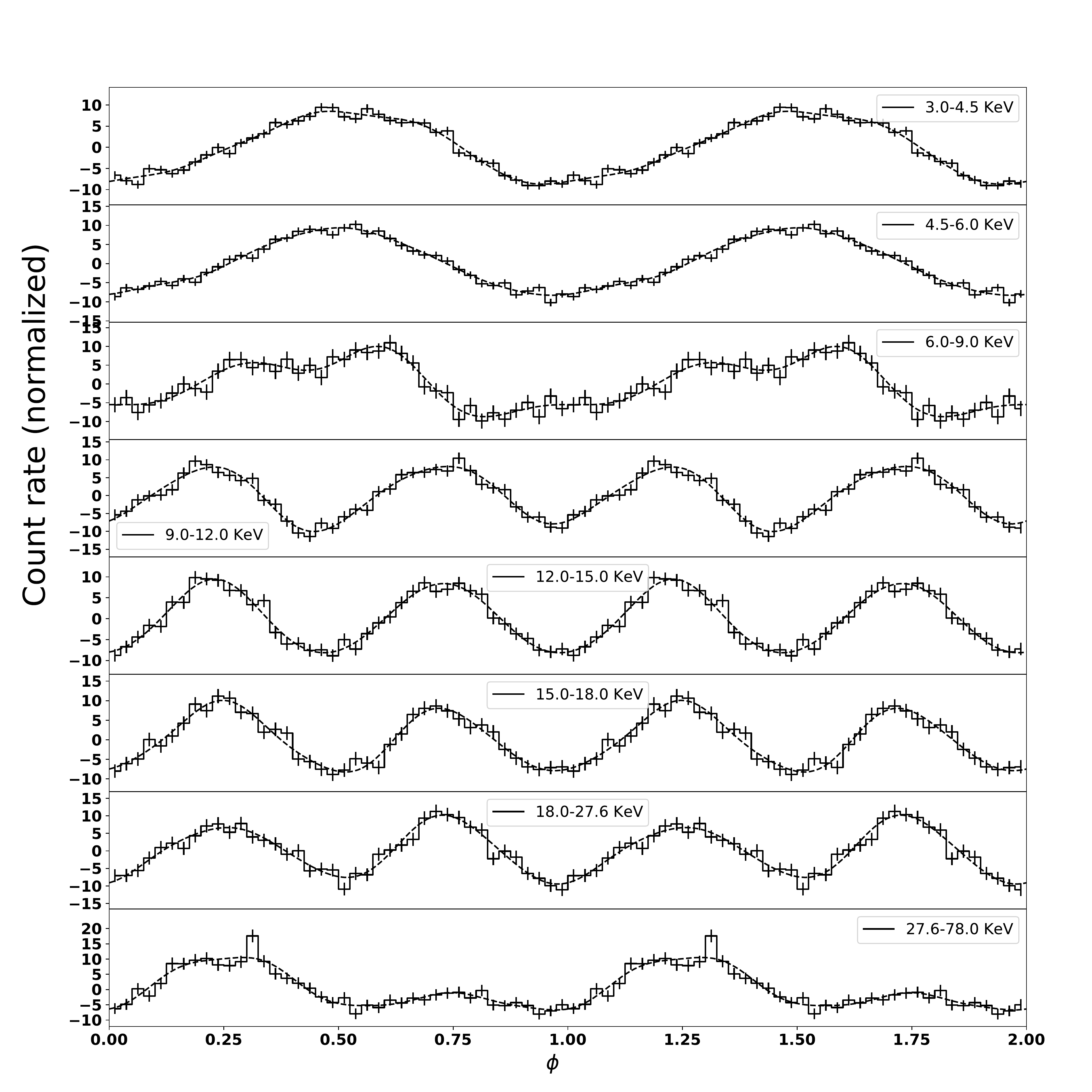} %
    \qquad
    \includegraphics[width=0.47\linewidth]{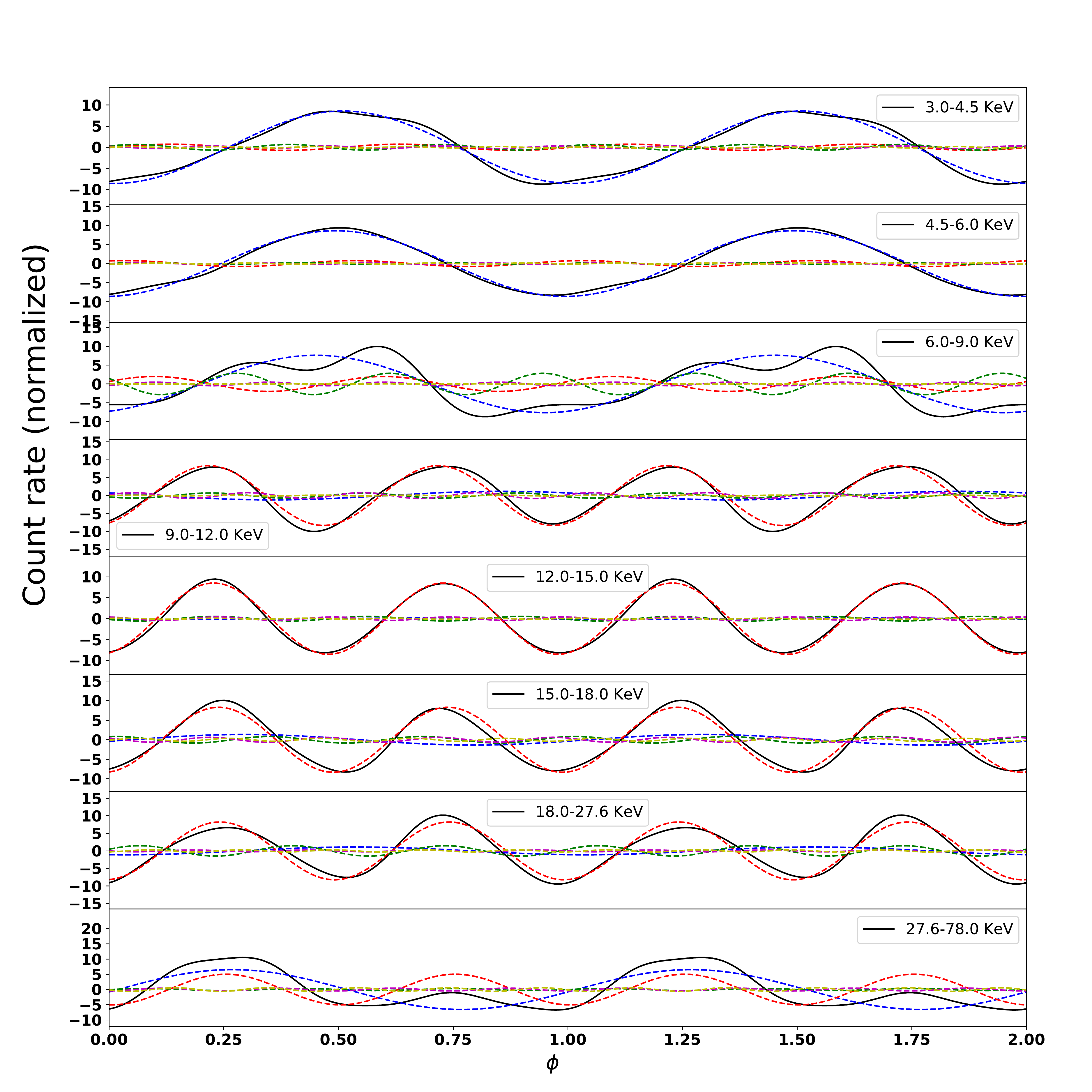} 
    \caption{Left: Normalized pulse profiles of V~0332+53 (the observation with OBS ID=80102002002 and a luminosity value of $(1.24-1.74)\cdot 10^{38} \mathrm{erg\cdot s^{-1}}$) at different energy ranges based on \nustar data, the measured profiles are represented by solid lines whereas the Fourier fits are represented by dashed lines, the boundaries of the energy ranges are [3, 4.5, 6, 9, 12, 15, 18, 27.6 and 78] keV. Right: The black solid line represents the overall approximate pulse profile and the colour dashed lines represent the different Fourier components. Blue: 1st, red: 2nd, green: 3rd, purple: 4th and yellow: 5th.}%
    \label{fig:V0332+53_80102002002_NUSTAR_pulse_profiles}%
\end{figure*}

\begin{figure*}
	\centering
	\includegraphics[width=\linewidth]{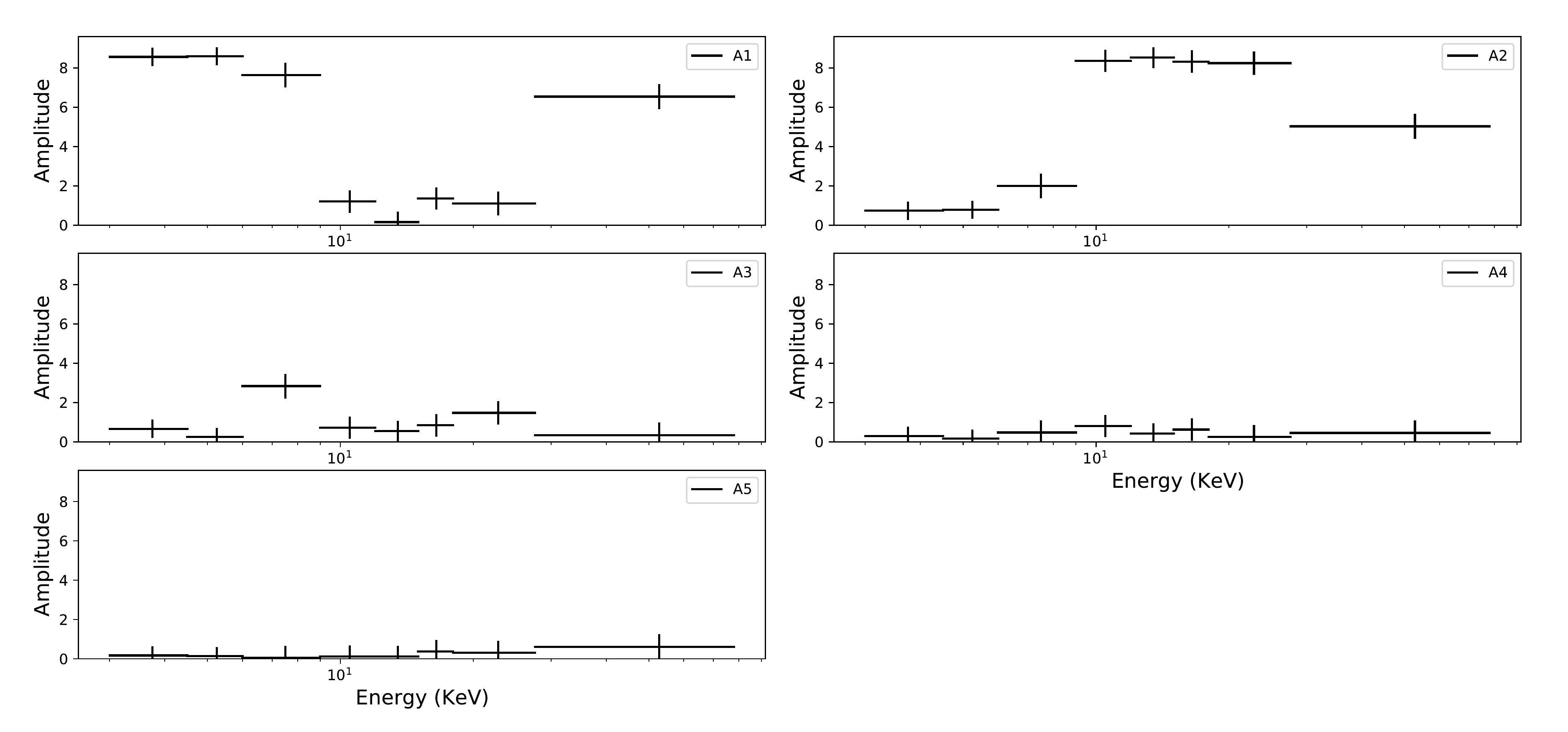}
	\caption{Variation of the harmonic amplitudes for V~0332+53 (the observation with OBS ID=80102002002 and a luminosity value of $(1.24-1.74)\cdot 10^{38} \mathrm{erg\cdot s^{-1}}$), the boundaries of the energy ranges are [3, 4.5, 6, 9, 12, 15, 18, 27.6 and 78] keV. Only \nustar data is represented by black markers.}
	\label{fig:V0332+53_80102002002fourier_variation}
\end{figure*}

\begin{figure*}%
    \centering
    \includegraphics[width=0.47\linewidth]{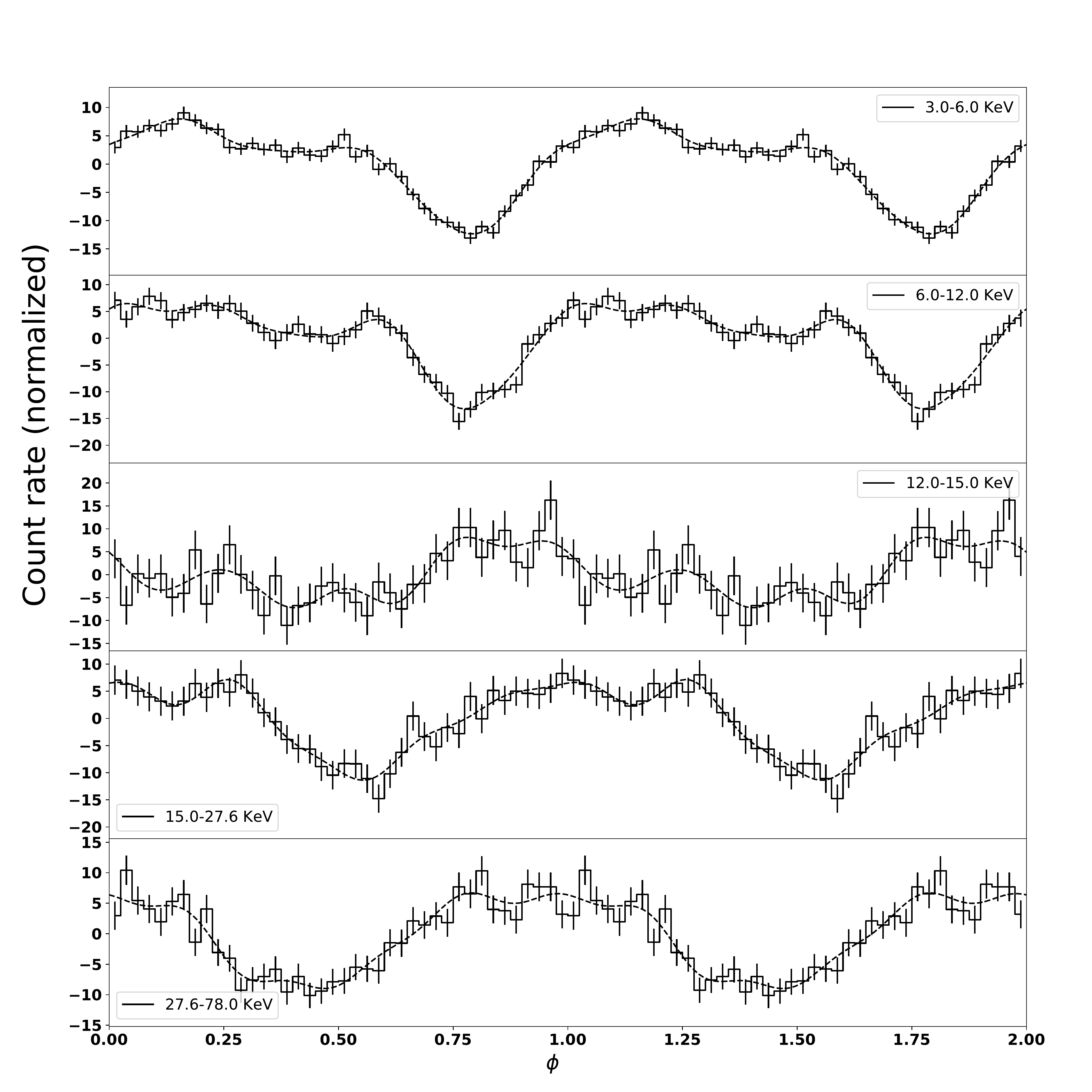} %
    \qquad
    \includegraphics[width=0.47\linewidth]{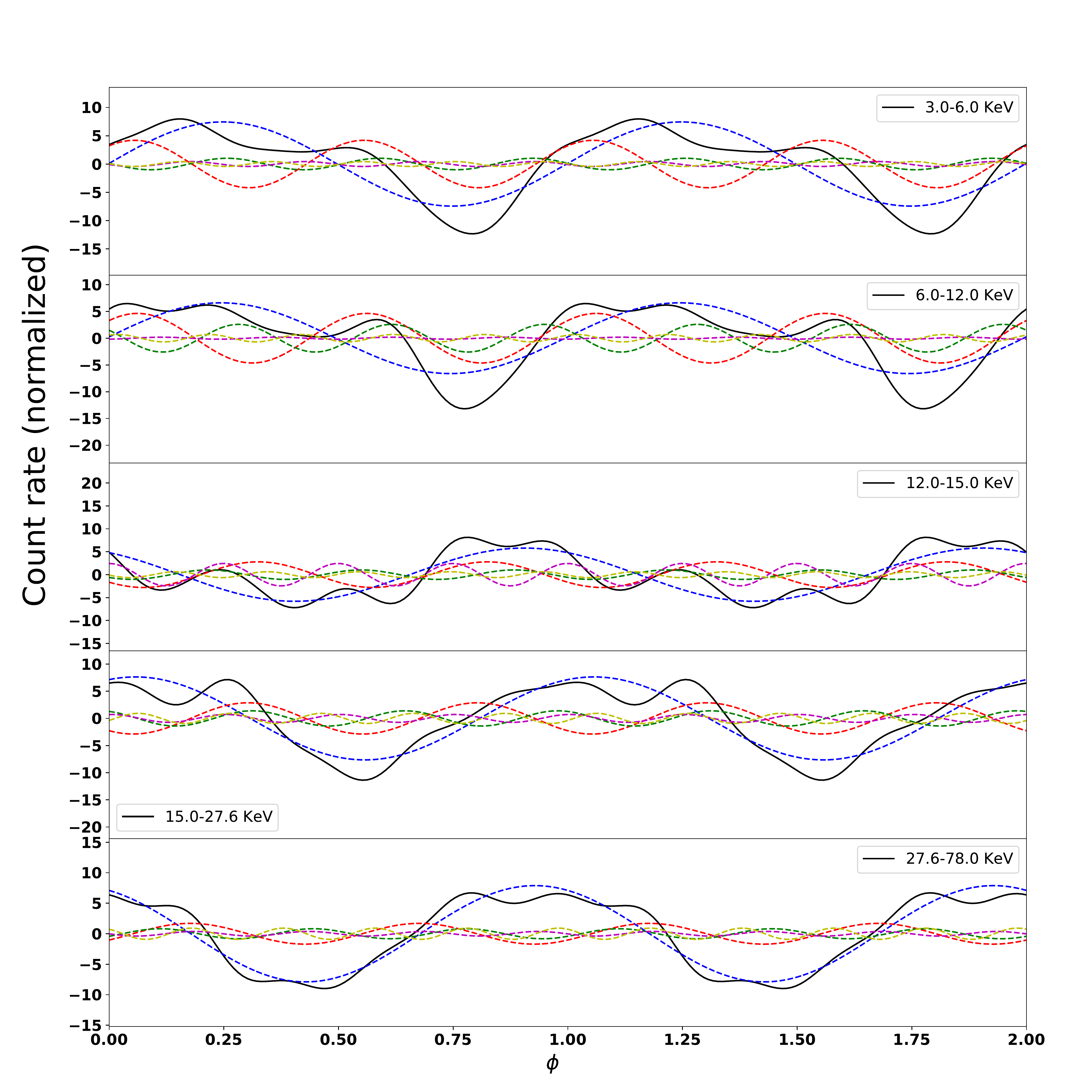} 
    \caption{Left: Normalized pulse profiles of V~0332+53 (the observation with OBS ID=80102002004 and a luminosity value of $(5.7-9.5)\cdot 10^{37} \mathrm{erg\cdot s^{-1}}$) at different energy ranges based on \nustar data, the measured profiles are represented by solid lines whereas the Fourier fits are represented by dashed lines, the boundaries of the energy ranges are [3, 4.5, 6, 9, 12, 15, 18, 27.6 and 78] keV. Right: The black solid line represents the overall approximate pulse profile and the colour dashed lines represent the different Fourier components. Blue: 1st, red: 2nd, green: 3rd, purple: 4th and yellow: 5th.}%
    \label{fig:V0332+53_80102002004_NUSTAR_pulse_profiles}%
\end{figure*}

\begin{figure*}
	\centering
	\includegraphics[width=\linewidth]{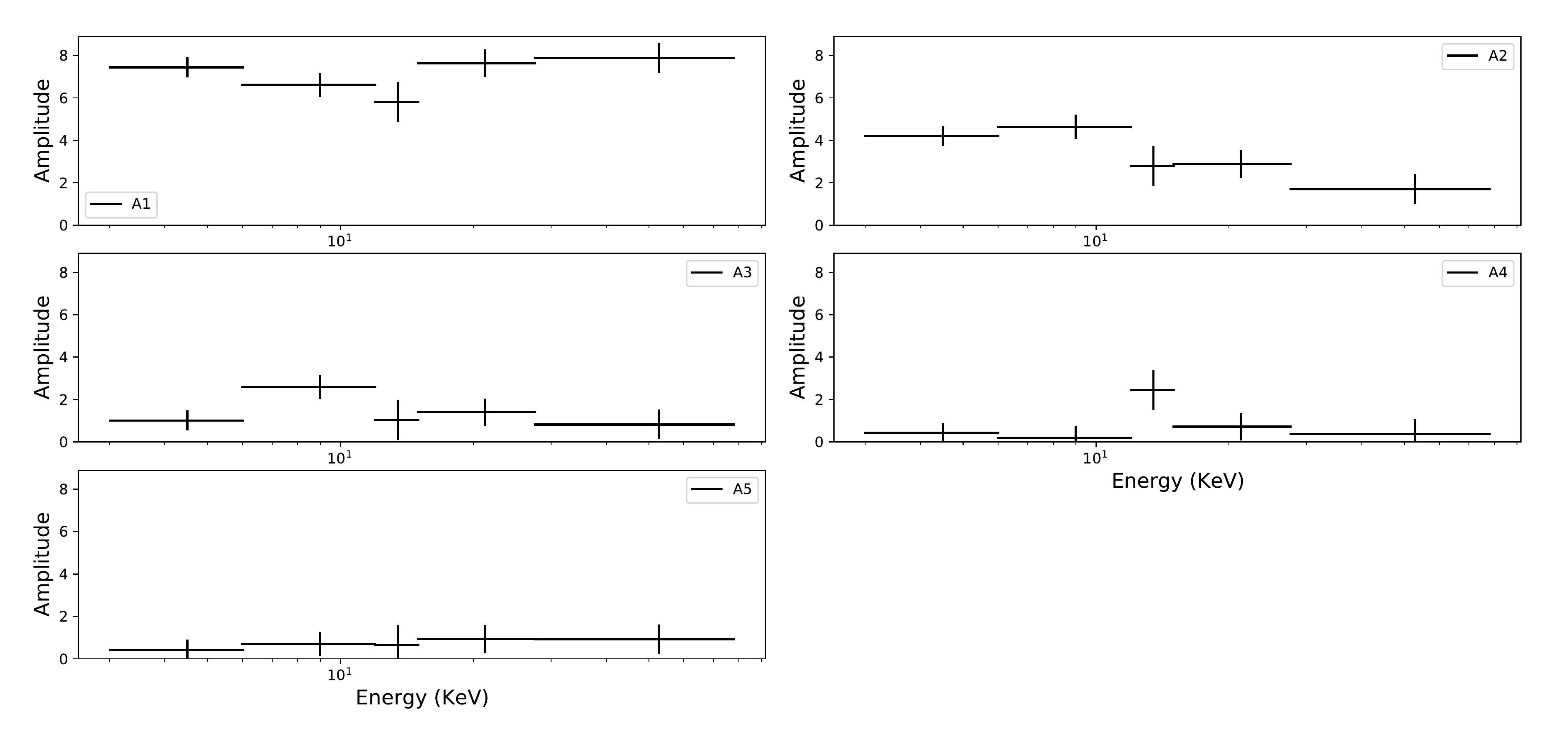}
	\caption{Variation of the harmonic amplitudes for V~0332+53 (the observation with OBS ID=80102002004 and a luminosity value of $(5.7-9.5)\cdot 10^{37} \mathrm{erg\cdot s^{-1}}$), the boundaries of the energy ranges are [3, 4.5, 6, 9, 12, 15, 18, 27.6 and 78] keV. Only \nustar data is represented by black markers.}
	\label{fig:V0332+53_80102002004fourier_variation}
\end{figure*}

\begin{figure*}%
    \centering
    \includegraphics[width=0.47\linewidth]{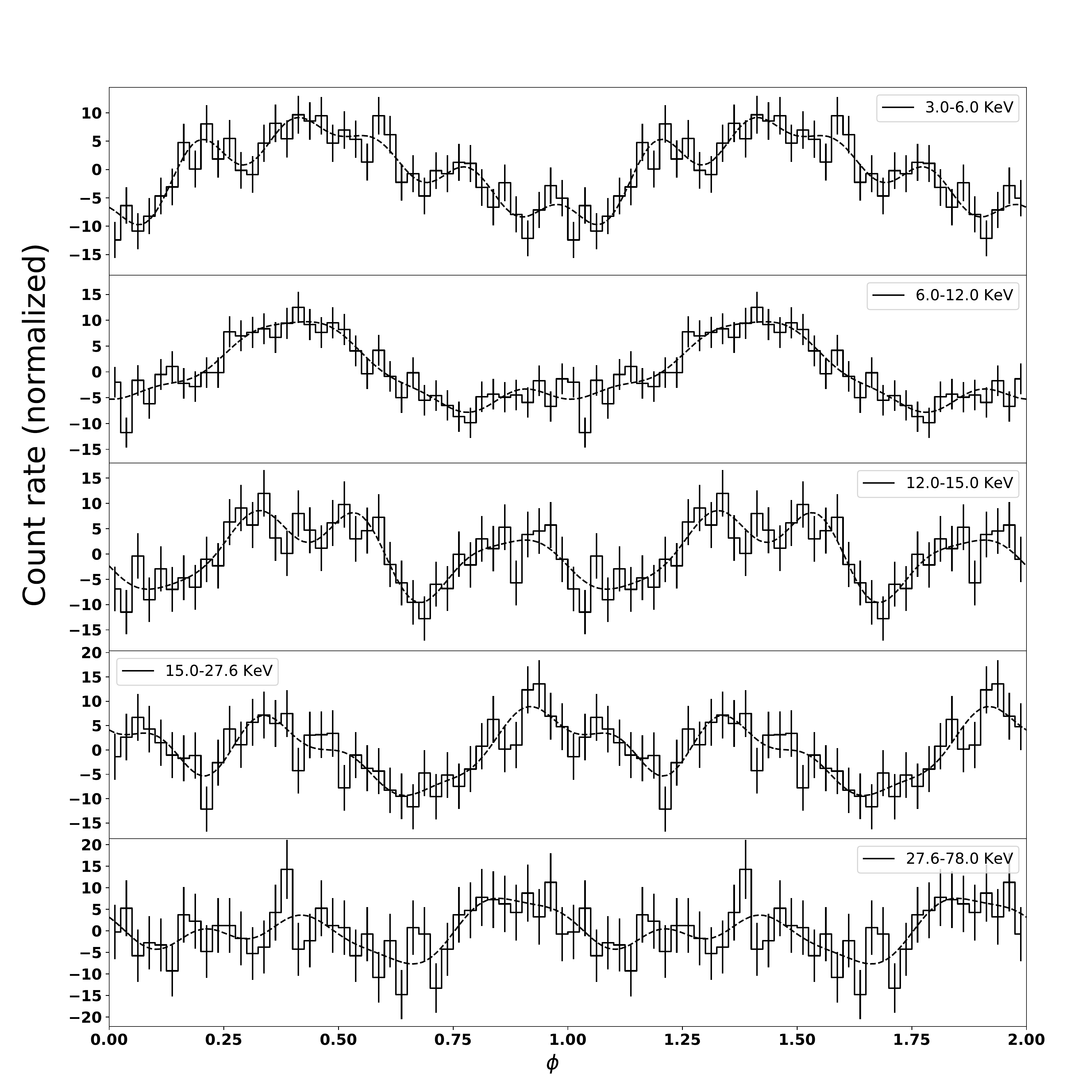} %
    \qquad
    \includegraphics[width=0.47\linewidth]{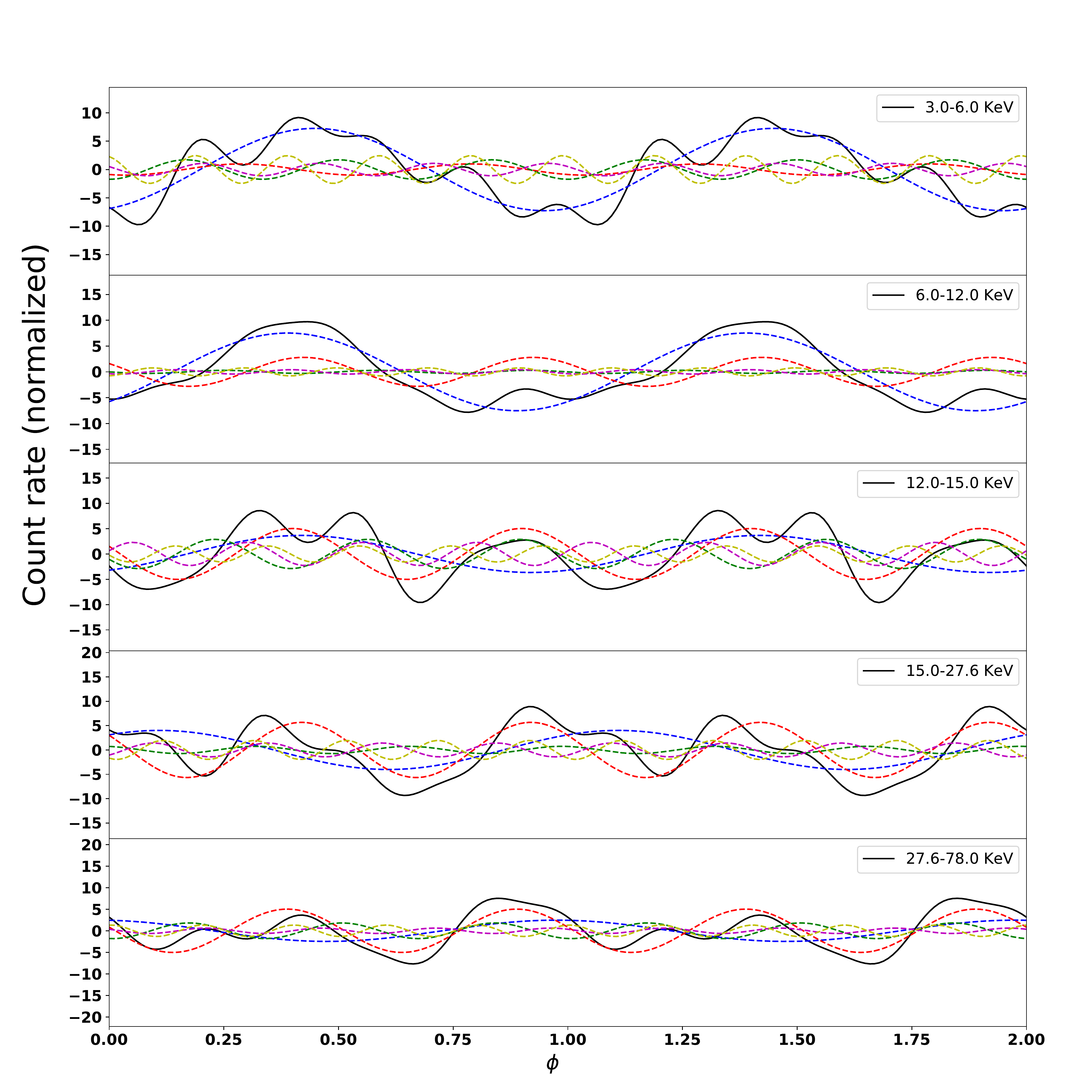} 
    \caption{Left: Normalized pulse profiles of V~0332+53 (the observation with OBS ID=80102002010 and a luminosity value of $(5.3-14)\cdot 10^{36} \mathrm{erg\cdot s^{-1}}$) at different energy ranges based on \nustar data, the measured profiles are represented by solid lines whereas the Fourier fits are represented by dashed lines, the boundaries of the energy ranges are [3, 4.5, 6, 9, 12, 15, 18, 27.6 and 78] keV. Right: The black solid line represents the overall approximate pulse profile and the colour dashed lines represent the different Fourier components. Blue: 1st, red: 2nd, green: 3rd, purple: 4th and yellow: 5th.}%
    \label{fig:V0332+53_80102002010_NUSTAR_pulse_profiles}%
\end{figure*}

\begin{figure*}
	\centering
	\includegraphics[width=\linewidth]{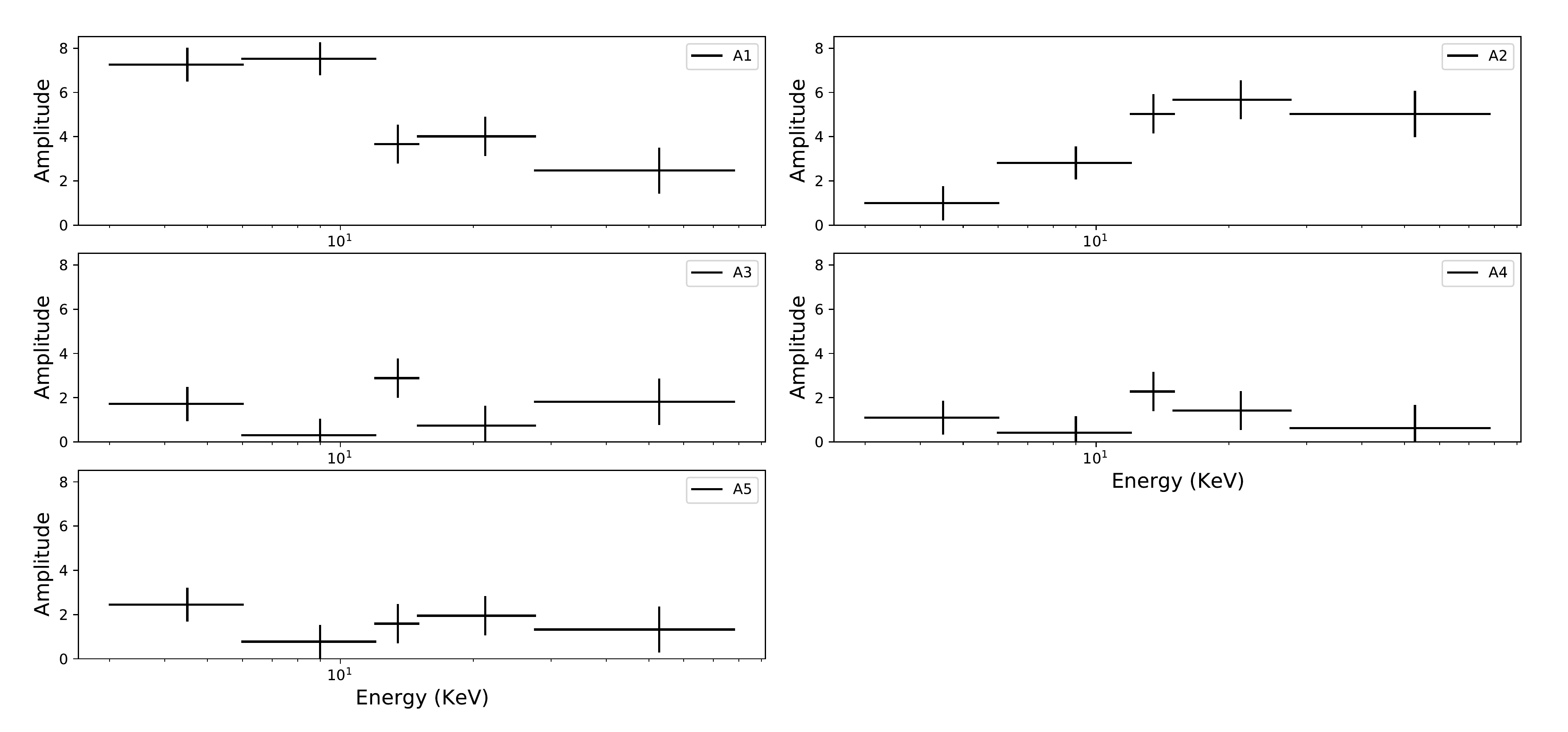}
	\caption{Variation of the harmonic amplitudes for V~0332+53 (the observation with OBS ID=80102002010 and a luminosity value of $(5.3-14)\cdot 10^{36} \mathrm{erg\cdot s^{-1}}$), the boundaries of the energy ranges are [3, 4.5, 6, 9, 12, 15, 18, 27.6 and 78] keV. Only \nustar data is represented by black markers.}
	\label{fig:V0332+53_80102002010fourier_variation}
\end{figure*}


\begin{figure*}%
    \centering
    \includegraphics[width=0.47\linewidth]{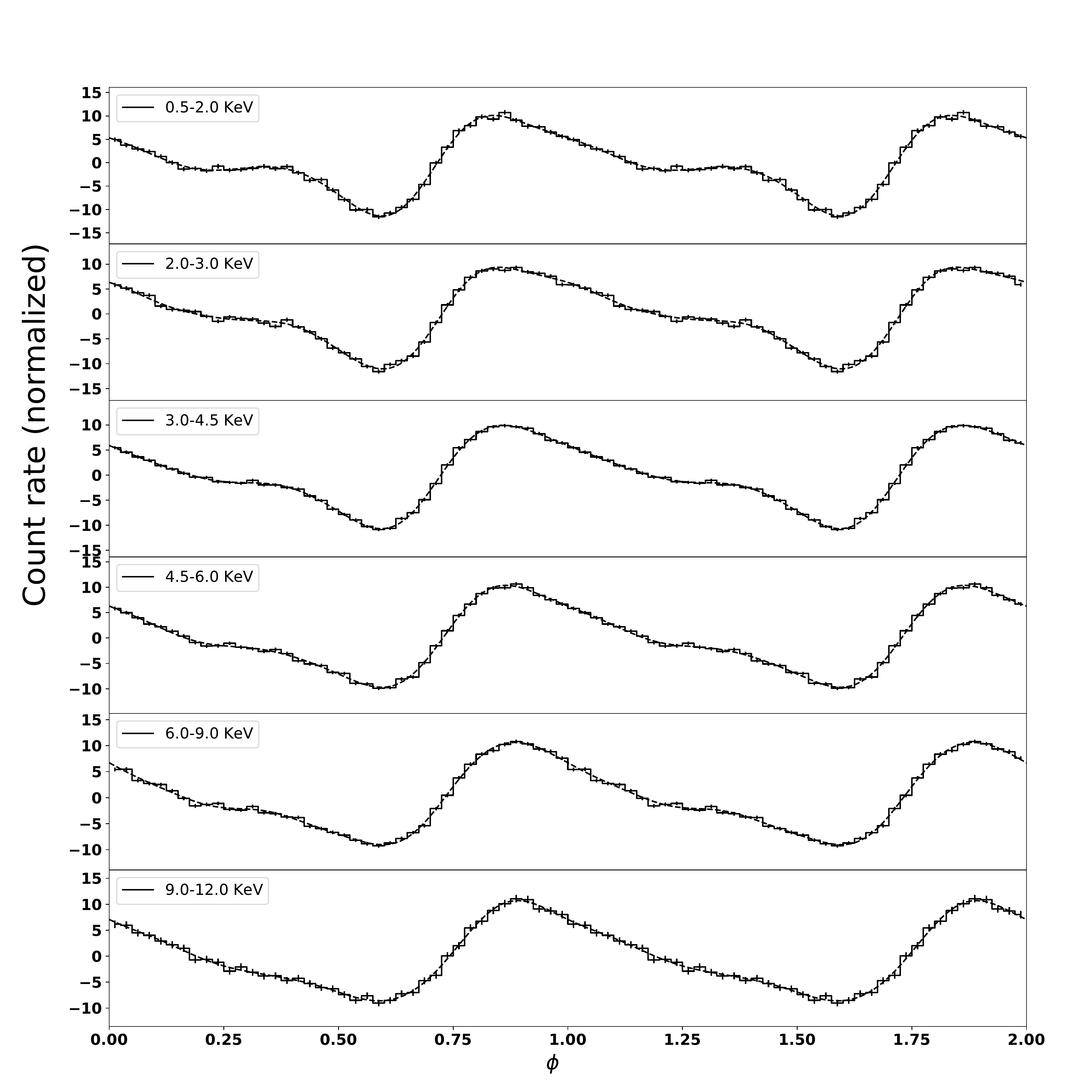} %
    \qquad
    \includegraphics[width=0.47\linewidth]{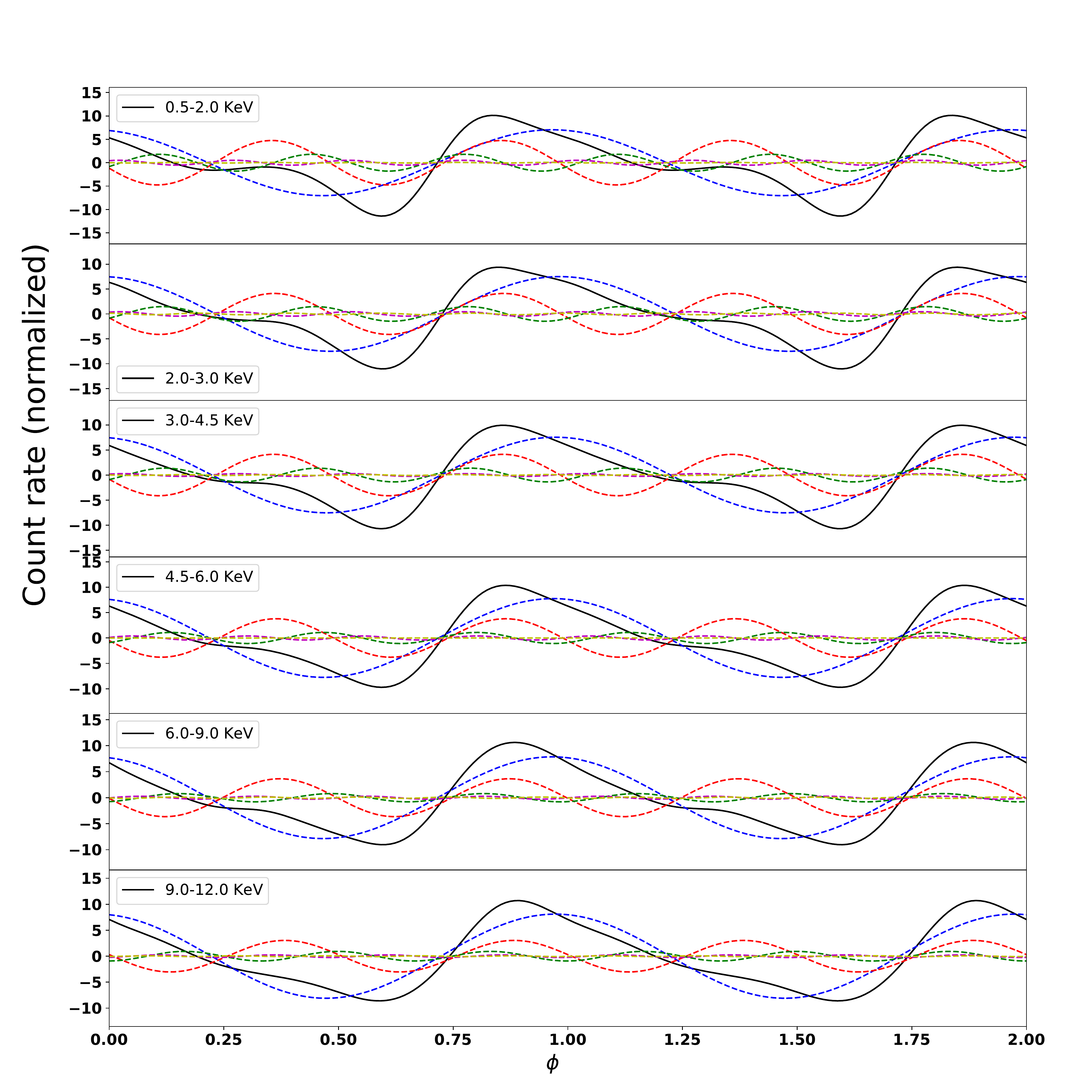} 
    \caption{Left: Normalized pulse profiles of Cen~X-3 at different energy ranges based on \xmm data, the measured profiles are represented by solid lines whereas the Fourier fits are represented by dashed lines, the boundaries of the energy ranges are [0.5, 2, 3, 4.5, 6, 9, 12] keV. Right: The black solid line represents the overall approximate pulse profile and the colour dashed lines represent the different Fourier components. Blue: 1st, red: 2nd, green: 3rd, purple: 4th and yellow: 5th.}%
    \label{fig:Cen_X-3_XMM_pulse_profiles}%
\end{figure*}

\begin{figure*}%
    \centering
    \includegraphics[width=0.47\linewidth]{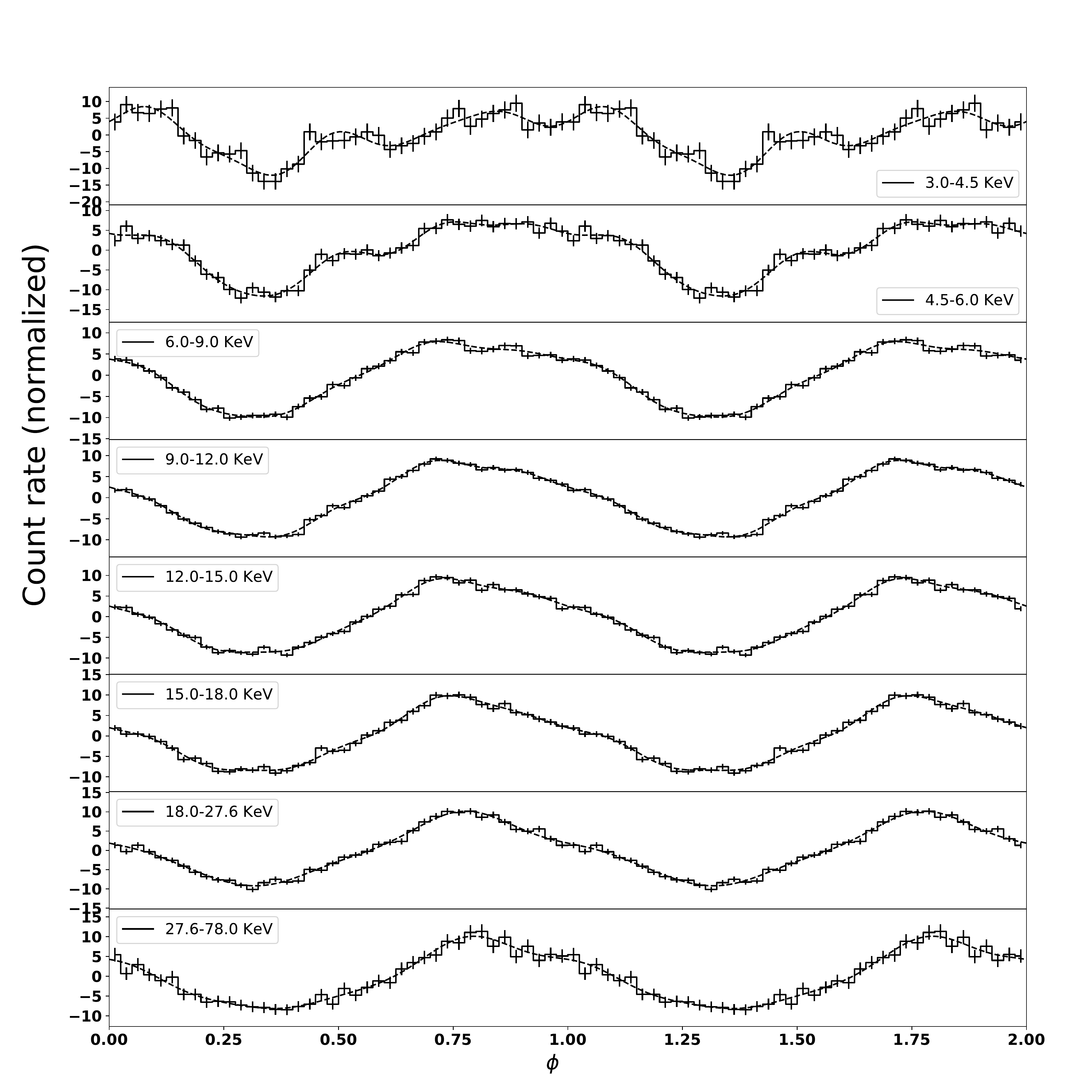} %
    \qquad
    \includegraphics[width=0.47\linewidth]{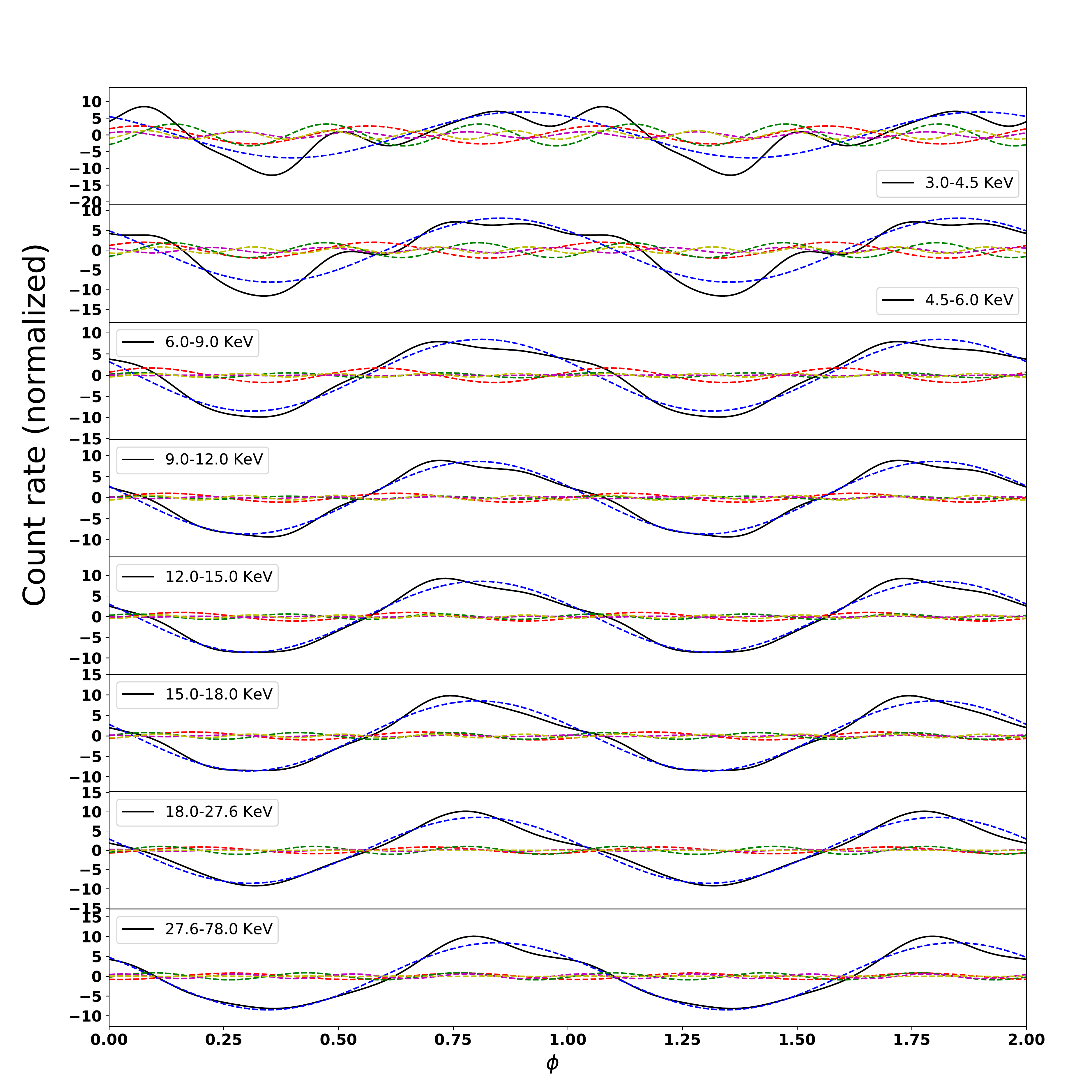} 
    \caption{Left: Normalized pulse profiles of Cen~X-3 at different energy ranges based on \nustar data, the measured profiles are represented by solid lines whereas the Fourier fits are represented by dashed lines, the boundaries of the energy ranges are [3, 4.5, 6, 9, 12, 15, 18, 27.6 and 78] keV. Right: The black solid line represents the overall approximate pulse profile and the colour dashed lines represent the different Fourier components. Blue: 1st, red: 2nd, green: 3rd, purple: 4th and yellow: 5th.}%
    \label{fig:Cen_X-3_NUSTAR_pulse_profiles}%
\end{figure*}

\begin{figure*}
	\centering
	\includegraphics[width=\linewidth]{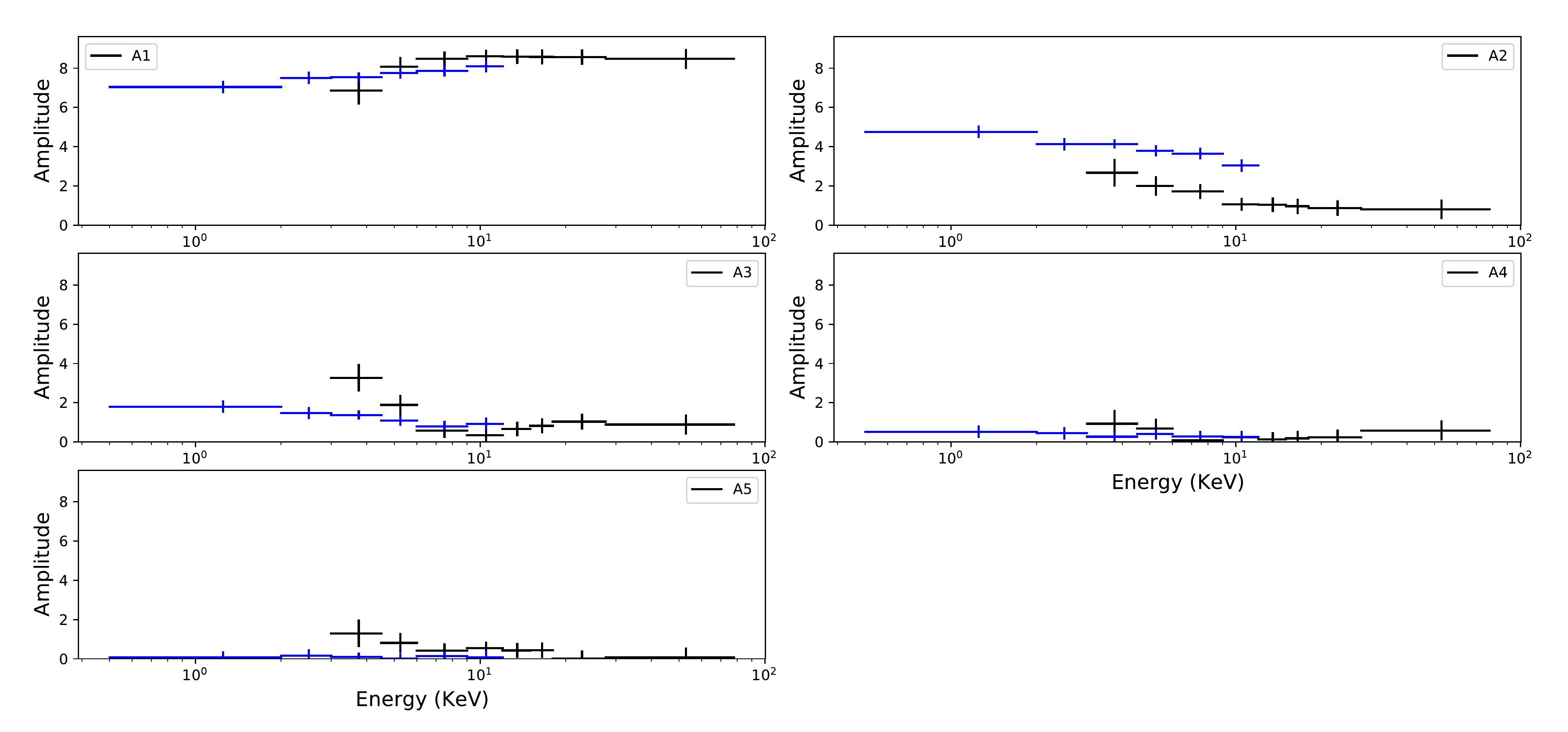}
	\caption{Variation of the harmonic amplitudes for Cen~X-3, the boundaries of the energy ranges are [0.5, 2, 3, 4.5, 6, 9, 12, 15, 18, 27.6 and 78] keV. \xmm data is represented by blue markers and \nustar data is represented by black markers.}
	\label{fig:Cen_X-3_fourier_variation}
\end{figure*}


\begin{figure*}%
    \centering
    \includegraphics[width=0.47\linewidth]{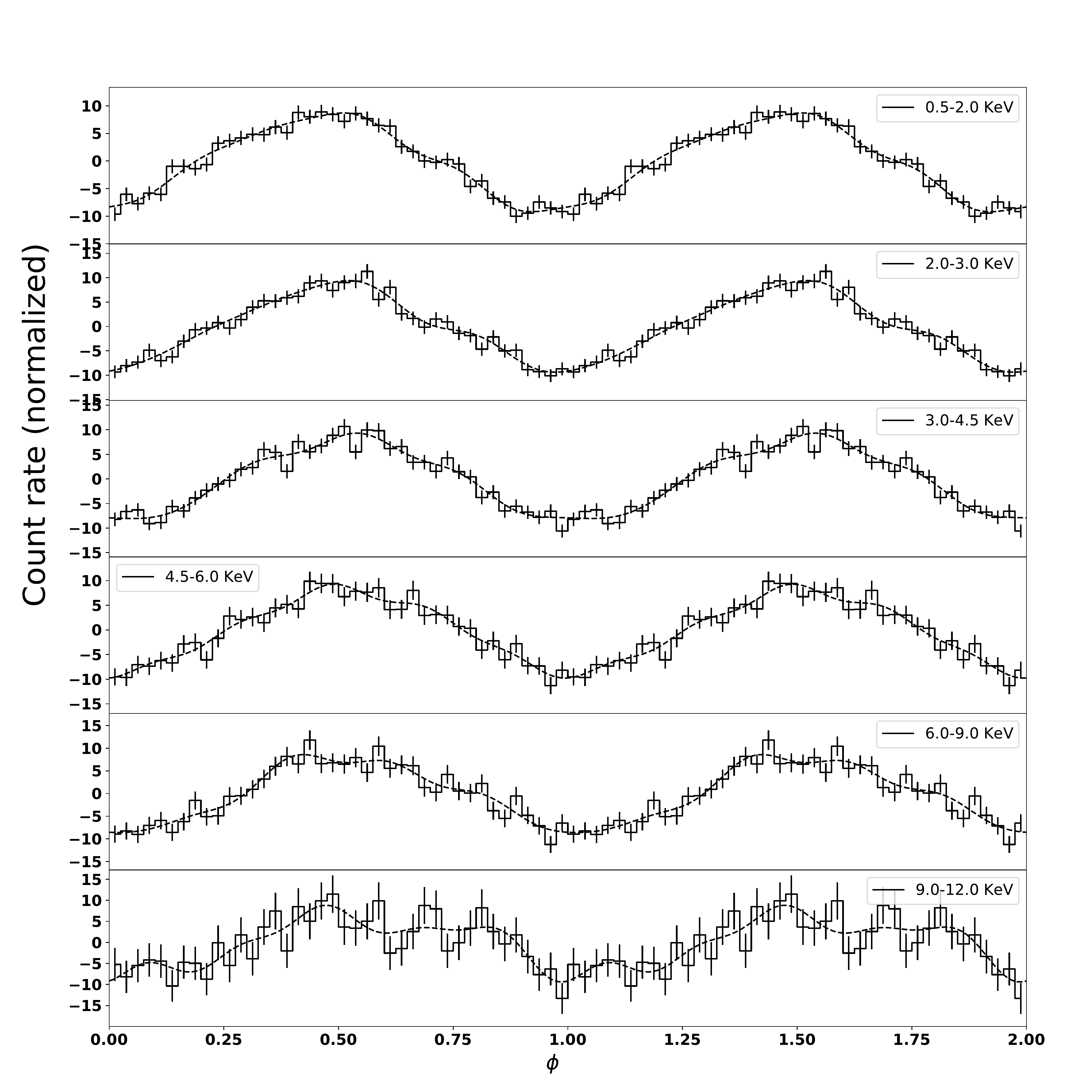} %
    \qquad
    \includegraphics[width=0.47\linewidth]{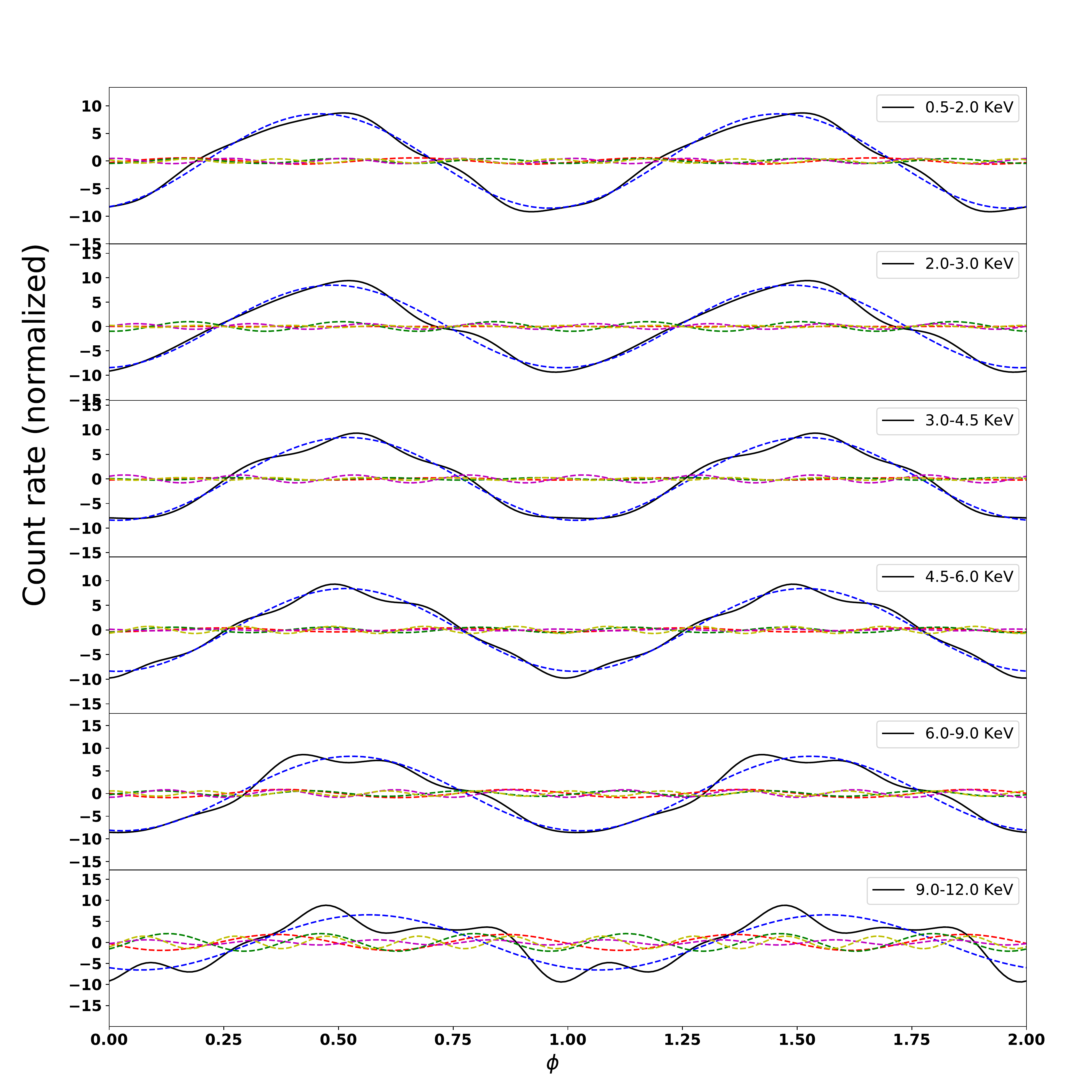} 
    \caption{Left: Normalized pulse profiles of LMC~X-4 at different energy ranges based on \xmm data, the measured profiles are represented by solid lines whereas the Fourier fits are represented by dashed lines, the boundaries of the energy ranges are [0.5, 2, 3, 4.5, 6, 9, 12] keV. Right: The black solid line represents the overall approximate pulse profile and the colour dashed lines represent the different Fourier components. Blue: 1st, red: 2nd, green: 3rd, purple: 4th and yellow: 5th.}%
    \label{fig:LMC_X-4_XMM_pulse_profiles}%
\end{figure*}

\begin{figure*}%
    \centering
    \includegraphics[width=0.47\linewidth]{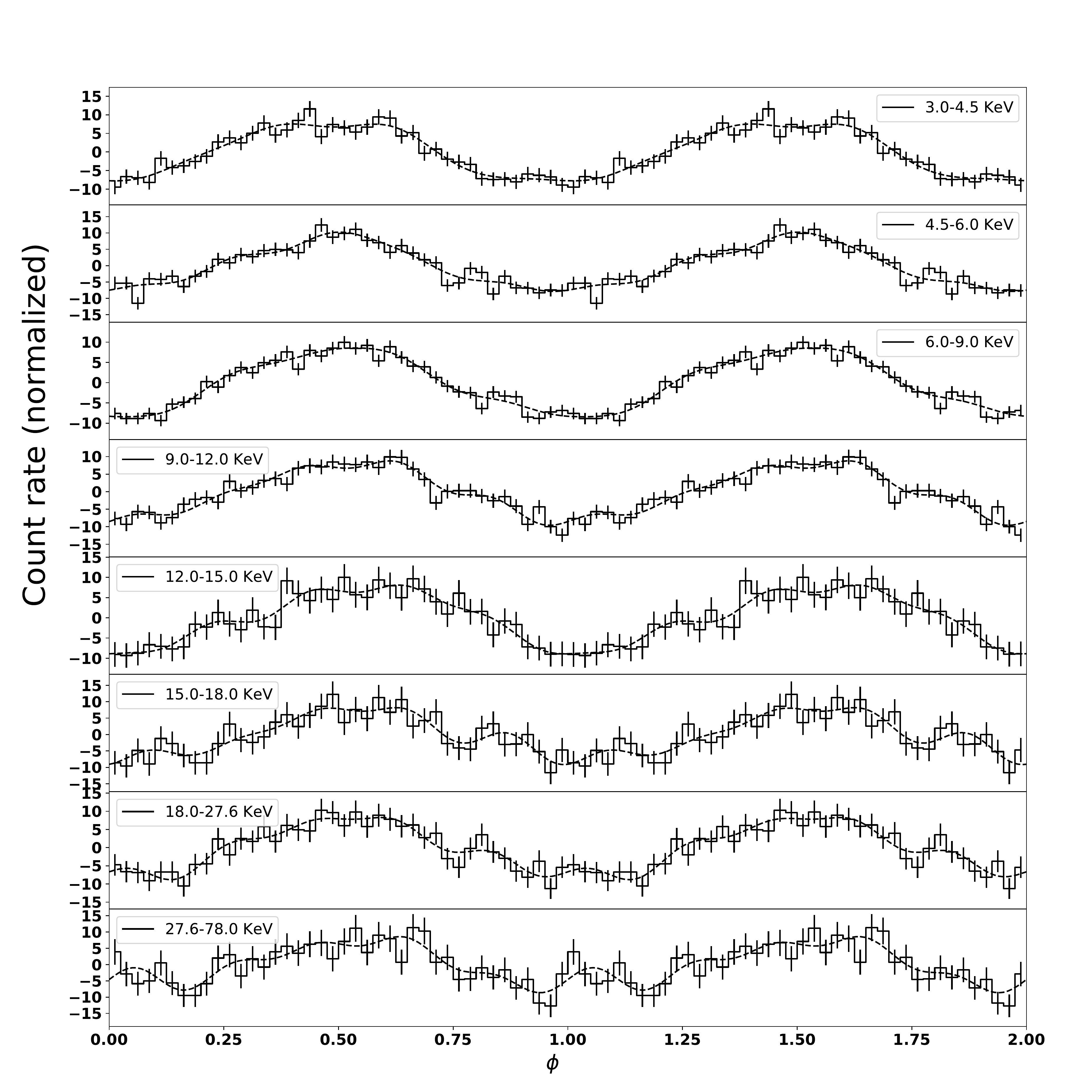} %
    \qquad
    \includegraphics[width=0.47\linewidth]{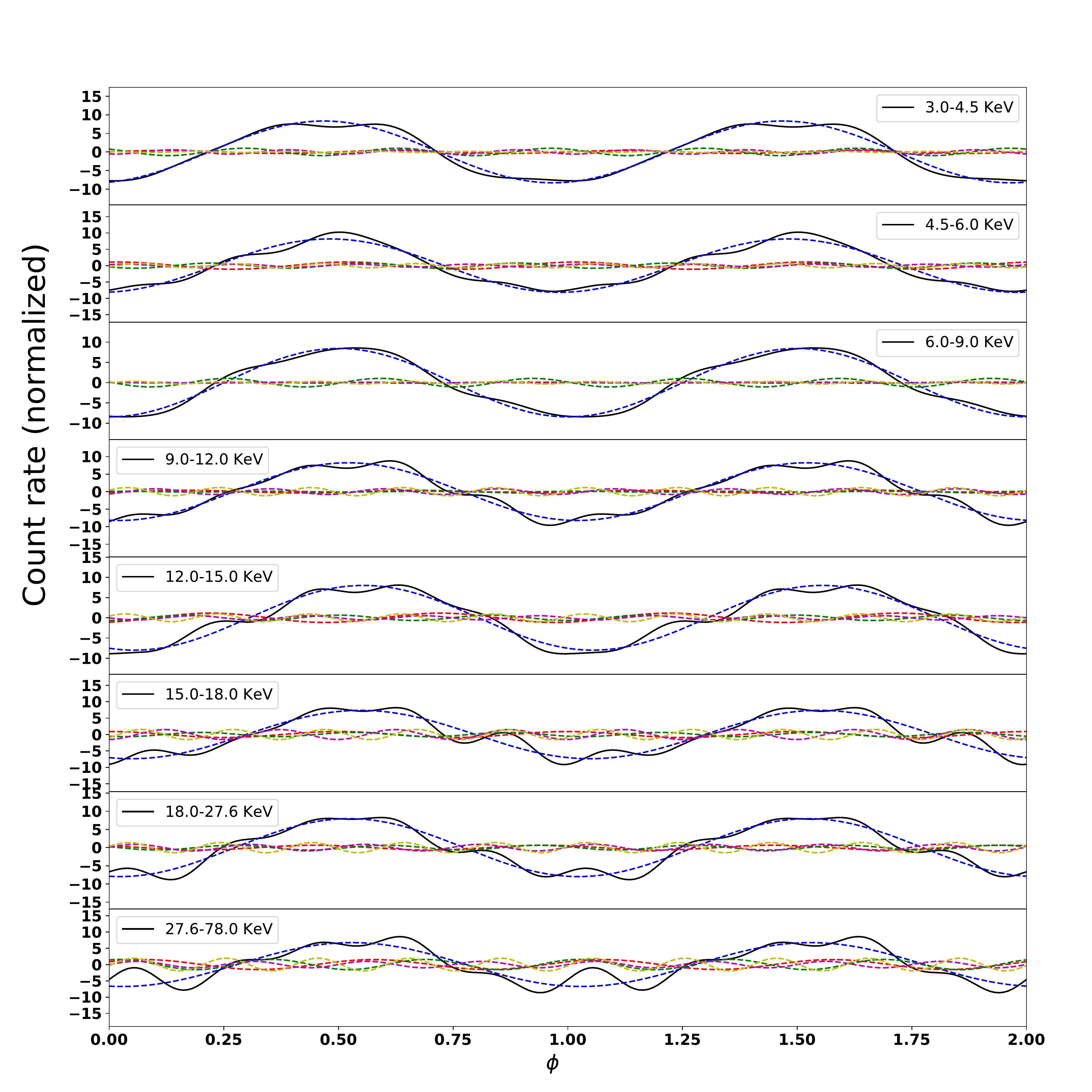} 
    \caption{Left: Normalized pulse profiles of LMC~X-4 at different energy ranges based on \nustar data, the measured profiles are represented by solid lines whereas the Fourier fits are represented by dashed lines, the boundaries of the energy ranges are [3, 4.5, 6, 9, 12, 15, 18, 27.6 and 78] keV. Right: The black solid line represents the overall approximate pulse profile and the colour dashed lines represent the different Fourier components. Blue: 1st, red: 2nd, green: 3rd, purple: 4th and yellow: 5th.}%
    \label{fig:LMC_X-4_NUSTAR_pulse_profiles}%
\end{figure*}

\begin{figure*}
	\centering
	\includegraphics[width=\linewidth]{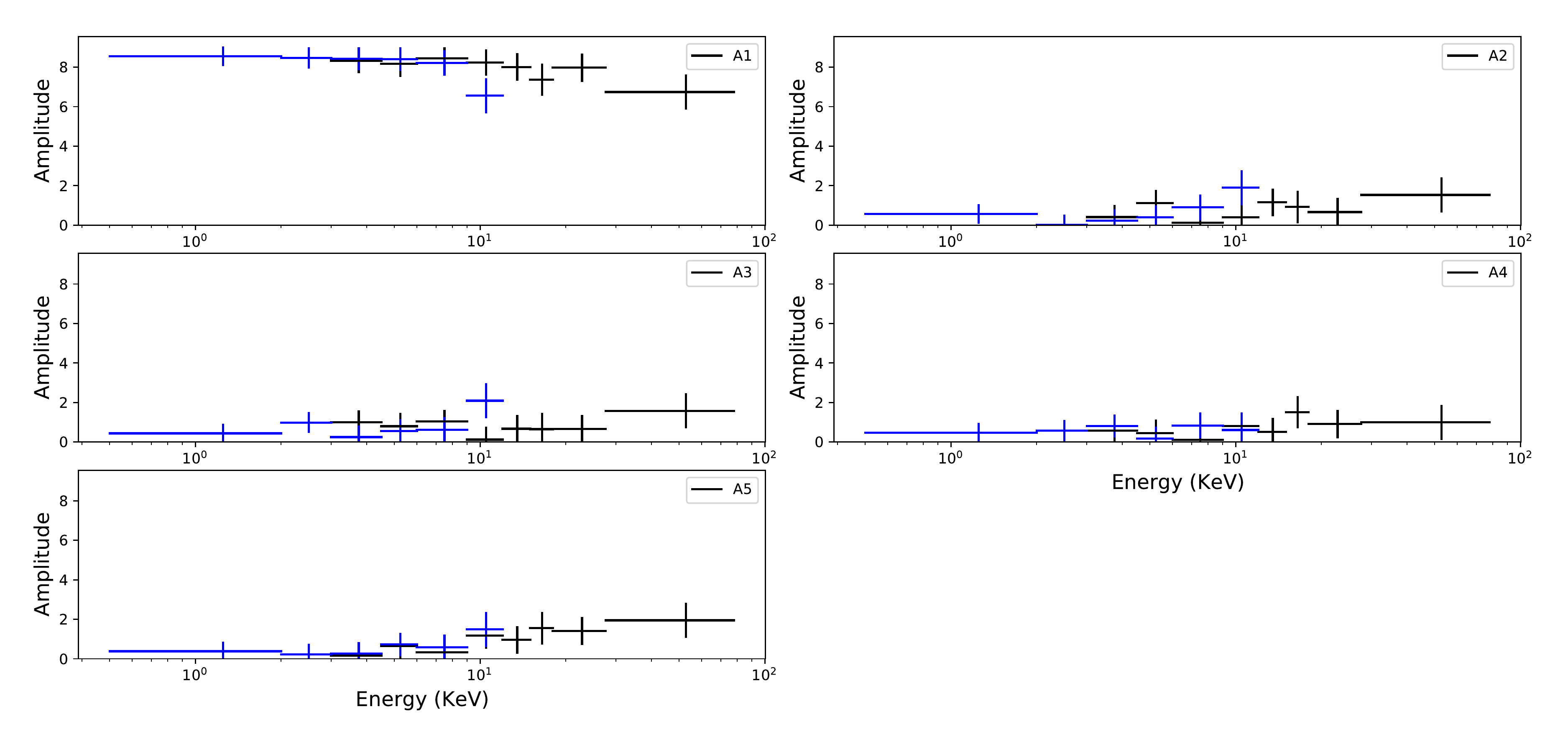}
	\caption{Variation of the harmonic amplitudes for LMC~X-4, the boundaries of the energy ranges are [0.5, 2, 3, 4.5, 6, 9, 12, 15, 18, 27.6 and 78] keV. \xmm data is represented by blue markers and \nustar data is represented by black markers.}
	\label{fig:LMC_X-4_fourier_variation}
\end{figure*}


\begin{figure*}%
    \centering
    \includegraphics[width=0.47\linewidth]{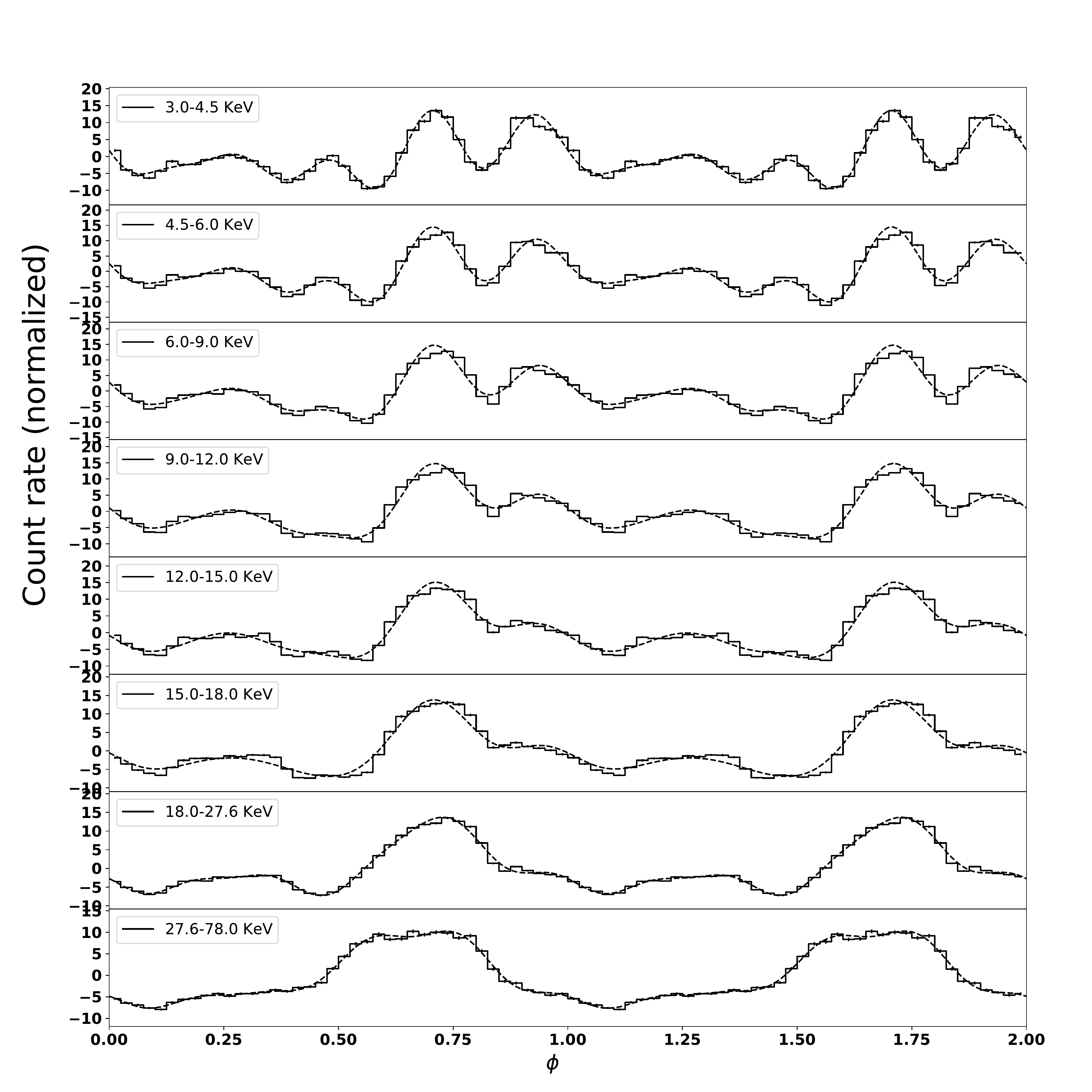} %
    \qquad
    \includegraphics[width=0.47\linewidth]{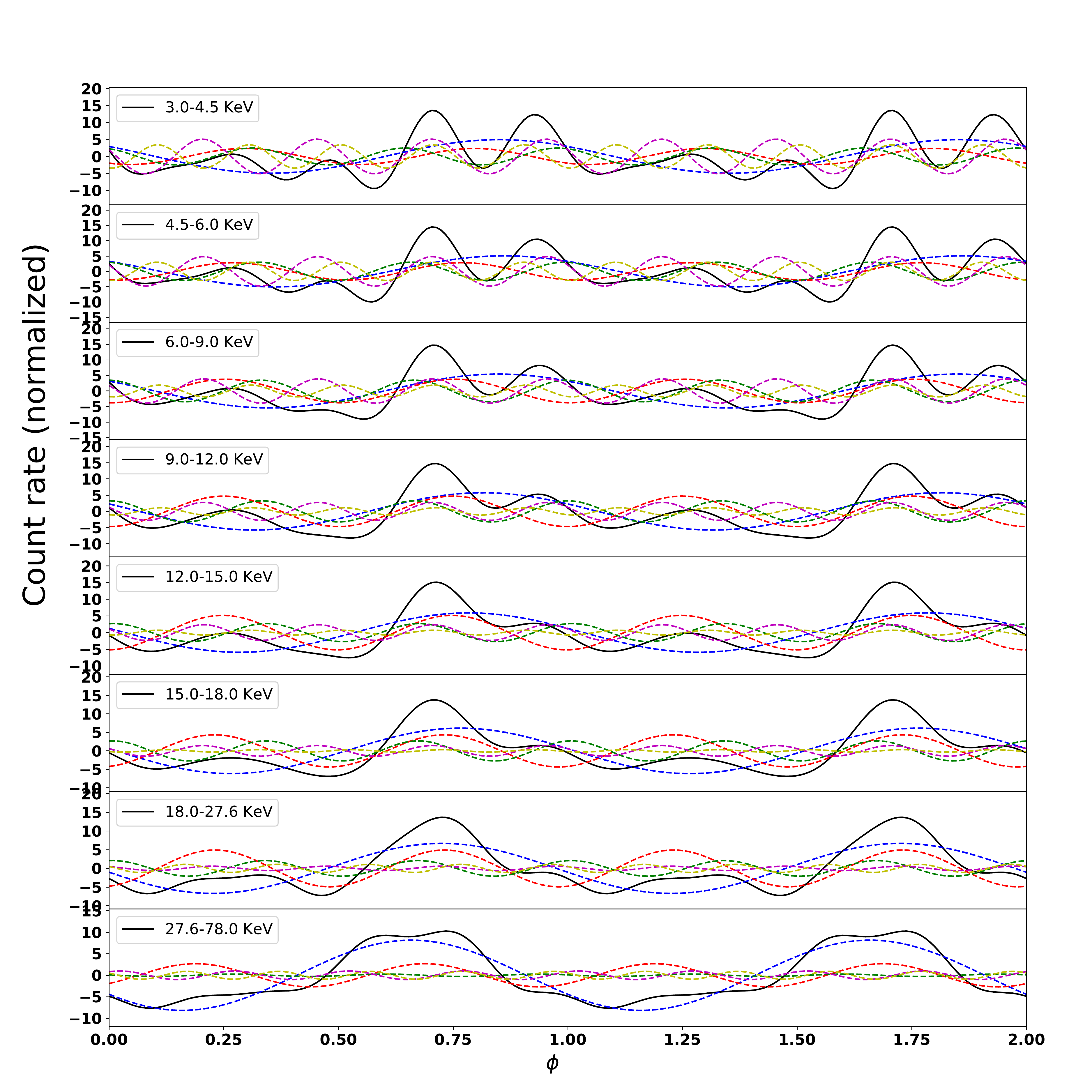} 
    \caption{Left: Normalized pulse profiles of GRO~J2058+42 at different energy ranges based on \nustar data, the measured profiles are represented by solid lines whereas the Fourier fits are represented by dashed lines, the boundaries of the energy ranges are [3, 4.5, 6, 9, 12, 15, 18, 27.6 and 78] keV. Right: The black solid line represents the overall approximate pulse profile and the colour dashed lines represent the different Fourier components. Blue: 1st, red: 2nd, green: 3rd, purple: 4th and yellow: 5th.}%
    \label{fig:GRO_J2058+42_NUSTAR_pulse_profiles}%
\end{figure*}

\begin{figure*}
	\centering
	\includegraphics[width=\linewidth]{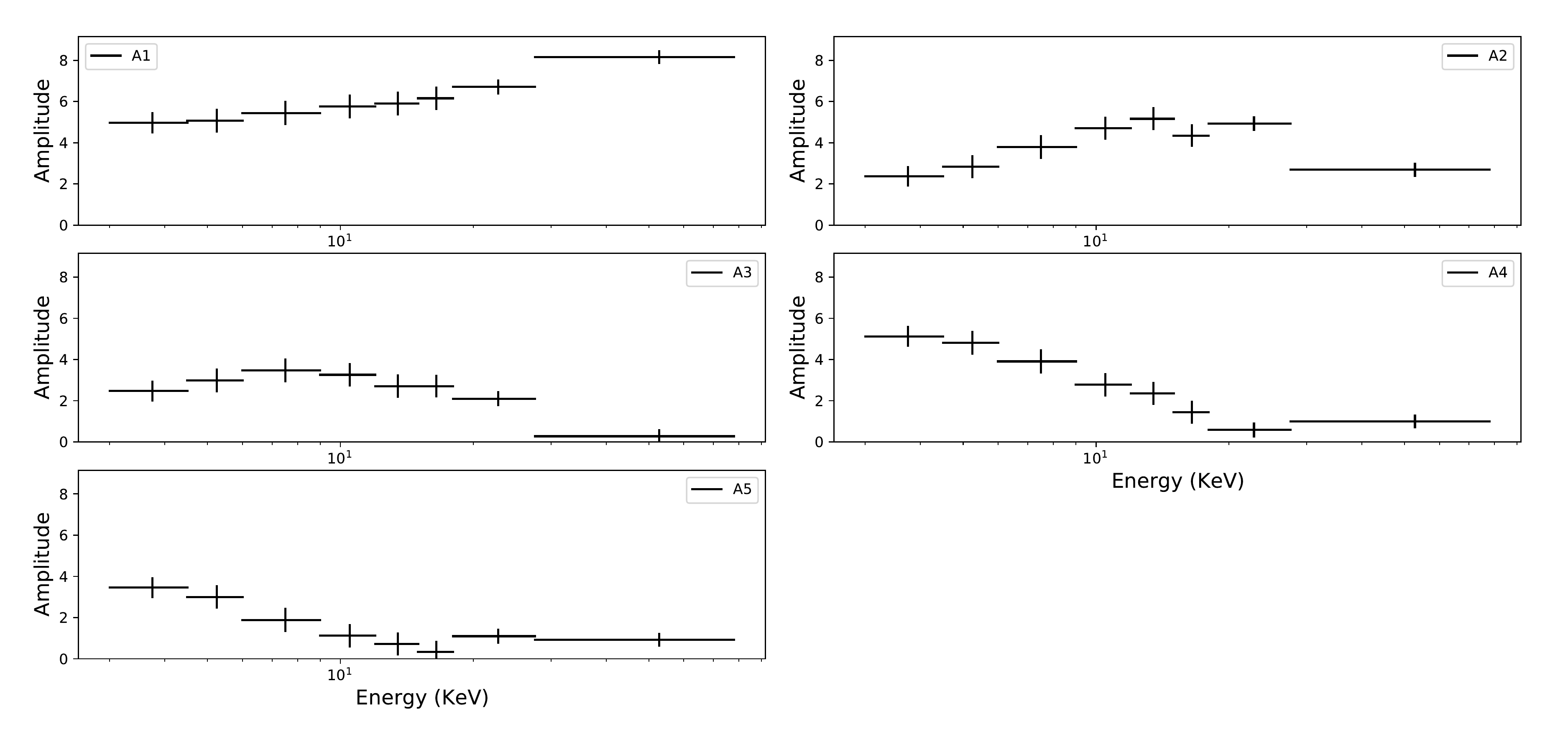}
	\caption{Variation of the harmonic amplitudes for GRO~J2058+42, the boundaries of the energy ranges are [3, 4.5, 6, 9, 12, 15, 18, 27.6 and 78] keV. Only \nustar data is represented by black markers.}
	\label{fig:GRO_J2058+42_fourier_variation}
\end{figure*}


\begin{figure*}%
    \centering
    \includegraphics[width=0.47\linewidth]{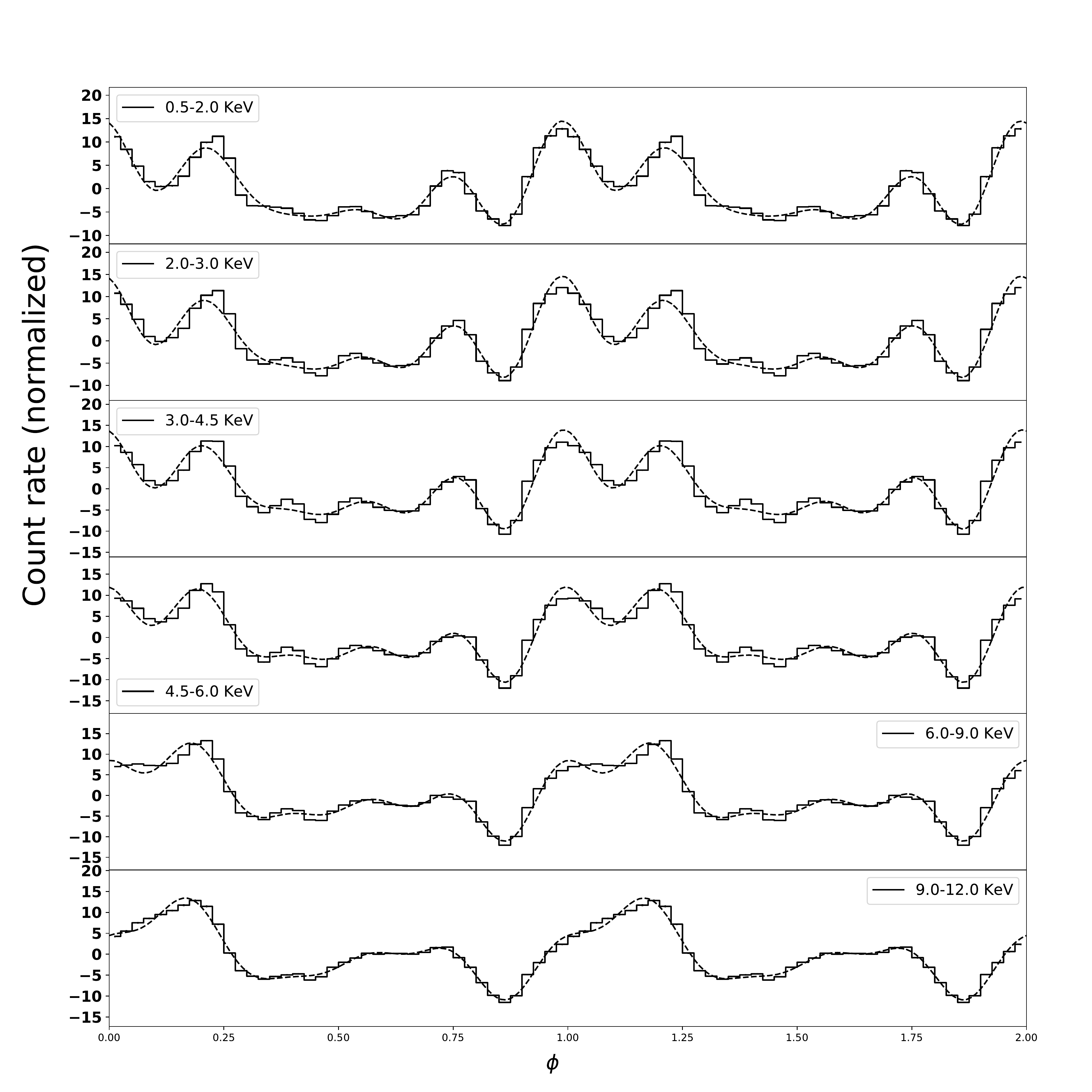} %
    \qquad
    \includegraphics[width=0.47\linewidth]{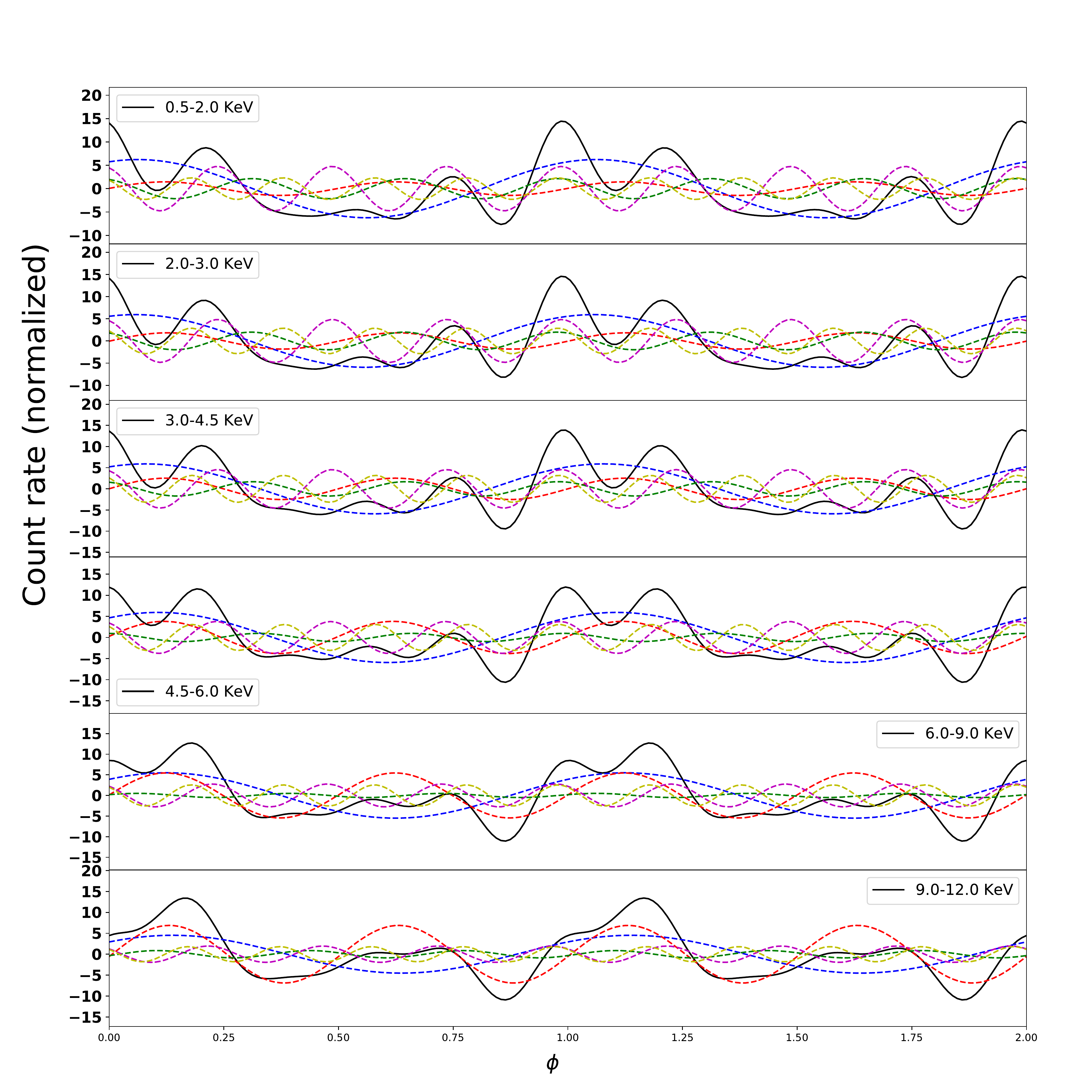} 
    \caption{Left: Normalized pulse profiles of Vela~X-1 at different energy ranges based on \xmm data, the measured profiles are represented by solid lines whereas the Fourier fits are represented by dashed lines, the boundaries of the energy ranges are [0.5, 2, 3, 4.5, 6, 9, 12] keV. Right: The black solid line represents the overall approximate pulse profile and the colour dashed lines represent the different Fourier components. Blue: 1st, red: 2nd, green: 3rd, purple: 4th and yellow: 5th.}%
    \label{fig:Vela_X-1_XMM_pulse_profiles}%
\end{figure*}

\begin{figure*}%
    \centering
    \includegraphics[width=0.47\linewidth]{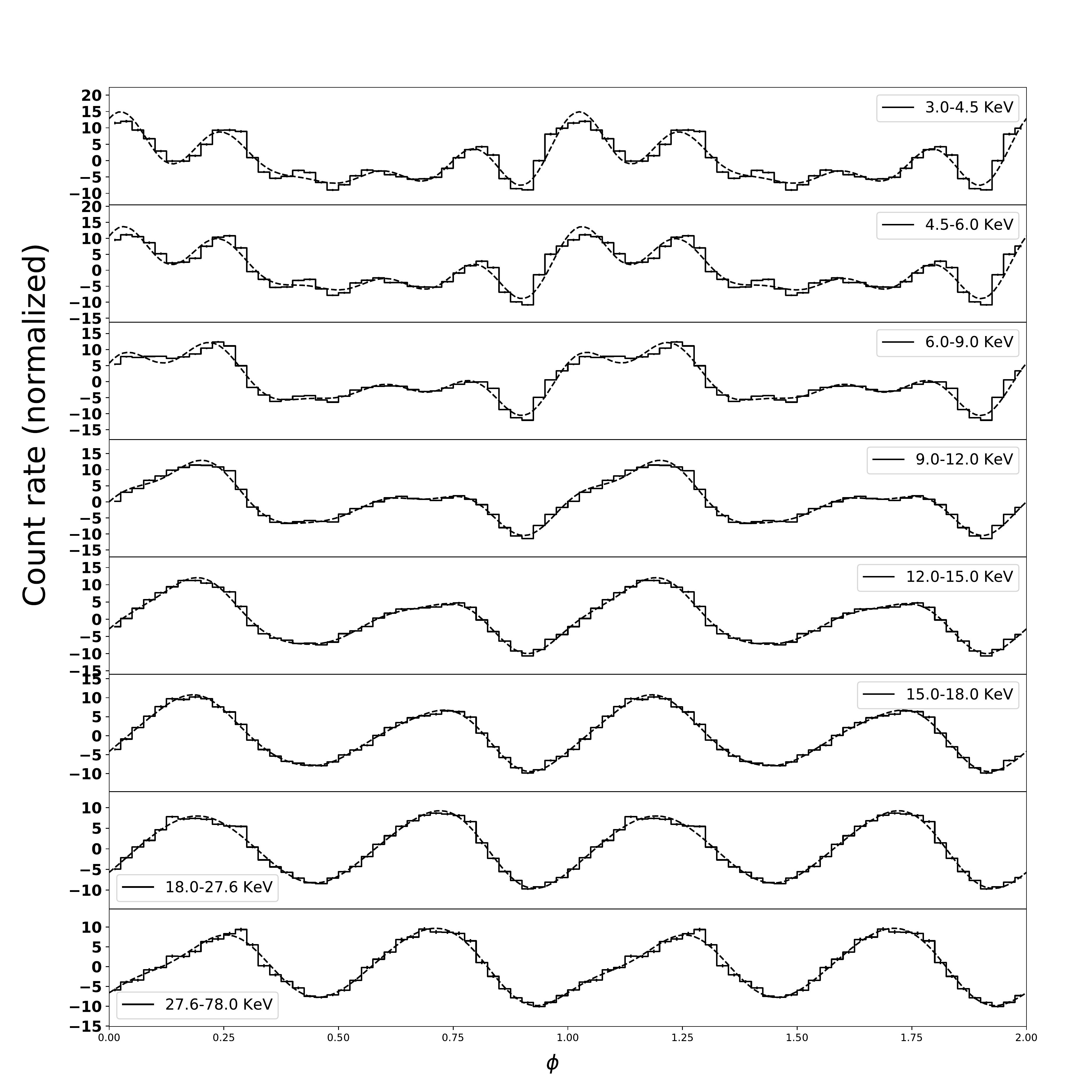} %
    \qquad
    \includegraphics[width=0.47\linewidth]{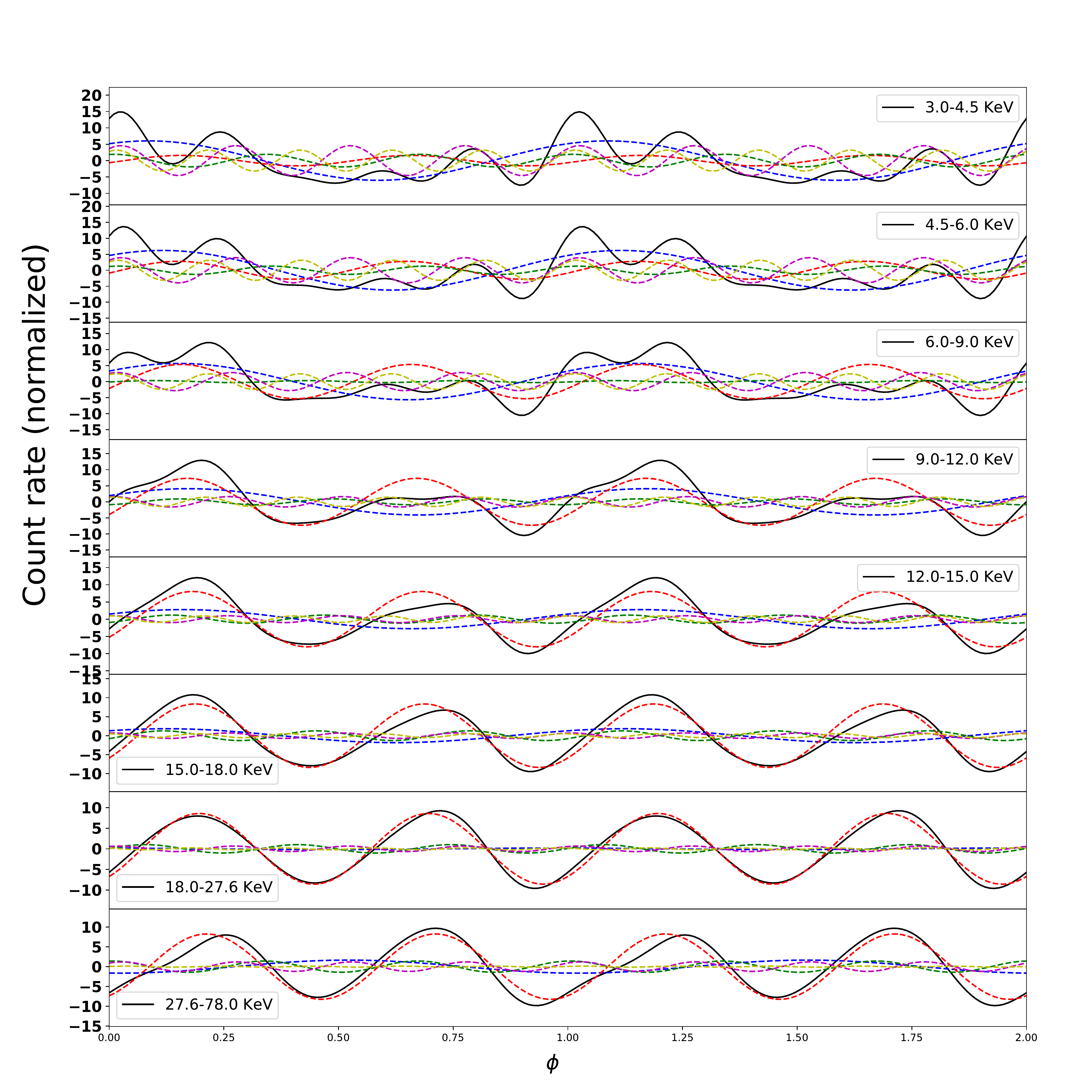} 
    \caption{Left: Normalized pulse profiles of Vela~X-1 at different energy ranges based on \nustar data, the measured profiles are represented by solid lines whereas the Fourier fits are represented by dashed lines, the boundaries of the energy ranges are [3, 4.5, 6, 9, 12, 15, 18, 27.6 and 78] keV. Right: The black solid line represents the overall approximate pulse profile and the colour dashed lines represent the different Fourier components. Blue: 1st, red: 2nd, green: 3rd, purple: 4th and yellow: 5th.}%
    \label{fig:Vela_X-1_NUSTAR_pulse_profiles}%
\end{figure*}

\begin{figure*}
	\centering
	\includegraphics[width=\linewidth]{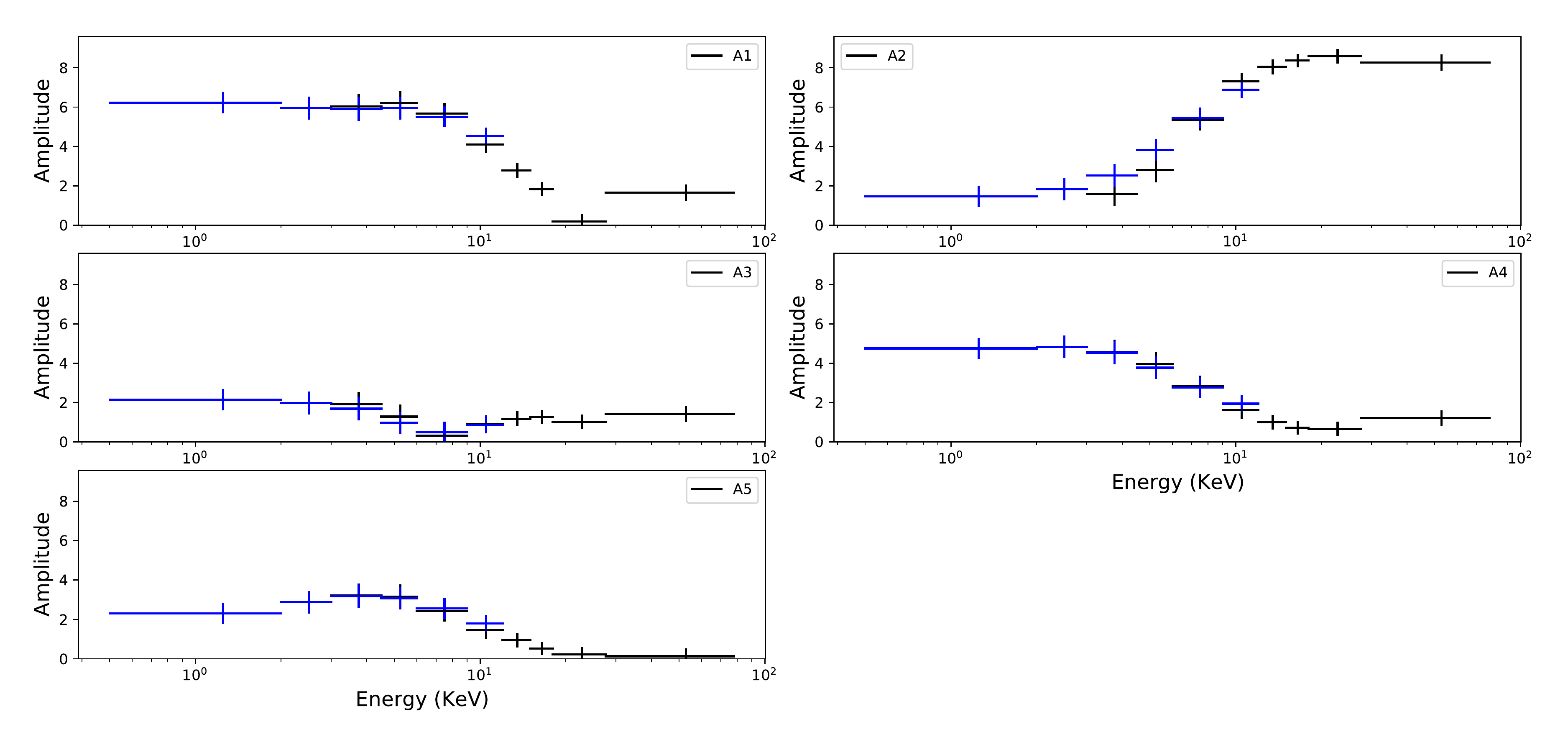}
	\caption{Variation of the harmonic amplitudes for Vela~X-1, the boundaries of the energy ranges are [0.5, 2, 3, 4.5, 6, 9, 12, 15, 18, 27.6 and 78] keV. \xmm data is represented by blue markers and \nustar data is represented by black markers.}
	\label{fig:Vela_X-1_fourier_variation}
\end{figure*}


\begin{figure*}%
    \centering
    \includegraphics[width=0.47\linewidth]{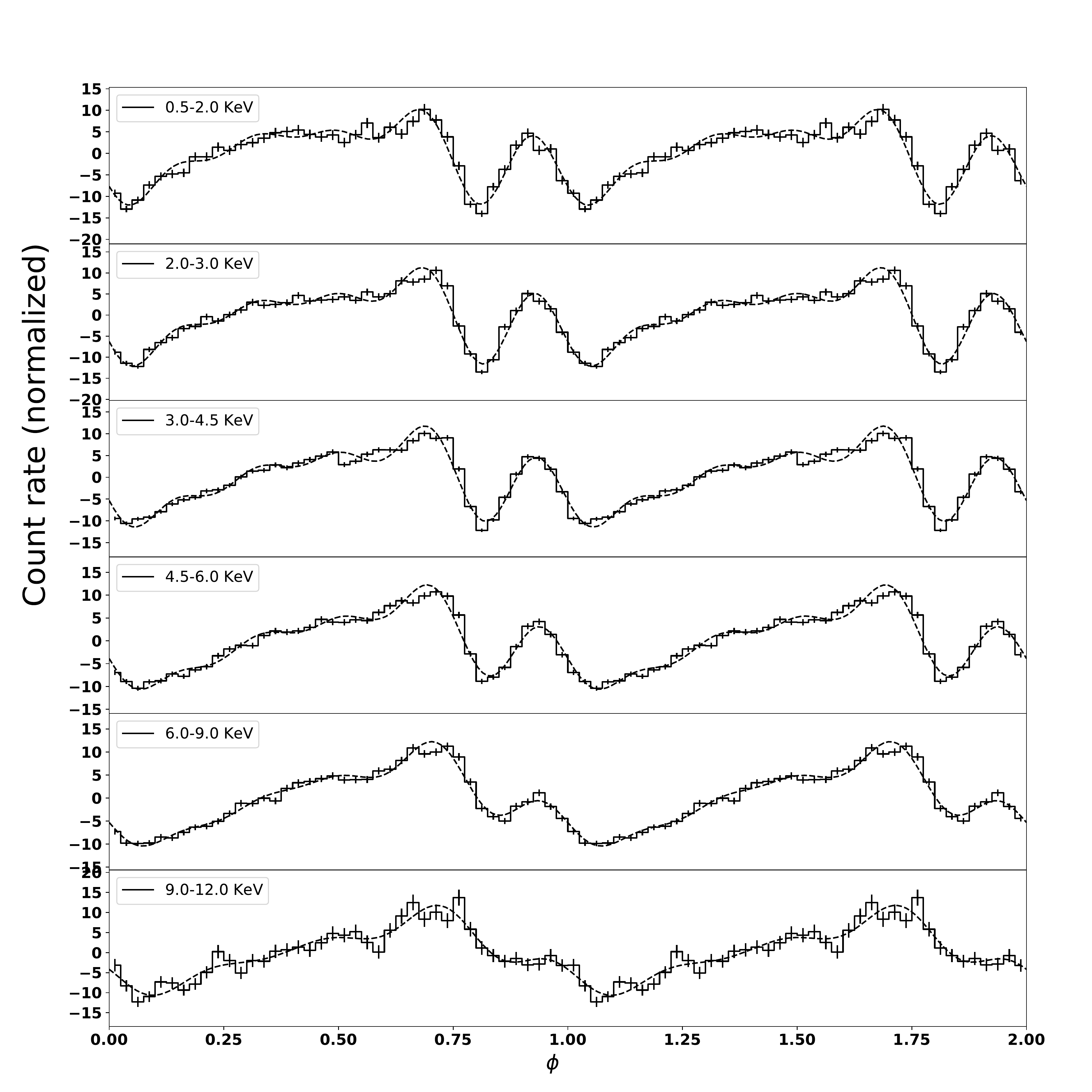} %
    \qquad
    \includegraphics[width=0.47\linewidth]{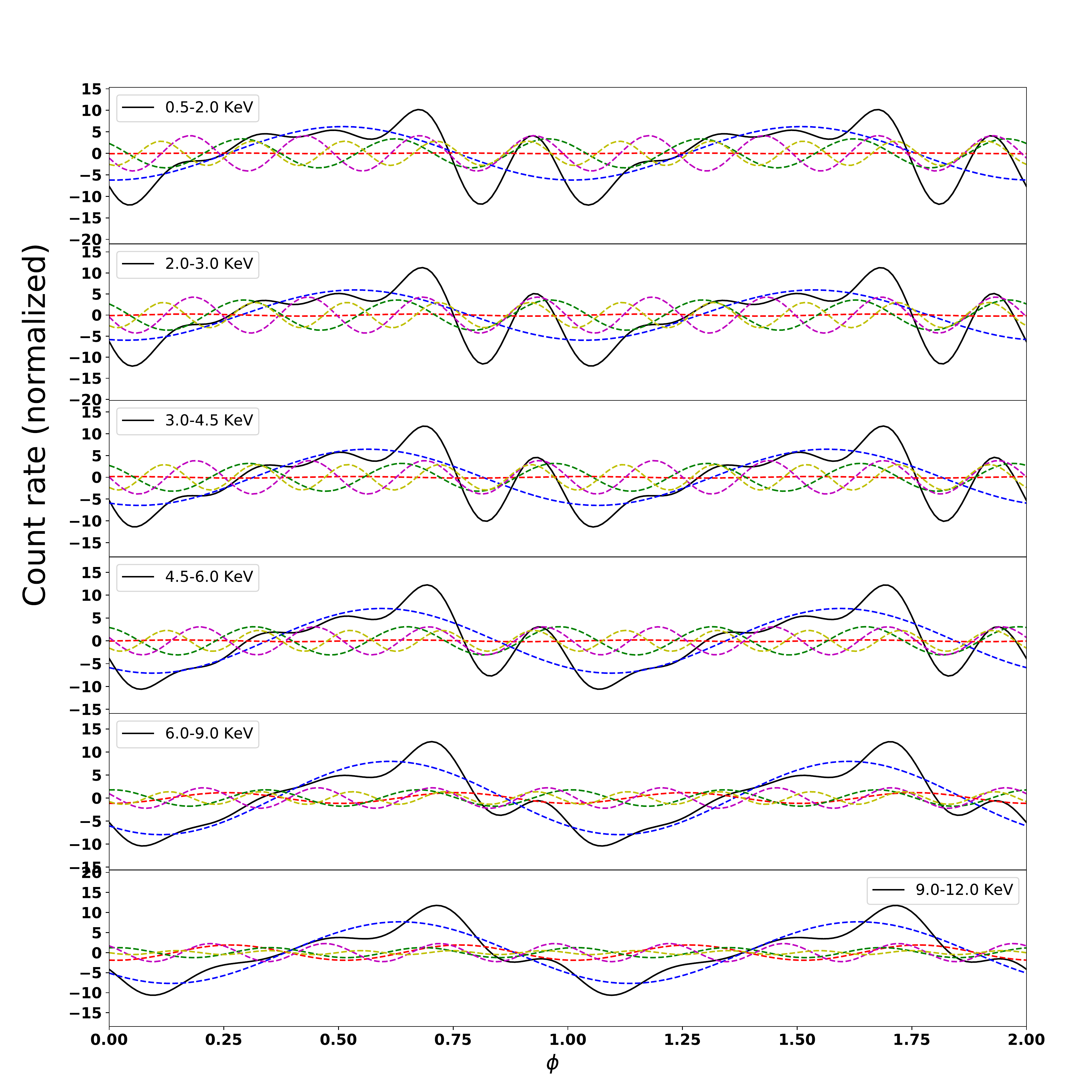} 
    \caption{Left: Normalized pulse profiles of 1E1145.1$-$6141 at different energy ranges based on \xmm data, the measured profiles are represented by solid lines whereas the Fourier fits are represented by dashed lines, the boundaries of the energy ranges are [0.5, 2, 3, 4.5, 6, 9, 12] keV. Right: The black solid line represents the overall approximate pulse profile and the colour dashed lines represent the different Fourier components. Blue: 1st, red: 2nd, green: 3rd, purple: 4th and yellow: 5th.}%
    \label{fig:1E1145.1-6141_XMM_pulse_profiles}%
\end{figure*}

\begin{figure*}%
    \centering
    \includegraphics[width=0.47\linewidth]{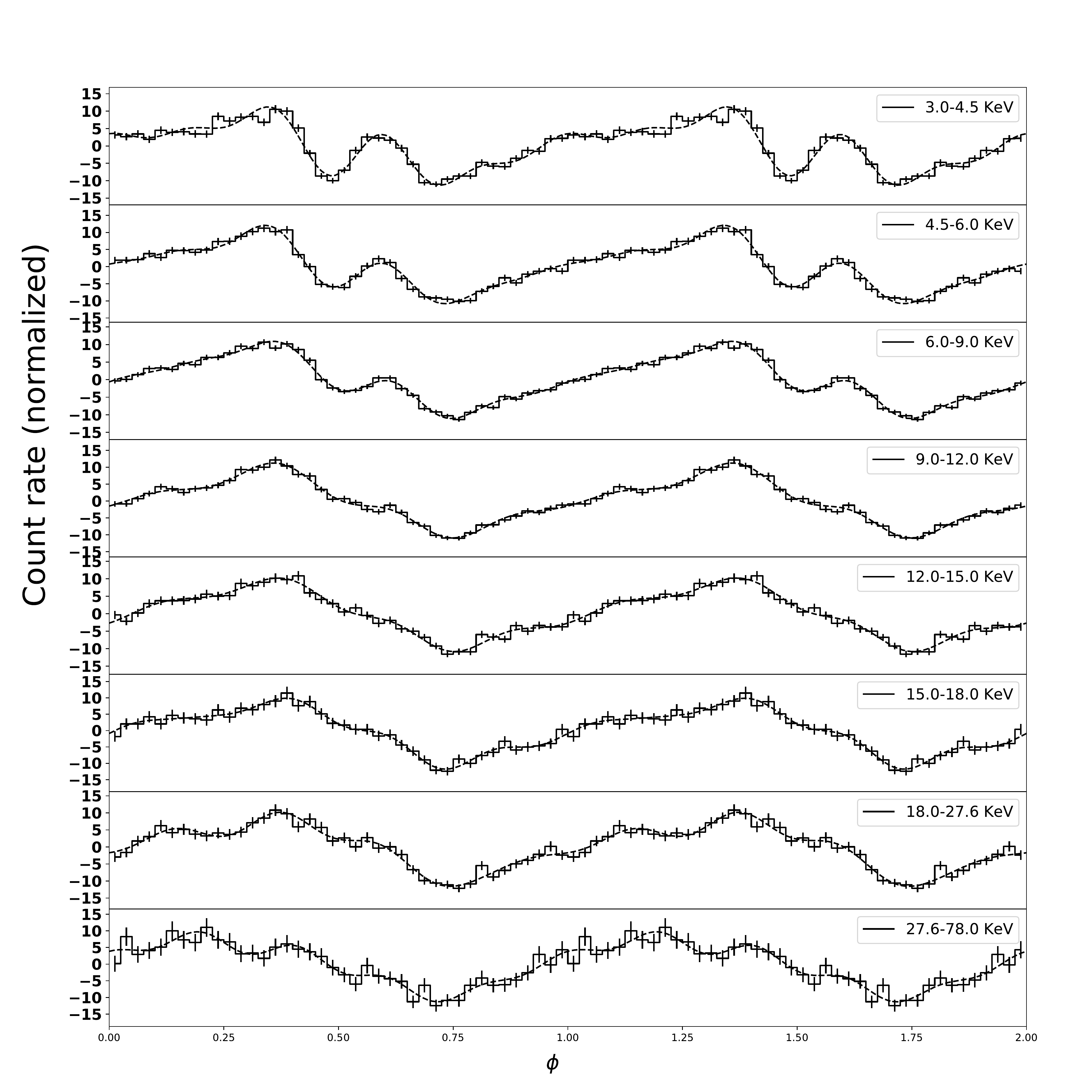} %
    \qquad
    \includegraphics[width=0.47\linewidth]{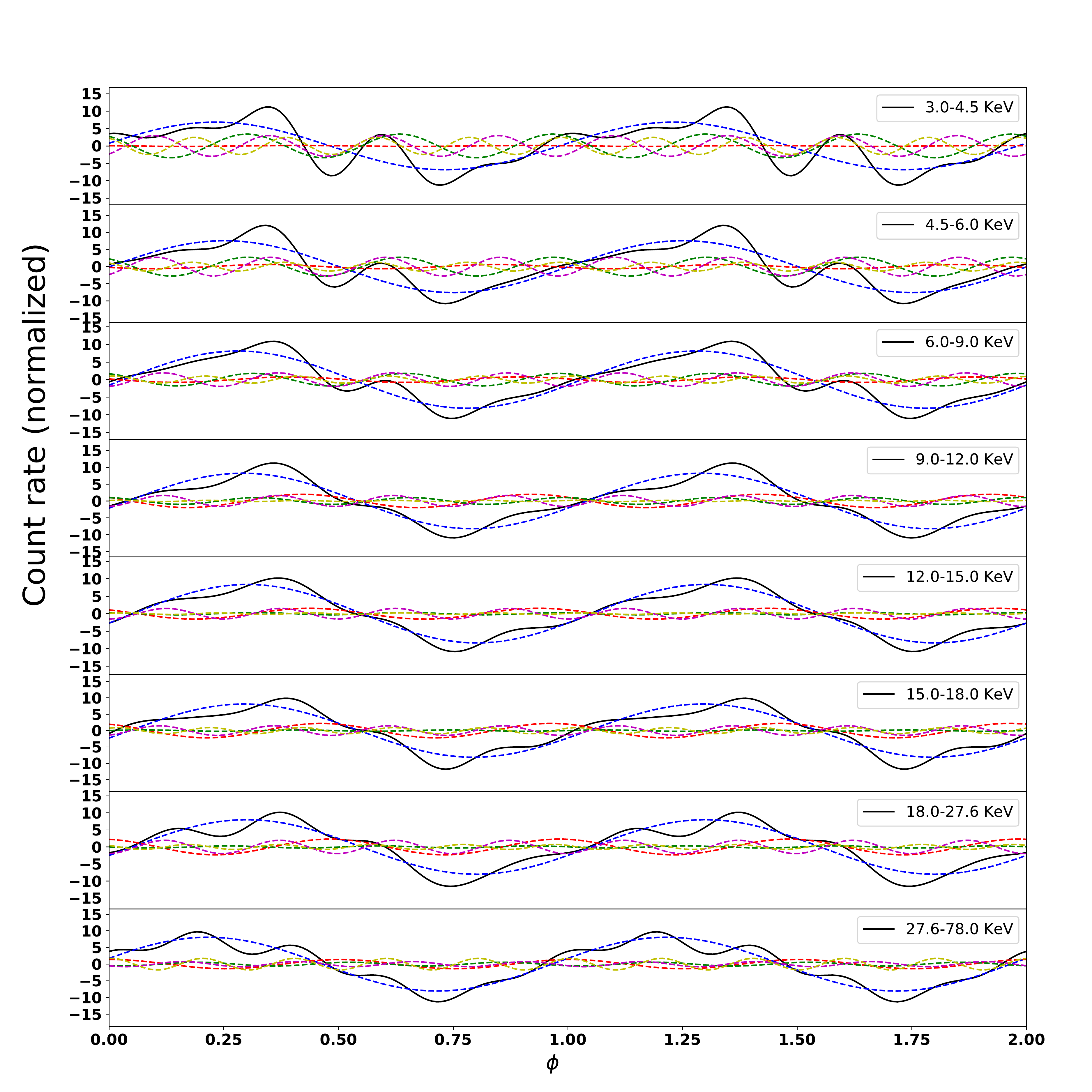} 
    \caption{Left: Normalized pulse profiles of 1E1145.1$-$6141 at different energy ranges based on \nustar data, the measured profiles are represented by solid lines whereas the Fourier fits are represented by dashed lines, the boundaries of the energy ranges are [3, 4.5, 6, 9, 12, 15, 18, 27.6 and 78] keV. Right: The black solid line represents the overall approximate pulse profile and the colour dashed lines represent the different Fourier components. Blue: 1st, red: 2nd, green: 3rd, purple: 4th and yellow: 5th.}%
    \label{fig:1E1145.1-6141_NUSTAR_pulse_profiles}%
\end{figure*}

\begin{figure*}
	\centering
	\includegraphics[width=\linewidth]{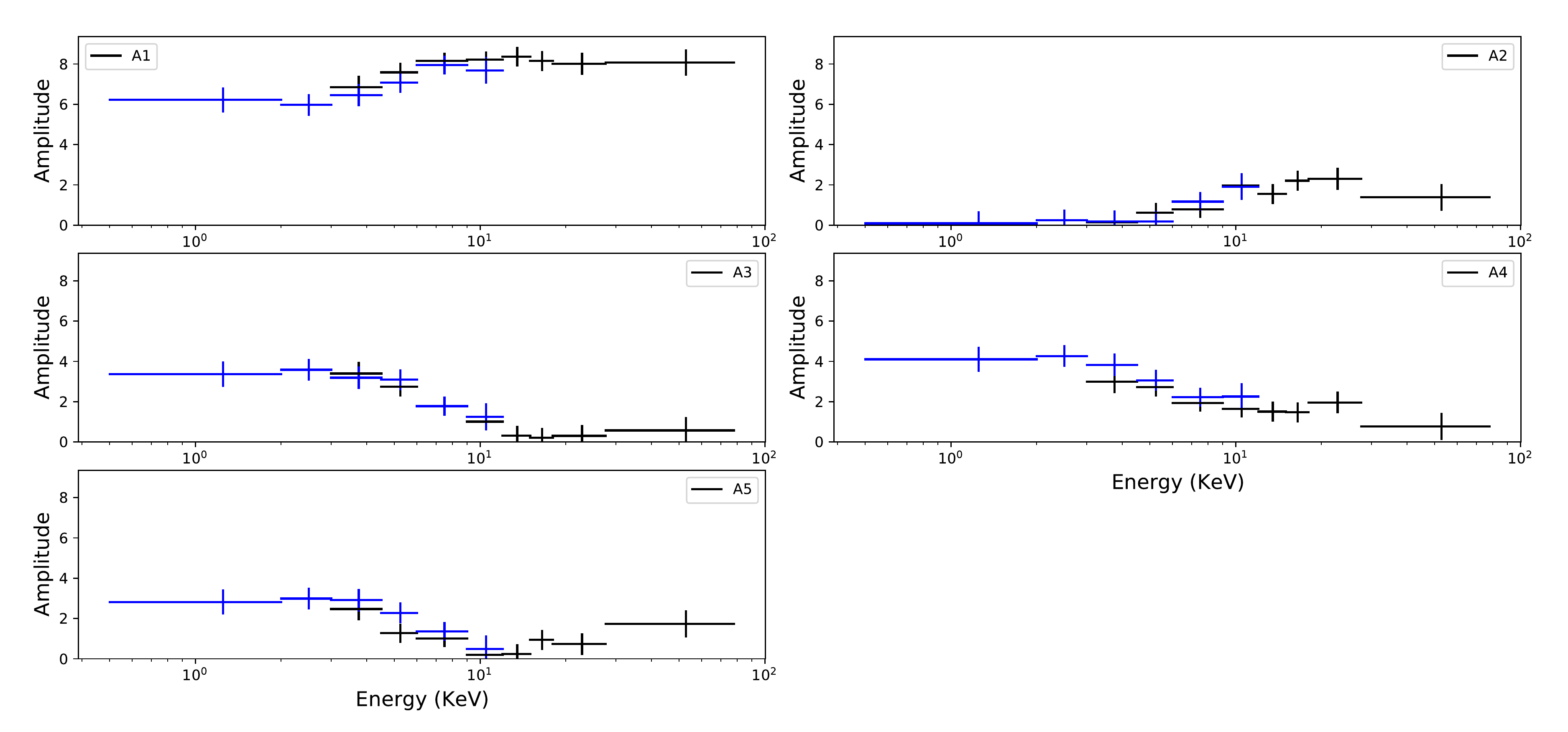}
	\caption{Variation of the harmonic amplitudes for 1E1145.1$-$6141, the boundaries of the energy ranges are [0.5, 2, 3, 4.5, 6, 9, 12, 15, 18, 27.6 and 78] keV. \xmm data is represented by blue markers and \nustar data is represented by black markers.}
	\label{fig:1E1145.1-6141_fourier_variation}
\end{figure*}


\begin{figure*}%
    \centering
    \includegraphics[width=0.47\linewidth]{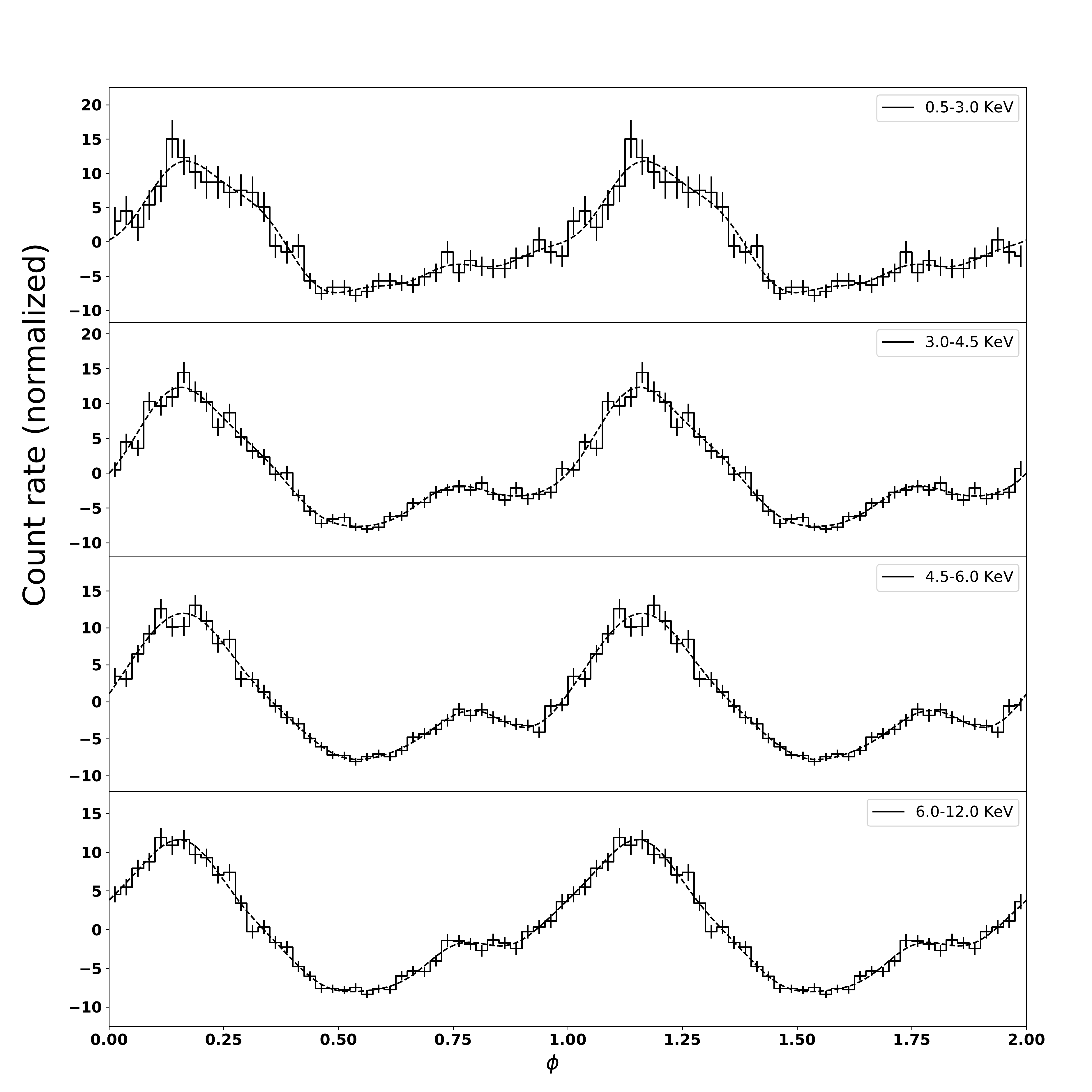} %
    \qquad
    \includegraphics[width=0.47\linewidth]{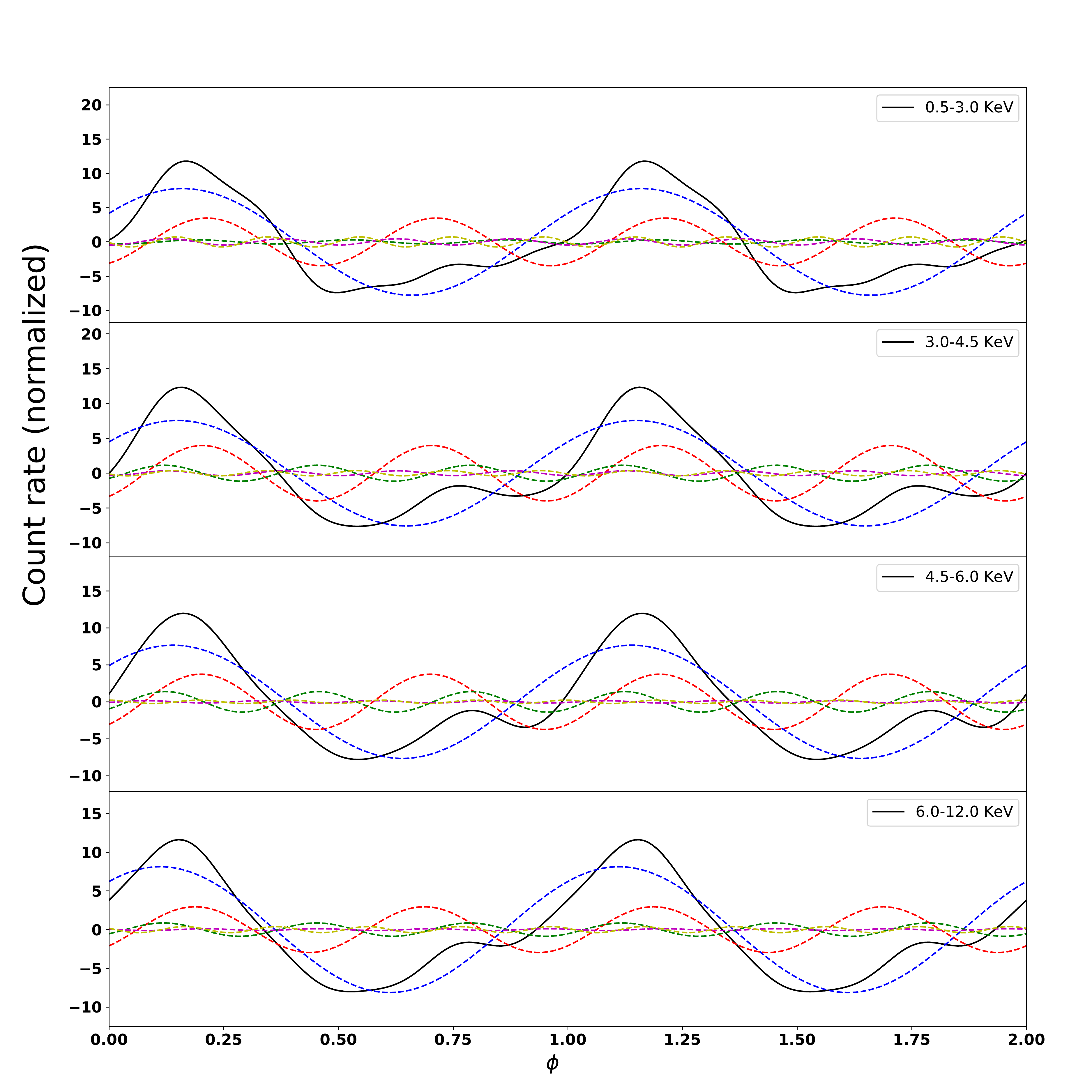} 
    \caption{Left: Normalized pulse profiles of IGR~J17252$-$3616 at different energy ranges based on \xmm data, the measured profiles are represented by solid lines whereas the Fourier fits are represented by dashed lines, the boundaries of the energy ranges are [0.5, 3, 4.5, 6, 12] keV. Right: The black solid line represents the overall approximate pulse profile and the colour dashed lines represent the different Fourier components. Blue: 1st, red: 2nd, green: 3rd, purple: 4th and yellow: 5th.}%
    \label{fig:IGR_J17252-361_XMM_pulse_profiles}%
\end{figure*}

\begin{figure*}
	\centering
	\includegraphics[width=\linewidth]{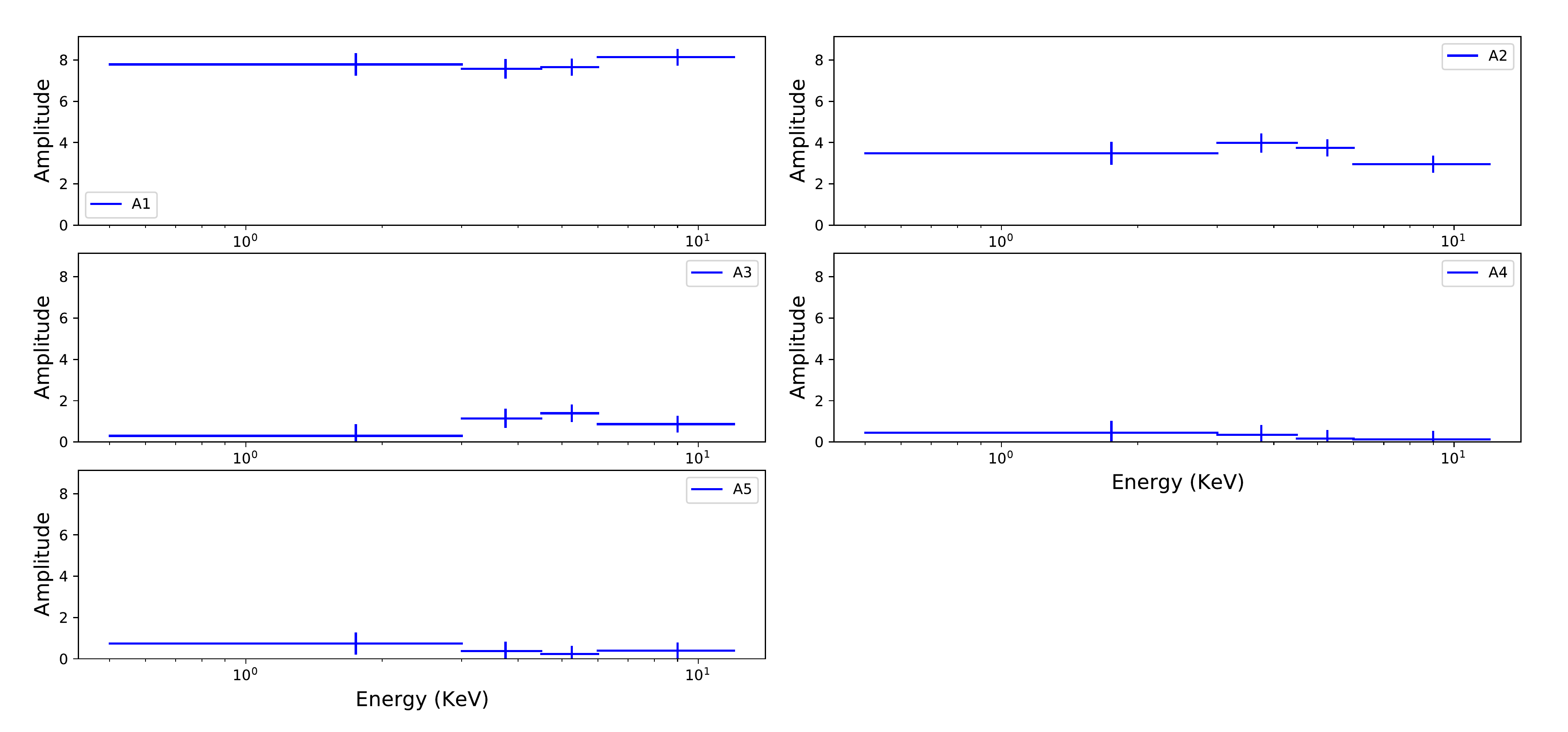}
	\caption{Variation of the harmonic amplitudes for IGR~J17252$-$3616, the boundaries of the energy ranges are [0.5, 3, 4.5, 6, 12] keV. Only \xmm data is represented by blue markers.}
	\label{fig:IGR_J17252-361_fourier_variation}
\end{figure*}


\begin{figure*}%
    \centering
    \includegraphics[width=0.47\linewidth]{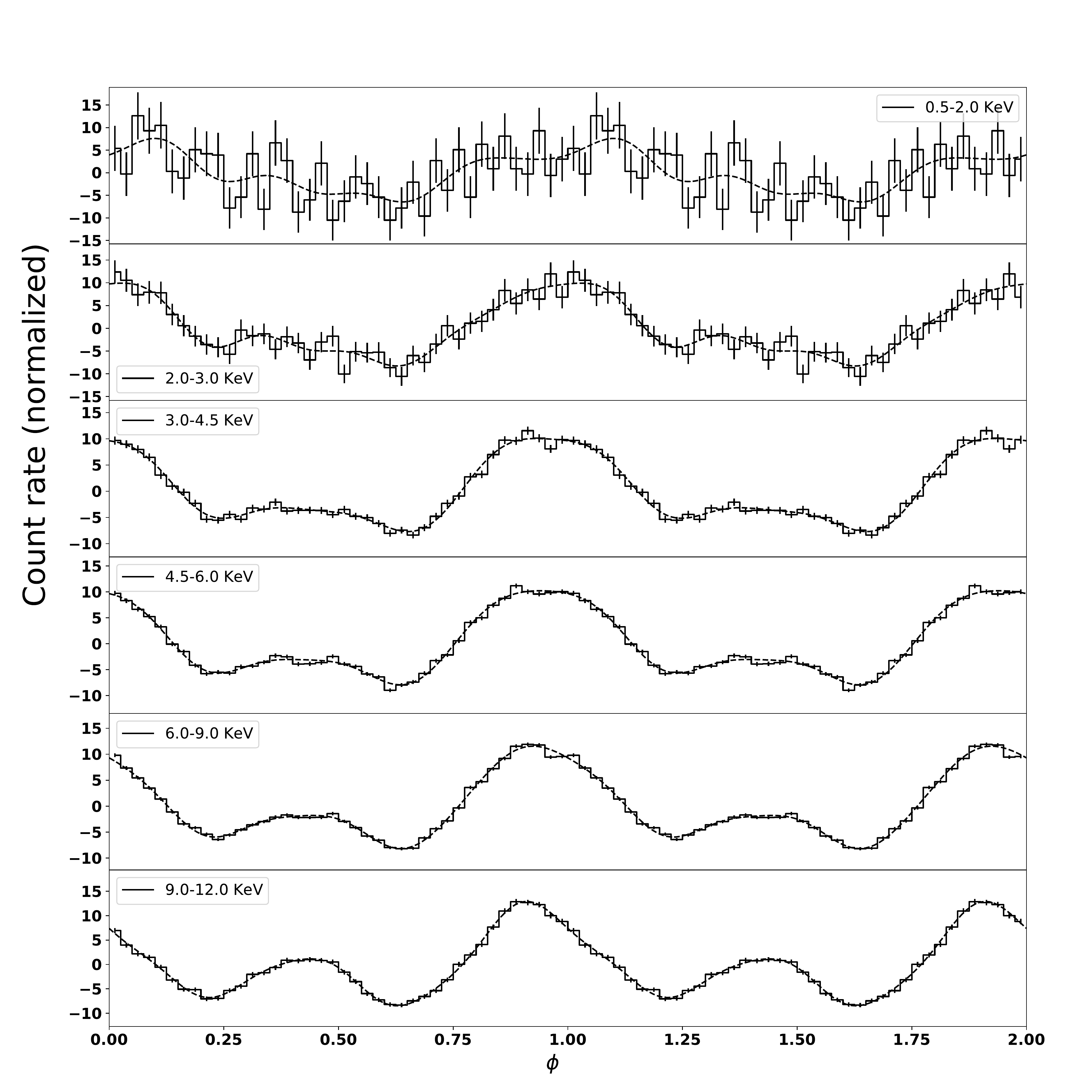} %
    \qquad
    \includegraphics[width=0.47\linewidth]{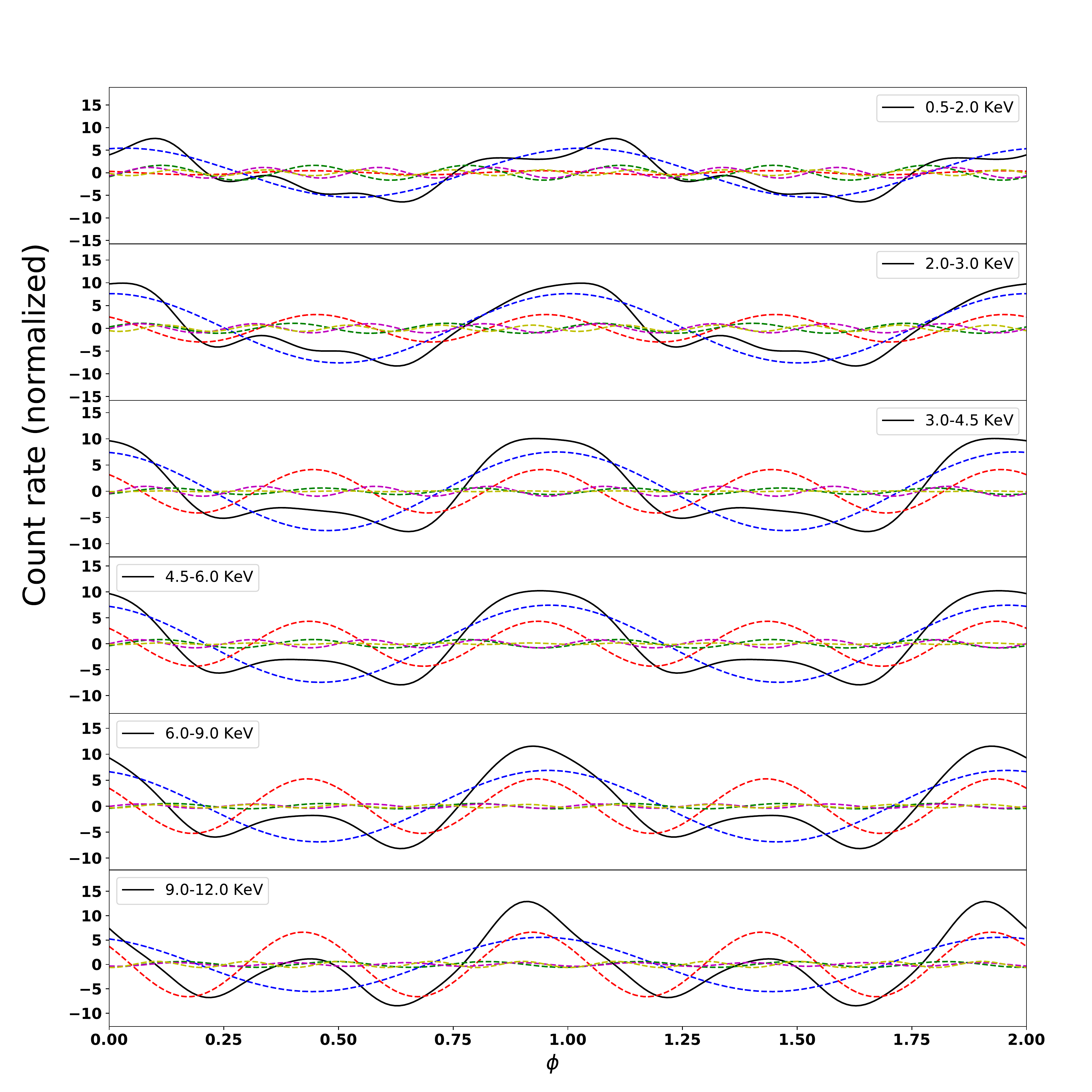} 
    \caption{Left: Normalized pulse profiles of GX301$-$2 at different energy ranges based on \xmm data, the measured profiles are represented by solid lines whereas the Fourier fits are represented by dashed lines, the boundaries of the energy ranges are [0.5, 2, 3, 4.5, 6, 9, 12] keV. Right: The black solid line represents the overall approximate pulse profile and the colour dashed lines represent the different Fourier components. Blue: 1st, red: 2nd, green: 3rd, purple: 4th and yellow: 5th.}%
    \label{fig:GX301-2_XMM_pulse_profiles}%
\end{figure*}

\begin{figure*}%
    \centering
    \includegraphics[width=0.47\linewidth]{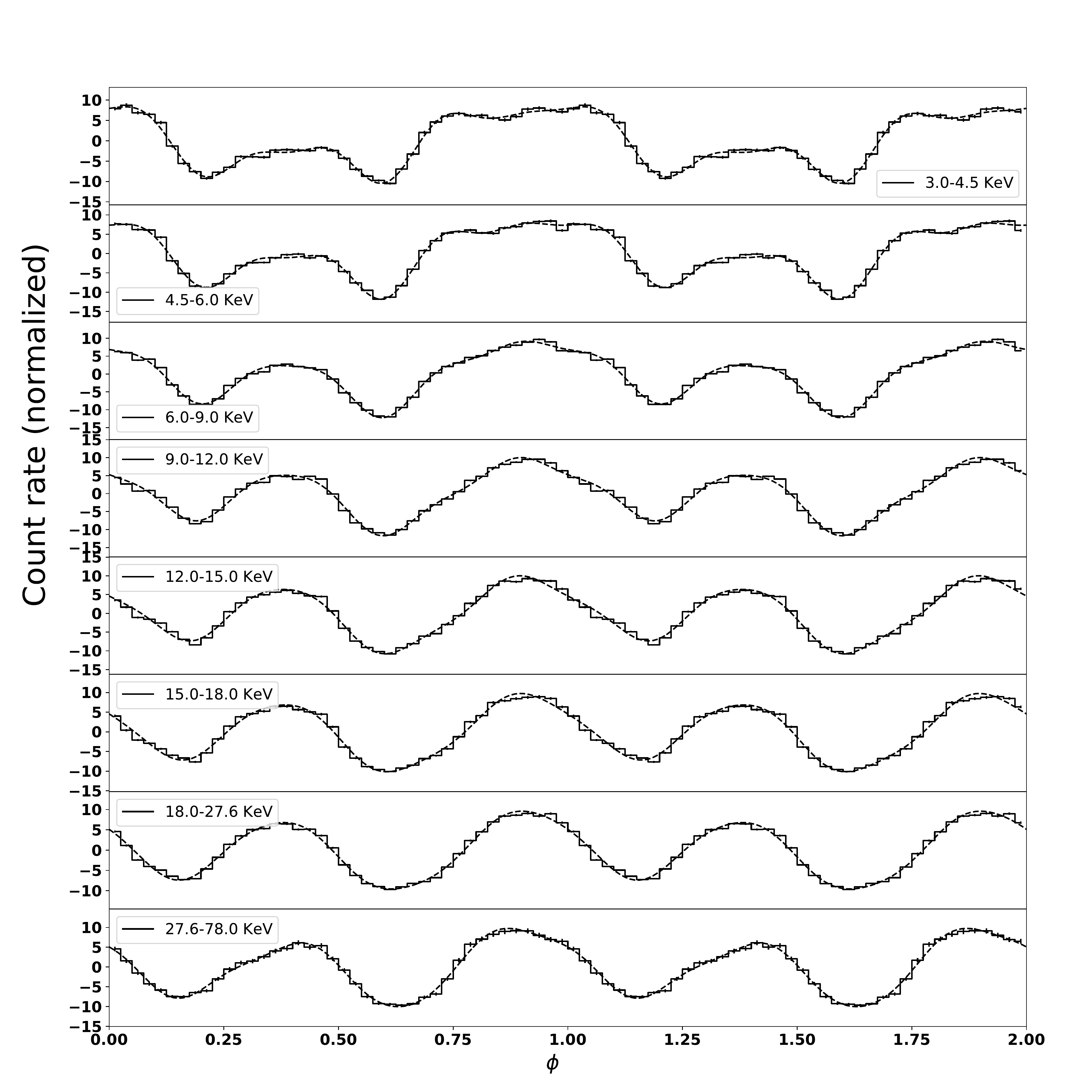} %
    \qquad
    \includegraphics[width=0.47\linewidth]{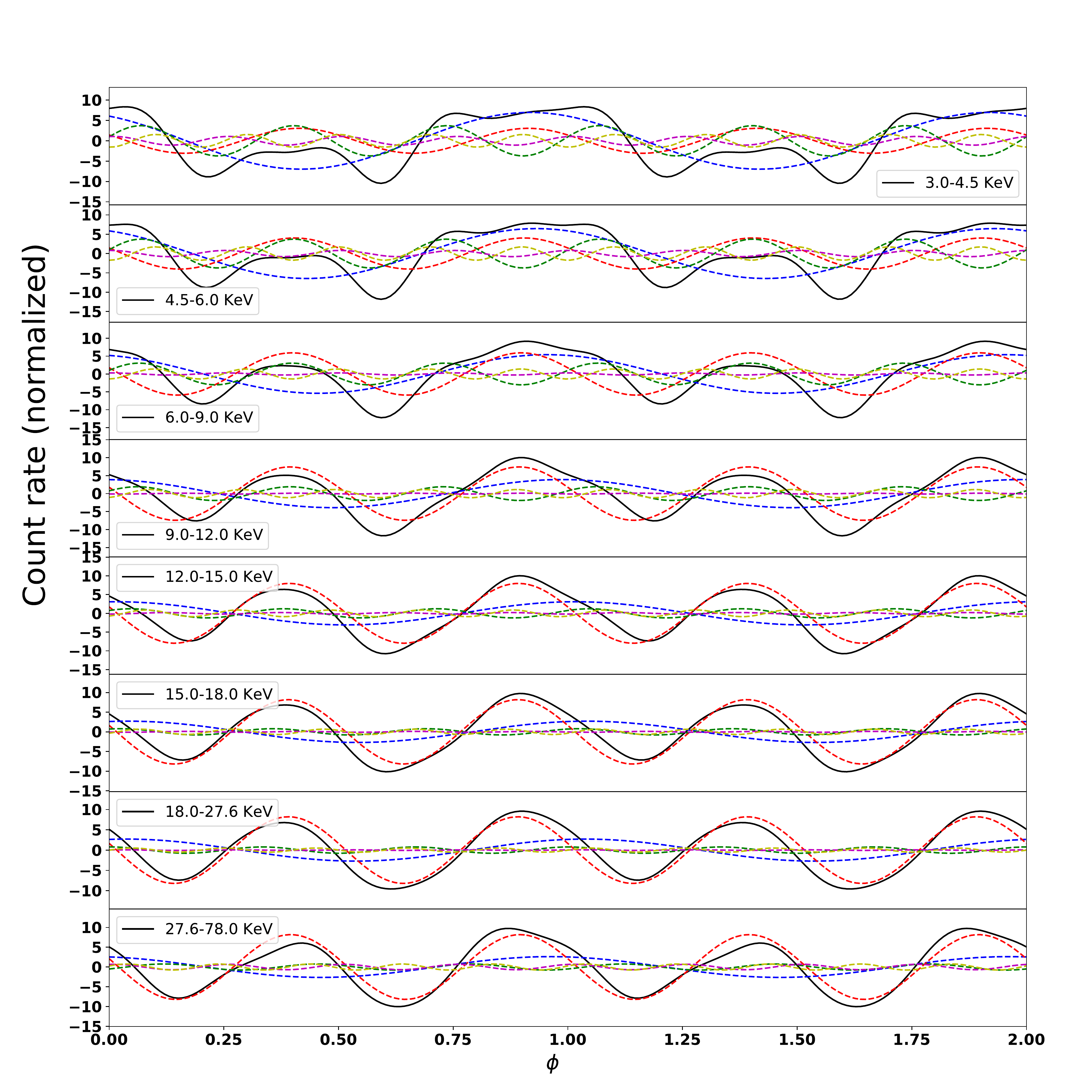} 
    \caption{Left: Normalized pulse profiles of GX301$-$2 at different energy ranges based on \nustar data, the measured profiles are represented by solid lines whereas the Fourier fits are represented by dashed lines, the boundaries of the energy ranges are [3, 4.5, 6, 9, 12, 15, 18, 27.6 and 78] keV. Right: The black solid line represents the overall approximate pulse profile and the colour dashed lines represent the different Fourier components. Blue: 1st, red: 2nd, green: 3rd, purple: 4th and yellow: 5th.}%
    \label{fig:GX301-2_NUSTAR_pulse_profiles}%
\end{figure*}

\begin{figure*}
	\centering
	\includegraphics[width=\linewidth]{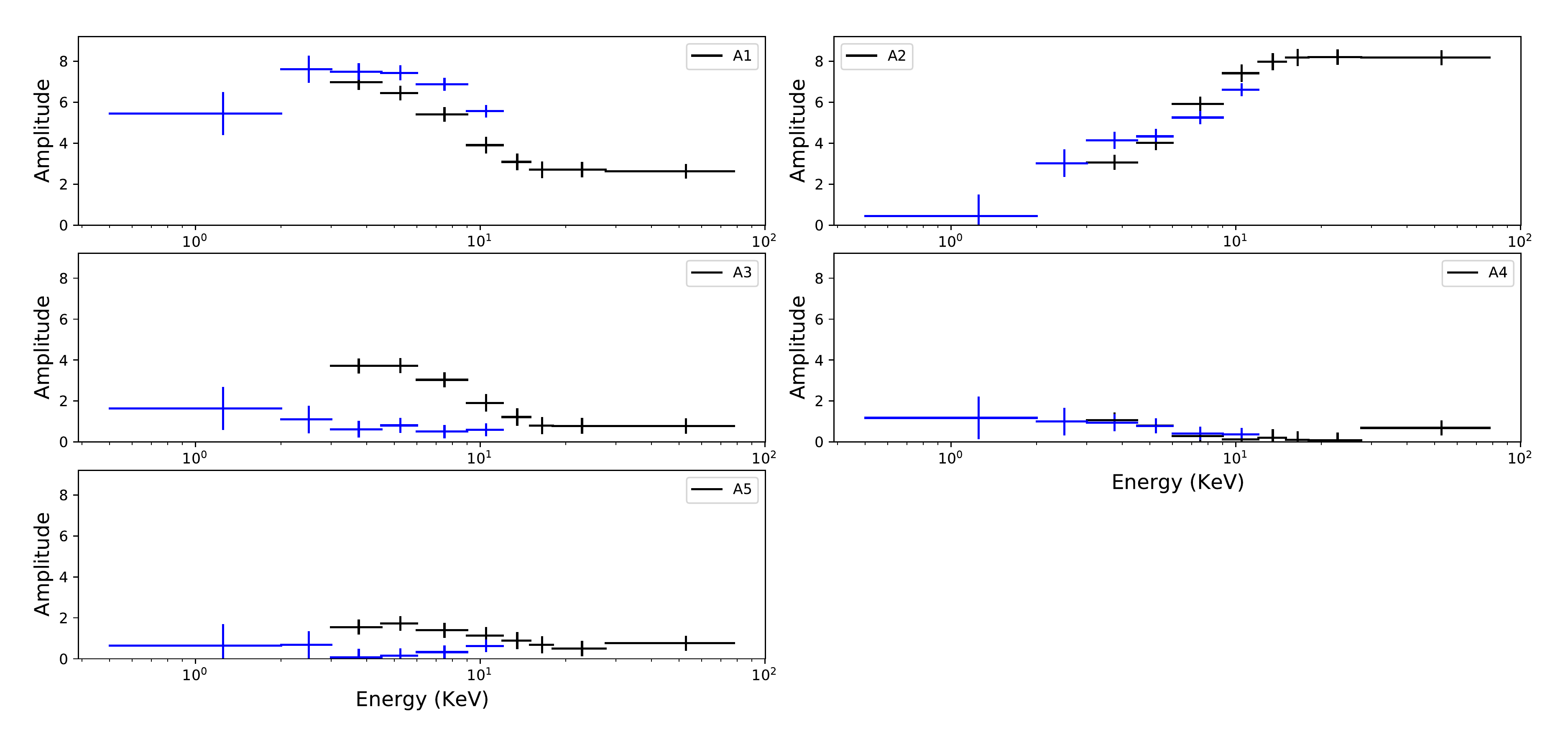}
	\caption{Variation of the harmonic amplitudes for GX301$-$2, the boundaries of the energy ranges are [0.5, 2, 3, 4.5, 6, 9, 12, 15, 18, 27.6 and 78] keV. \xmm data is represented by blue markers and \nustar data is represented by black markers.}
	\label{fig:GX301-2_fourier_variation}
\end{figure*}


\begin{figure*}%
    \centering
    \includegraphics[width=0.47\linewidth]{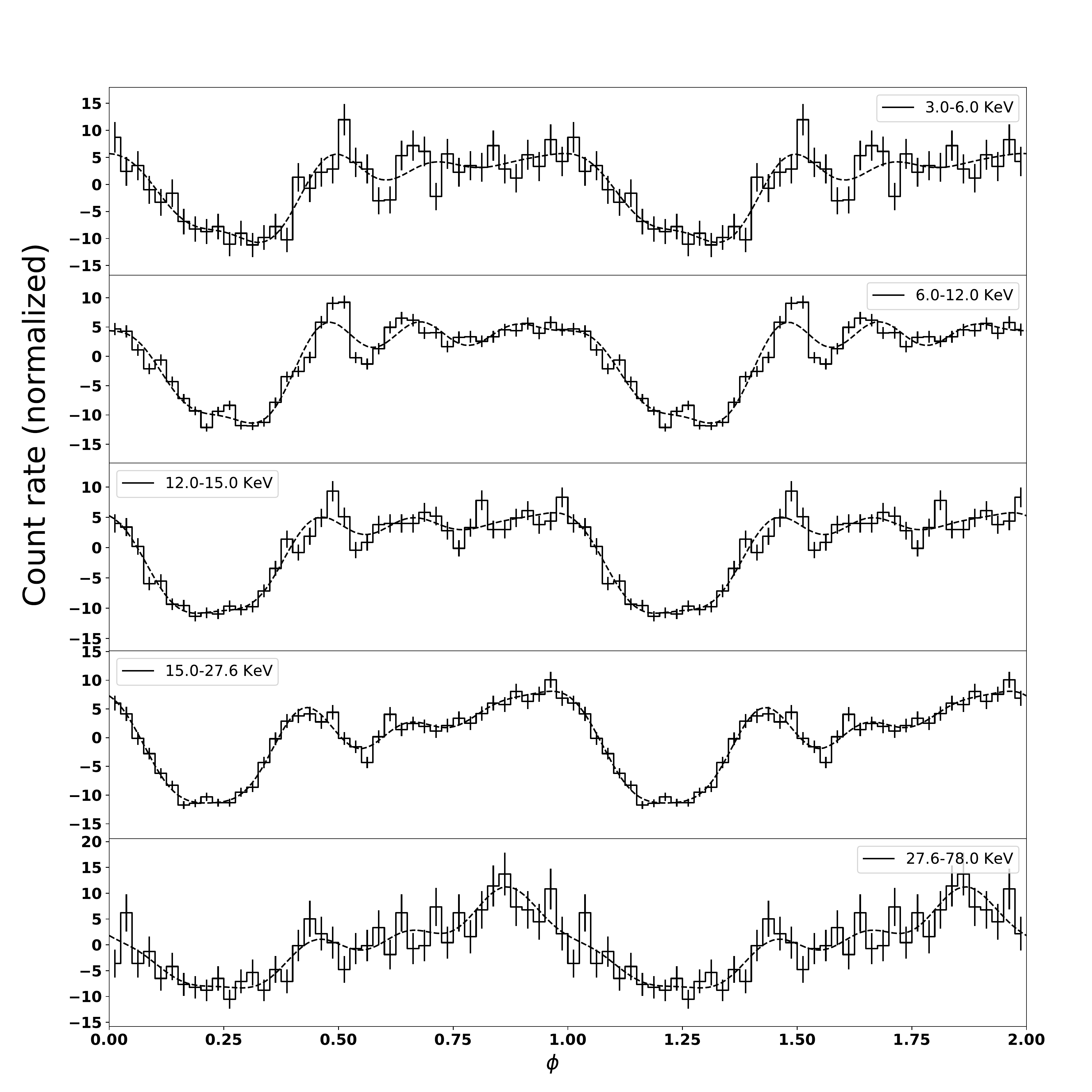} %
    \qquad
    \includegraphics[width=0.47\linewidth]{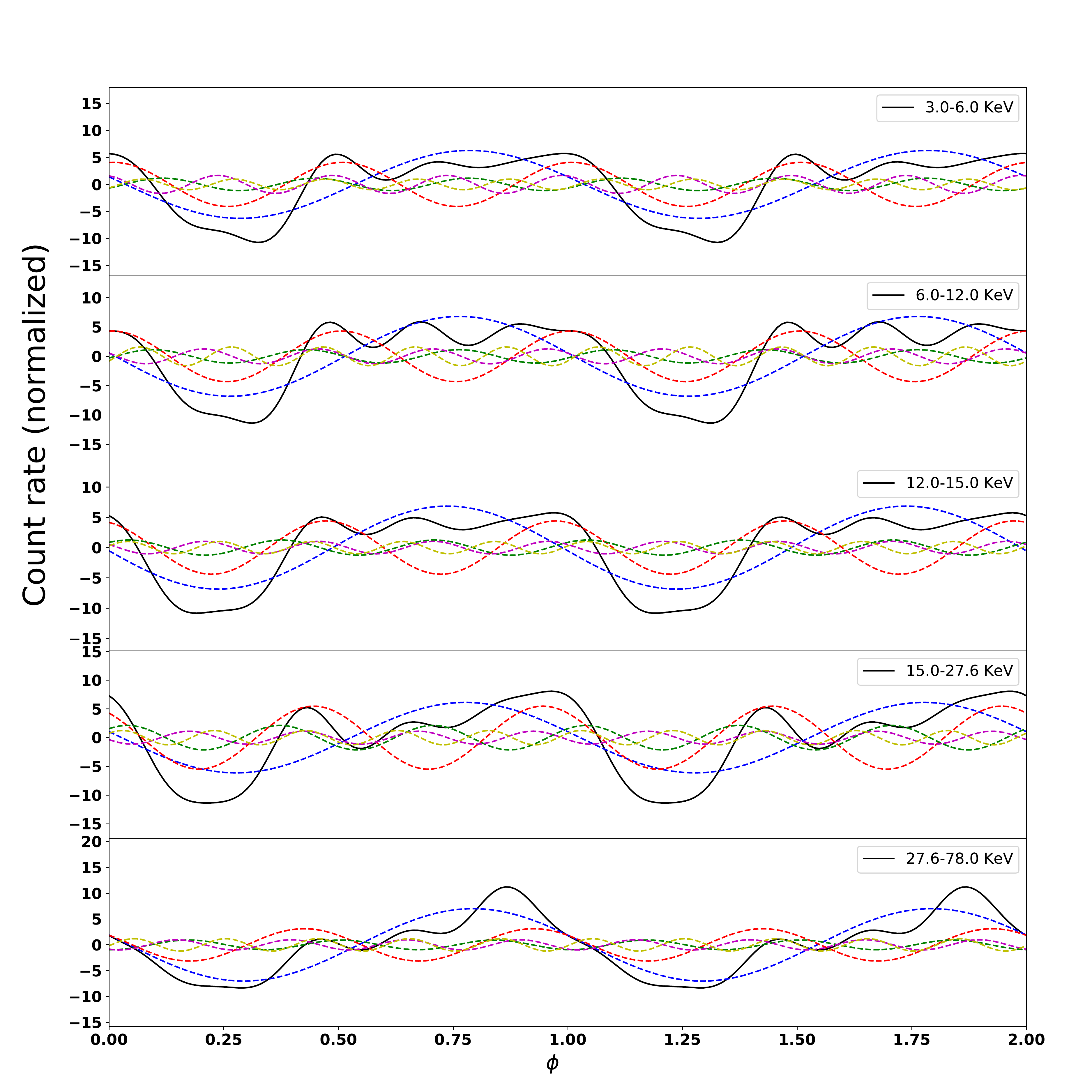} 
    \caption{Left: Normalized pulse profiles of IGR~J16393$-$4643 at different energy ranges based on \nustar data, the measured profiles are represented by solid lines whereas the Fourier fits are represented by dashed lines, the boundaries of the energy ranges are [3, 6, 12, 15, 27.6 and 78] keV. Right: The black solid line represents the overall approximate pulse profile and the colour dashed lines represent the different Fourier components. Blue: 1st, red: 2nd, green: 3rd, purple: 4th and yellow: 5th.}%
    \label{fig:IGR_J16393-4643_NUSTAR_pulse_profiles}%
\end{figure*}

\begin{figure*}
	\centering
	\includegraphics[width=\linewidth]{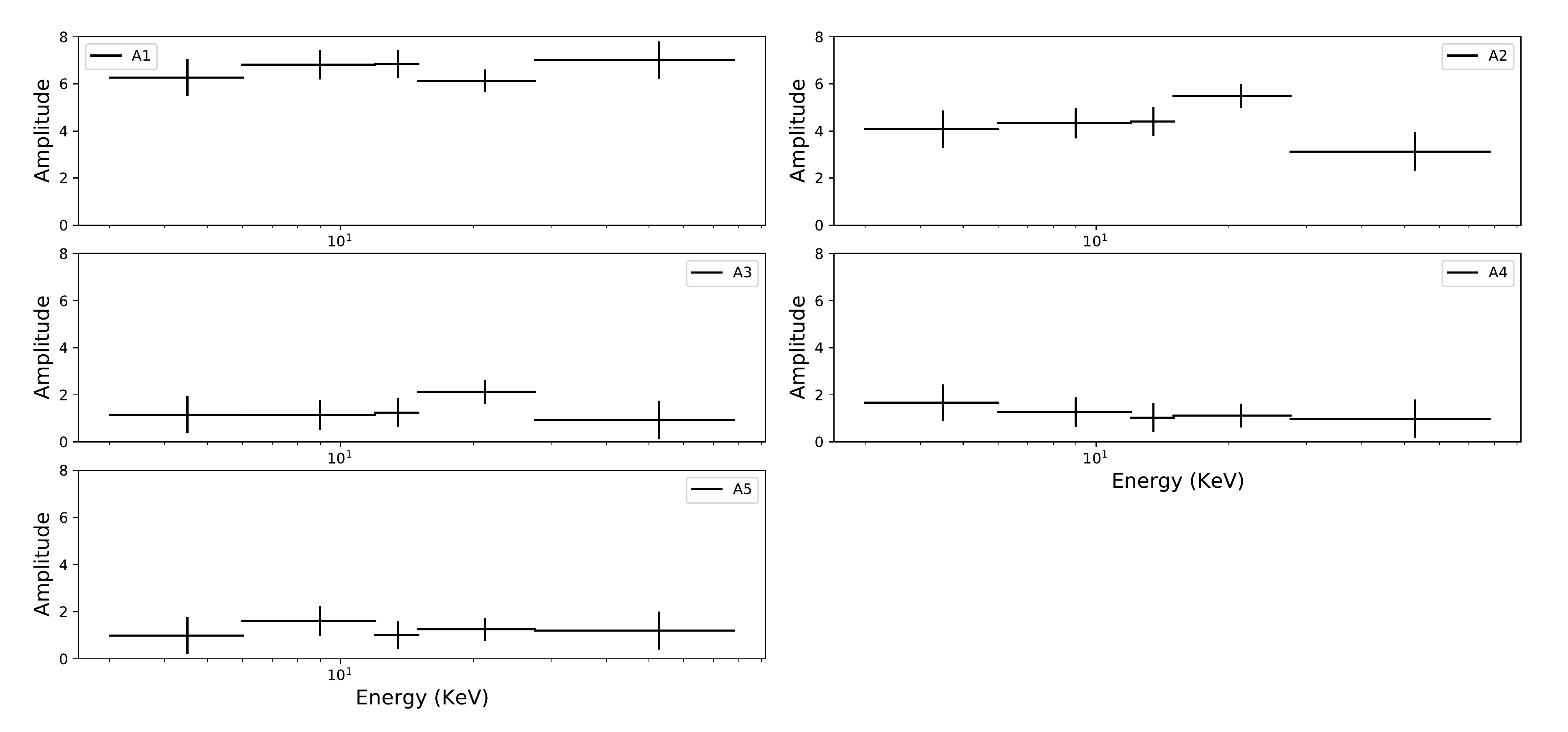}
	\caption{Variation of the harmonic amplitudes for IGR~J16393$-$4643, the boundaries of the energy ranges are [3, 6, 12, 15, 27.6 and 78] keV. Only \nustar data is represented by black markers.}
	\label{fig:IGR_J16393-4643_fourier_variation}
\end{figure*}


\end{document}